\pdfoutput=1

\documentclass[11pt,twoside,a4paper,cmspaper,final,collab]{cms-tdr}
\def\svnVersion{183613}\def\cmsCernNo{2011-170}\def\cmsCernDate{\today}\def\cmsMessage{Submitted to the Journal of High Energy Physics}

\begin{document}\cmsNoteHeader{HIN-10-006}

\hyphenation{had-ron-i-za-tion}
\hyphenation{cal-or-i-me-ter}
\hyphenation{de-vices}
\RCS$Revision: 183612 $
\RCS$HeadURL: svn+ssh://alverson@svn.cern.ch/reps/tdr2/papers/HIN-10-006/trunk/HIN-10-006.tex $
\RCS$Id: HIN-10-006.tex 183612 2013-04-29 18:47:09Z alverson $
\newcommand{\ee}{\Pep\Pem\xspace}
\newcommand{\mumu}{\Pgmp\Pgmm\xspace}
\newcommand{\eexp}[1]{\ensuremath{\text{e}^{#1}}\xspace}

\newcommand{\qqbar}{\cPq\cPaq\xspace}
\newcommand{\QQbar}{\ensuremath{{\cmsSymbolFace{Q}\overline{\cmsSymbolFace{Q}}}}\xspace}
\newcommand{\Jpsi}{\JPsi}
\newcommand{\B}{\PB\xspace}
\newcommand{\D}{\PD\xspace}

\newcommand{\PgUbc}{\ensuremath{\PgU\text{(2S+3S)}}\xspace}
\newcommand{\PgUn}{\ensuremath{\PgU\text{(nS)}}\xspace}

\newcommand{\z}{\cPZ\xspace}

\newcommand{\dndy}{\ensuremath{\mathrm{d}N/\mathrm{d}y}\xspace}
\newcommand{\dnchdy}{\ensuremath{\mathrm{d}N_{\text{ch}}/\mathrm{d}y}\xspace}
\newcommand{\dndeta}{\ensuremath{\mathrm{d}N/\mathrm{d}\eta}\xspace}
\newcommand{\dnchdeta}{\ensuremath{\mathrm{d}N_{\text{ch}}/\mathrm{d}\eta}\xspace}
\newcommand{\dndpt}{\ensuremath{\mathrm{d}N/\mathrm{d}\pt}\xspace}
\newcommand{\dnchdpt}{\ensuremath{\mathrm{d}N_{\text{ch}}/\mathrm{d}\pt}\xspace}
\newcommand{\deta}{\ensuremath{\Delta\eta}\xspace}
\newcommand{\dphi}{\ensuremath{\Delta\phi}\xspace}

\newcommand {\npart}  {\ensuremath{N_{\text{part}}}\xspace}
\newcommand {\ncoll}  {\ensuremath{N_{\text{coll}}}\xspace}

\newcommand{\raa}{\ensuremath{R_{AA}}\xspace}
\newcommand{\taa}{\ensuremath{T_{AA}}\xspace}

\newcommand{\eq}[1]{Eq.~\eqref{#1}\xspace}
\newcommand{\fig}[1]{Fig.~\ref{#1}\xspace}
\newcommand{\tab}[1]{Table~\ref{#1}\xspace}

\newcommand{\pp}{{\ensuremath{\Pp\Pp}}\xspace}
\newcommand{\ppbar}{\ensuremath{\Pp\Pap}\xspace}
\newcommand{\PbPb}{\ensuremath{\text{PbPb}}\xspace}
\newcommand{\AuAu}{\ensuremath{\text{AuAu}}\xspace}

\newcommand{\sqrts}{\ensuremath{\sqrt{s}}\xspace}
\newcommand{\sqrtsnn}{\ensuremath{\sqrt{s_{_{\text{NN}}}}}\xspace}

\newcommand{\mbinv} {\mbox{\ensuremath{\,\text{mb}^\text{$-$1}}}\xspace}
\providecommand{\mubinv} {\mbox{\ensuremath{\,\mu\text{b}^\text{$-$1}}}\xspace}

\providecommand{\CASCADE} {{\textsc{cascade}}\xspace}
\providecommand{\HYDJET} {{\textsc{hydjet}}\xspace}

\cmsNoteHeader{HIN-10-006} 
\title{Suppression of non-prompt
  \JPsi, prompt
  \JPsi, and \PgUa\\
  in PbPb collisions at
  $\sqrt{s_{_{\text{NN}}}}=2.76\TeV$}

\author[cern]{Abdulla Abdulsalam, Jacob Anderson, Andrey Belyaev, Torsten Dahms, Raphael Granier de
  Cassagnac, Michael Gardner, Zhen Hu, Mihee Jo, Hyunchul Kim, Vineet
  Kumar, Nuno Leonardo,  Marco De Mattia, Camelia Mironov, Dong-Ho Moon, Guillermo Rangel, Jorge
  Robles, Ian Shipsey, Prashant Shukla, Catherine Silvestre}

\date{\today}

\abstract{Yields of prompt and non-prompt
  \JPsi, as well as
  \PgUa\ mesons, are measured by the CMS experiment via their
  $\Pgmp\Pgmm$ decays in PbPb and pp collisions at
  $\sqrtsnn = 2.76\TeV$ for
  quarkonium rapidity $|y|<2.4$. Differential cross sections and
  nuclear modification factors are reported as functions of $y$ and
  transverse momentum \pt, as well as collision
  centrality. For prompt
  \JPsi\ with relatively
  high \pt
  ($6.5<\pt<30\GeVc$),
  a strong, centrality-dependent suppression is observed in PbPb
  collisions, compared to the yield in pp collisions scaled by the
  number of inelastic nucleon-nucleon collisions. In the same
  kinematic range, a suppression of non-prompt
  $\JPsi$, which is
  sensitive to the in-medium $\cPqb$-quark energy loss, is
  measured for the first time. Also the low-\pt
  \PgUa\ mesons are suppressed in PbPb collisions.}

\hypersetup{%
  pdfauthor={CMS Collaboration},%
  pdftitle={Suppression of non-prompt J/psi, prompt J/psi, and Y(1S)
    in PbPb collisions at sqrt(sNN) = 2.76 TeV},%
  pdfsubject={CMS},%
  pdfkeywords={CMS, physics, heavy ions, dimuons, quarkonia}}

\maketitle 

\section{Introduction}
\label{sec:introduction}
At large energy densities and high temperatures, strongly interacting
matter consists of a deconfined and chirally-symmetric system of
quarks and gluons~\cite{Karsch:2003jg}. This state, often referred to
as ``quark-gluon plasma'' (QGP)~\cite{Shuryak:1977ut}, constitutes the
main object of the studies performed with relativistic heavy-ion
collisions~\cite{Arsene:2004fa,Back:2004je,Adcox:2004mh,Adams:2005dq}.

The formation of a QGP in high-energy nuclear collisions can be
evidenced in a variety of ways. One of its most striking expected
signatures is the suppression of quarkonium
states~\cite{Matsui:1986dk}, both of the charmonium (\Jpsi, $\psi'$,
$\chi_c$, etc.) and the bottomonium ($\PgU\text{(1S,\,2S,\,3S)}$,
$\chi_b$, etc.) families.  This is thought to be a direct effect of
deconfinement, when the binding potential between the constituents of
a quarkonium state, a heavy quark and its antiquark, is screened by
the colour charges of the surrounding light quarks and gluons.  The
suppression is predicted to occur above the critical temperature of
the medium ($T_c$) and depends on the \QQbar binding energy. Since the
\PgUa\ is the most tightly bound state among all quarkonia, it is
expected to be the one with the highest dissociation
temperature. Examples of dissociation temperatures are given in
Ref.~\cite{Mocsy:2007jz}: $T_{\text{dissoc}}\sim\!1\,T_c$, $1.2\,T_c,$
and $2\,T_c$ for the \PgUc, \PgUb, and \PgUa, respectively. Similarly,
in the charmonium family the dissociation temperatures are
$\leq1\,T_c$ and $1.2\,T_c$ for the $\psi'$ and \Jpsi,
respectively. However, there are further possible changes to the
quarkonium production in heavy-ion collisions. On the one hand,
modifications to the parton distribution functions inside the nucleus
(shadowing) and other cold-nuclear-matter effects can reduce the
production of quarkonia without the presence of a
QGP~\cite{Vogt:2010aa,Zhao:2011cv}. On the other hand, the large
number of heavy quarks produced in heavy-ion collisions, in particular
at the energies accessible by the Large Hadron Collider (LHC), could
lead to an increased production of quarkonia via statistical
recombination~\cite{Zhao:2010nk,Andronic:2006ky,Capella:2007jv,Thews:2005vj,Yan:2006ve,Grandchamp:2005yw}.

Charmonium studies in heavy-ion collisions have been carried out for
25 years, first at the Super Proton Synchrotron (SPS) by the
NA38~\cite{Baglin:1994ui},
NA50~\cite{Alessandro:2004ap,Alessandro:2006ju}, and
NA60~\cite{Arnaldi:2007zz} fixed-target experiments at 17.3--19.3\GeV
centre-of-mass energy per nucleon pair (\sqrtsnn), and then at the
Relativistic Heavy Ion Collider (RHIC) by the PHENIX experiment at
\sqrtsnn = 200\GeV~\cite{Adare:2006ns}. In all cases, \Jpsi
suppression was observed in the most central collisions. At the SPS,
the suppression of the $\psi'$ meson was also
measured~\cite{Alessandro:2006ju}. Experimentally, the suppression is
quantified by the ratio of the yield measured in heavy-ion collisions
and a reference. At RHIC, the reference was provided by the properly
scaled yield measured in \pp collisions. Such a ratio is called the
nuclear modification factor, \raa. In the absence of modifications,
one would expect \raa = 1 for hard processes, which scale with the
number of inelastic nucleon-nucleon collisions. For bottomonia, the
production cross section is too small at RHIC to make definitive
statements~\cite{Abelev:2010am}. With the higher energy and luminosity
available at the LHC, new studies for charmonia and bottomonia have
become possible: (i) ATLAS has reported a suppression of inclusive
\Jpsi with high transverse momenta \pt in central \PbPb collisions
compared to peripheral collisions at \sqrtsnn =
2.76\TeV~\cite{Aad:2010px}; (ii) ALICE has measured the \raa for
inclusive \Jpsi with low \pt and sees no centrality dependence of the
\Jpsi suppression~\cite{Abelev:2012rv}; (iii) a suppression of the
excited \PgU\ states with respect to the ground state has been
observed in \PbPb collisions at \sqrtsnn = 2.76\TeV compared to \pp
collisions at the same centre-of-mass energy by the Compact Muon
Solenoid (CMS) collaboration~\cite{Chatrchyan:2011pe}.

At LHC energies, the inclusive \Jpsi yield contains a significant
non-prompt contribution from \cPqb-hadron decays~\cite{Aaij:2011jh,
  Khachatryan:2010yr, Aad:2011sp}. Owing to the long lifetime of the
\cPqb\ hadrons (${\cal O}(500)\mum/c$), compared to the QGP lifetime
(${\cal O}(10)\,\text{fm}/c$), this contribution should not suffer
from colour screening, but instead may reflect the \cPqb-quark energy
loss in the medium. Such energy loss would lead to a reduction of the
\cPqb-hadron yield at high \pt in \PbPb collisions compared to the
binary-collision-scaled \pp yield. In heavy-ion collisions, only
indirect measurements of this effect exist, through single electrons
from semileptonic open heavy-flavour
decays~\cite{Adare:2006nq,Abelev:2006db,Abelev:2006dbErratum}; to
date, the contributions from charm and bottom have not been
disentangled. The importance of an unambiguous measurement of open
bottom flavour is driven by the lack of knowledge regarding key
features of the dynamics of parton energy loss in the QGP, such as its
colour-charge and parton-mass
dependencies~\cite{Dokshitzer:2001zm,Armesto:2005iq} and the relative
role of radiative and collisional energy
loss~\cite{Peigne:2008nd}. CMS is well equipped to perform direct
measurements of \cPqb-hadron production in heavy-ion collisions by
identifying non-prompt \Jpsi from \cPqb-hadron decays via the
reconstruction of secondary \mumu vertices.

The paper is organised as follows: the CMS detector is briefly
described in Section~\ref{sec:cms}. Section~\ref{sec:data-selection}
presents the data collection, the \PbPb event selection, the muon
reconstruction and selection, and the Monte Carlo (MC)
simulations. The methods employed for signal extraction are detailed
in Section~\ref{sec:signal-extraction}. Section~\ref{sec:acceff}
describes the acceptance correction factors and the estimation of the
reconstruction efficiencies. The \pp baseline measurements are
summarized in Section~\ref{sec:ppRef}. The results are presented in
Section~\ref{sec:results}, followed by their discussion in
Section~\ref{sec:discussion}.

\section{The CMS Detector}
\label{sec:cms}
A detailed description of the CMS experiment can be found in
Ref.~\cite{Adolphi:2008zzk}. The central feature of the CMS apparatus
is a superconducting solenoid of 6\,m internal diameter. Within the
field volume are the silicon tracker, the crystal electromagnetic
calorimeter, and the brass/scintillator hadron calorimeter.

CMS uses a right-handed coordinate system, with the origin at the
nominal interaction point, the $x$ axis pointing to the centre of the
LHC, the $y$ axis pointing up (perpendicular to the LHC plane), and
the $z$ axis along the counterclockwise-beam direction. The polar
angle $\theta$ is measured from the positive $z$ axis and the
azimuthal angle $\phi$ is measured in the $x$-$y$ plane. The
pseudorapidity is defined as $\eta = -\ln[\tan(\theta/2)]$.

Muons are detected in the interval $|\eta|< 2.4$ by gaseous detectors
made of three technologies: drift tubes, cathode strip chambers, and
resistive plate chambers, embedded in the steel return yoke.  The
silicon tracker is composed of pixel detectors (three barrel layers
and two forward disks on either side of the detector, made of
66~million $100\times150\mum^2$ pixels) followed by microstrip
detectors (ten barrel layers plus three inner disks and nine forward
disks on either side of the detector, with strips of pitch between 80
and 180\mum). The transverse momentum of muons matched to
reconstructed tracks is measured with a resolution better than
${\sim}1.5$\% for \pt smaller than 100\GeVc~\cite{TRK-10-004}.  The
good resolution is the result of the 3.8\,T magnetic field and the
high granularity of the silicon tracker.

In addition, CMS has extensive forward calorimetry, including two
steel/quartz-fibre Che\-ren\-kov forward hadron (HF) calorimeters,
which cover the pseudorapidity range $2.9 < |\eta|< 5.2$. These
detectors are used in the present analysis for the event selection and
\PbPb collision centrality determination, as described in the next
section. Two beam scintillator counters (BSC) are installed on the
inner side of the HF calorimeters for triggering and beam-halo
rejection.

\section{Data Selection}
\label{sec:data-selection}
\subsection{Event Selection}
\label{sec:event-sel}

Inelastic hadronic \PbPb collisions are selected using information
from the BSC and HF calorimeters, in coincidence with a bunch crossing
identified by the beam pick-up (one on each side of the interaction
point)~\cite{Adolphi:2008zzk}.  Events are further filtered offline by
requiring a reconstructed primary vertex based on at least two tracks,
and at least 3 towers on each HF with an energy deposit of more than
3\GeV per tower. These criteria reduce contributions from single-beam
interactions with the environment (e.g. beam-gas collisions and
collisions of the beam halo with the beam pipe), ultra-peripheral
electromagnetic interactions, and cosmic-ray muons. A small fraction
of the most peripheral PbPb collisions are not selected by these
\emph{minimum-bias} requirements, which accept $(97\pm3)$\% of the
inelastic hadronic cross section~\cite{Chatrchyan:2011sx}. A sample
corresponding to 55.7 M minimum-bias events passes all these
filters. Assuming an inelastic \PbPb cross section of $\sigma_{\PbPb}
= 7.65$~b~\cite{Chatrchyan:2011sx}, this sample corresponds to an
integrated luminosity of $\lumi_{\text{int}} = 7.28\mubinv$. This
value is only mentioned for illustration purposes; the final results
are normalized to the number of minimum-bias events.

The measurements reported here are based on dimuon events triggered by
the \Lone (L1) trigger, a hardware-based trigger that uses information
from the muon detectors. The CMS detector is also equipped with a
software-based high-level trigger (HLT). However, no further
requirements at the HLT level have been applied to the L1 muon objects
used for this analysis.

\begin{figure}[hbtp]
  \begin{center}
   \includegraphics[width=0.5\linewidth]{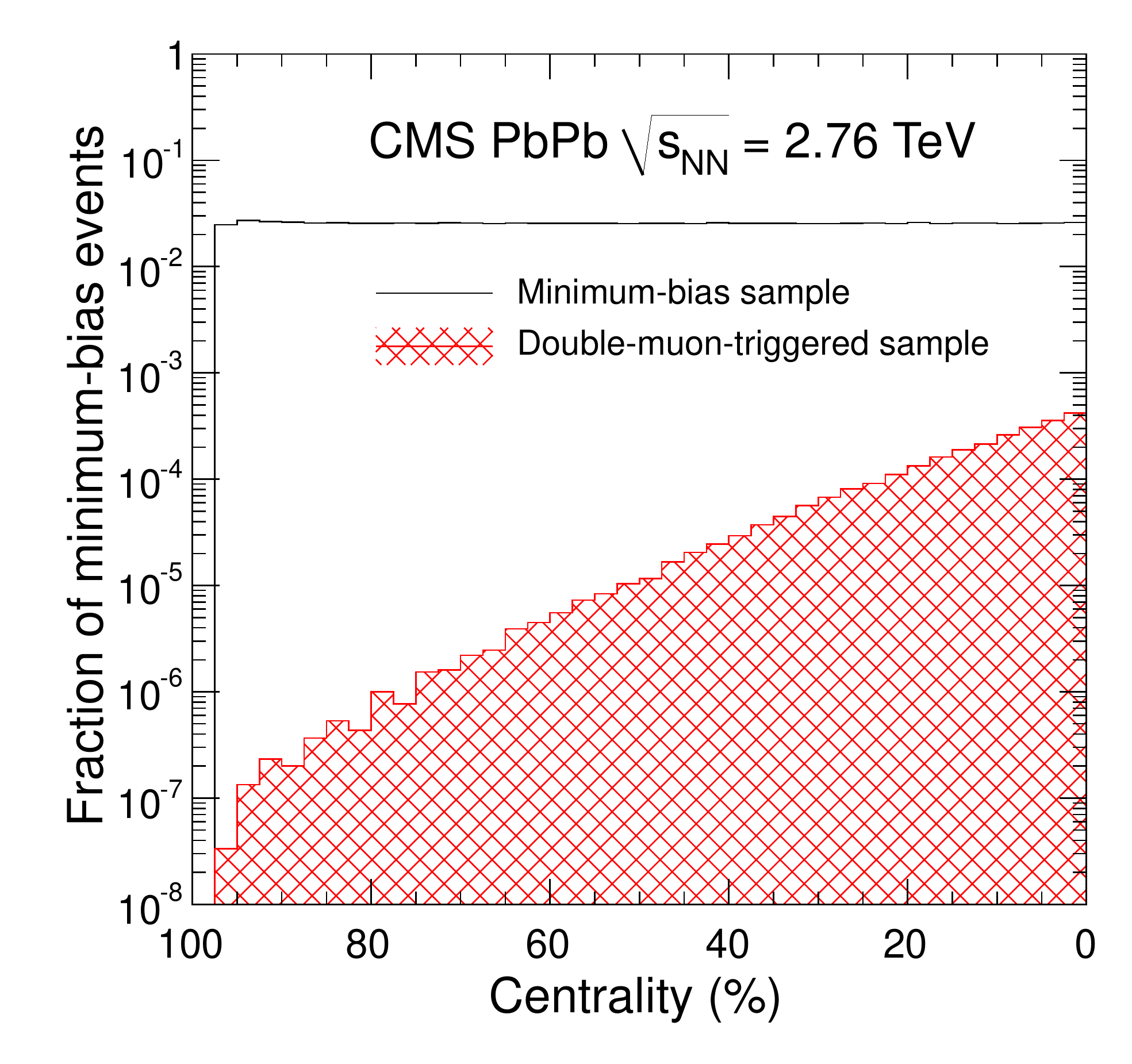}
   \caption{Centrality distribution of the minimum-bias sample (solid
     black line) overlaid with the double-muon triggered sample
     (hashed red) in bins of 2.5\%.}
    \label{fig:centrality}
  \end{center}
\end{figure}

The event centrality distribution of minimum-bias events is compared
to events selected by the double-muon trigger in
\fig{fig:centrality}. The centrality variable is defined as the
fraction of the total cross section, starting at 0\% for the most
central collisions. This fraction is determined from the distribution
of total energy measured in both HF
calorimeters~\cite{Chatrchyan:2011pb}. Using a Glauber-model
calculation as described in Ref.~\cite{Chatrchyan:2011sx}, one can
estimate variables related to the centrality, such as the number of
nucleons participating in the collisions (\npart) and the nuclear
overlap function (\taa), which is equal to the number of elementary
nucleon-nucleon (NN) binary collisions divided by the elementary NN
cross section and can be interpreted as the NN equivalent integrated
luminosity per heavy ion collision, at a given
centrality~\cite{Miller:2007ri}. The values of these variables are
presented in \tab{tab:glauber} for the centrality bins used in this
analysis. The double-muon-triggered events are more frequent in
central collisions since the main physics processes that generate
high-\pt muon pairs scale with the number of inelastic nucleon-nucleon
collisions. In the following, \npart will be the variable used to show
the centrality dependence of the measurements.

\begin{table}[htbp]
  \begin{center}
    \caption{Average and root-mean-square (RMS) values of the number of
      participating nucleons (\npart) and of the nuclear overlap function
      (\taa) for the centrality bins used in this
      analysis~\cite{Chatrchyan:2011sx}.}
    \label{tab:glauber}
    \begin{tabular}{rr@{.}lr@{.}lr@{.}lr@{.}lr@{.}lr@{.}lr@{.}lr@{.}l}
      \hline\vspace{0.1em}
      ~ & \multicolumn{4}{c}{$\npart$} &
      \multicolumn{4}{c}{$\taa$ (mb$^{-1}$)}\\
      Centrality (\%) & \multicolumn{2}{c}{Mean} & \multicolumn{2}{c}{RMS} & \multicolumn{2}{c}{Mean} & \multicolumn{2}{c}{RMS} \\\hline
      0--10	& 355&4 & 33&3 & 23&19 & 3&77 \\
      10--20	& 261&4 & 30&4 & 14&48 & 2&86 \\
      20--30	& 187&2 & 23&4 & 8&78  & 1&94 \\
      30--40	& 130&0 & 17&9 & 5&09  & 1&27 \\
      40--50	& 86&3  & 13&6 & 2&75  & 0&80 \\
      50--100	& 22&1  & 19&3 & 0&47  & 0&54 \\\hline
      0--20	& 308&4 & 56&8 & 18&83 & 5&49 \\
      20--100	& 64&2  & 63&0 & 2&37  & 3&05 \\\hline
      0--100	& 113&1 &115&6 & 5&66  & 7&54 \\\hline
    \end{tabular}
  \end{center}
\end{table}

Simulated MC events are used to tune the muon selection criteria, to
compute the acceptance and efficiency corrections, and to obtain
templates of the decay length distribution of \Jpsi from b-hadron
decays.  For the acceptance corrections described in
Section~\ref{sec:acc}, three separate MC samples, generated over full
phase space, are used: prompt \Jpsi, \Jpsi from \cPqb-hadron decays,
and \PgUa. Prompt \Jpsi and \PgUa\ are produced using \PYTHIA
6.424~\cite{Sjostrand:2006za} at \sqrts = 2.76\TeV, which generates
events based on the leading-order colour-singlet and colour-octet
mechanisms, with non-relativistic quantum chromodynamics (QCD) matrix
elements tuned~\cite{Bargiotti:2007zz} by comparison with CDF data
~\cite{Acosta:2004yw}. The colour-octet states undergo a shower
evolution. For the non-prompt \Jpsi studies, the \cPqb-hadron events
are produced with \PYTHIA in generic QCD 2$\rightarrow$2 processes. In
all three samples, the \Jpsi or \PgUa\ decay is simulated using the
\EVTGEN~\cite{Lange:2001uf} package. Prompt \Jpsi and \PgUa\ are
simulated assuming unpolarized production, while the non-prompt \Jpsi
polarization is determined by the sum of the exclusive states
generated by \EVTGEN. Final-state bremsstrahlung is implemented using
\PHOTOS~\cite{Barberio:1993qi}.

For some MC simulation studies, in particular the efficiency
corrections described in Section~\ref{sec:eff}, the detector response
to each \PYTHIA signal event is simulated with
\GEANTfour~\cite{Agostinelli:2002hh} and then embedded in a realistic
heavy-ion background event. The background events are produced with
the \HYDJET event generator~\cite{Lokhtin:2005px} and then simulated
with \GEANTfour as well. The \HYDJET parameters were tuned to
reproduce the particle multiplicities at all centralities seen in
data. The embedding is done at the level of detector hits and requires
that the signal and background production vertices match. The embedded
event is then processed through the trigger emulation and the full
event reconstruction chain. Collision data are used to validate the
efficiencies evaluated using MC simulations, as discussed in
Section~\ref{sec:eff}.

\subsection{Muon Selection}
\label{sec:muon-sel}

The muon offline reconstruction algorithm starts by reconstructing
tracks in the muon detectors, called \emph{standalone muons}. These
tracks are then matched to tracks reconstructed in the silicon tracker
by means of an algorithm optimized for the heavy-ion
environment~\cite{D'Enterria:2007xr,Roland:2006kz}. The final muon
objects, called \emph{global muons}, result from a global fit of the
standalone muon and tracker tracks. These are used to obtain the
results presented in this paper.

\begin{figure}[htbp]
  \centering
  \includegraphics[width=0.5\linewidth]{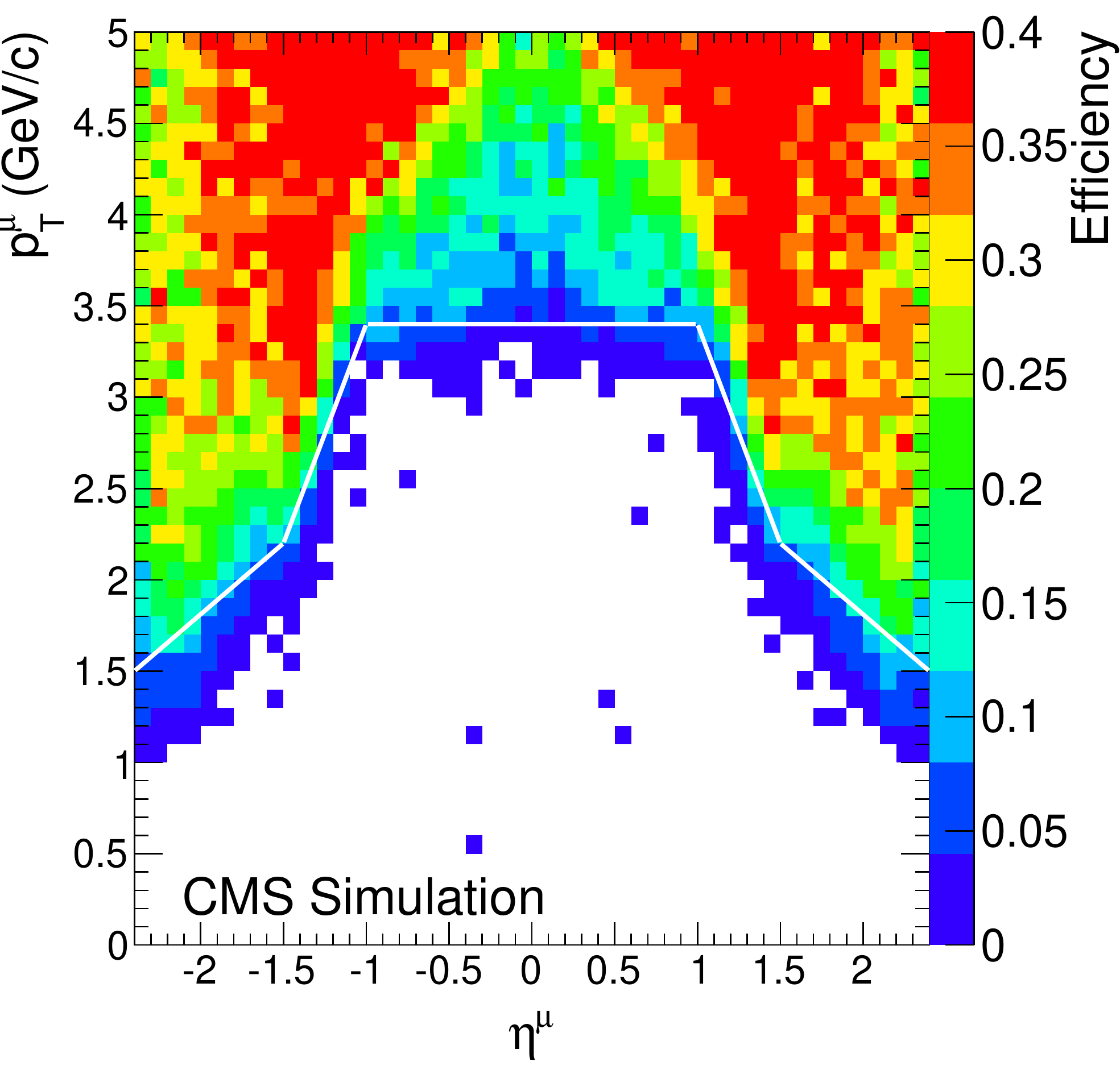}
  \caption{Reconstruction efficiency of global muons in the ($\eta^{\mu}$,
    $\pt^{\mu}$) space, illustrating the lower limits (white lines) of
    what is considered a detectable single muon for the \Jpsi
    analysis.}
  \label{fig:muPtEtaDoable}
\end{figure}

In \fig{fig:muPtEtaDoable}, the single-muon reconstruction efficiency
from MC simulations is presented as a function of the muon $\pt^\mu$
and $\eta^{\mu}$. The reconstruction efficiency is defined as the
number of all reconstructed global muons divided by the number of
generated muons in a given ($\eta^{\mu}$, $\pt^{\mu}$) bin. It takes
into account detector resolution effects, \ie reconstructed \pt and
$\eta$ values are used in the numerator and generated \pt and $\eta$
values in the denominator. To obtain a clear separation between
acceptance and efficiency corrections, a \emph{detectable} single-muon
acceptance is defined in the ($\eta^{\mu}$, $\pt^{\mu}$) space. For
the \Jpsi analysis this separation is defined by the contour that
roughly matches a global muon reconstruction efficiency of 10\%,
indicated by the white lines superimposed in \fig{fig:muPtEtaDoable},
which are described by the conditions
  \begin{align}\label{eq:singleMuonAcc}
    \pt^{\mu} &> 3.4\GeVc &\text{ for } |\eta^{\mu}| < 1.0, \notag\\
    \pt^{\mu} &> (5.8 - 2.4\times|\eta^{\mu}|)\GeVc &\text{ for } 1.0 < |\eta^{\mu}| < 1.5,\\
    \pt^{\mu} &> (3.4 - 0.78\times|\eta^{\mu}|)\GeVc &\text{ for } 1.5 < |\eta^{\mu}| < 2.4. \notag
  \end{align}
Muons failing these conditions are accounted for in the acceptance
corrections discussed in Section~\ref{sec:acc}. Muons that pass this
acceptance requirement can still fail to pass the trigger, track
reconstruction, or muon selection requirements. These losses are
accounted for by the efficiency corrections discussed in
Section~\ref{sec:eff}.

For the \PgUa\ analysis, where the signal-to-background ratio is less
favourable than in the \Jpsi mass range, a higher $\pt^{\mu}$ is
required than for the \Jpsi analysis,
\begin{equation}
  \label{eq:singleMuonAccUps}
    \pt^{\mu}>4\GeVc,
\end{equation}
independent of $\eta^{\mu}$.

Various additional global muon selection criteria are studied in MC
simulations. The MC distributions of the \Jpsi decay muons are in
agreement with those from data to better than 2\%, which is within the
systematic uncertainty of the data/MC efficiency ratio
(Section~\ref{sec:eff}). The transverse (longitudinal) distance of
closest approach to the measured vertex is required to be less than 3
(15)\,cm. Tracks are only kept if they have 11 or more hits in the
silicon tracker, and the $\chi^2$ per degree of freedom of the global
(inner) track fit is less than 20 (4). The $\chi^2$ probability of the
two tracks originating from a common vertex is required to be larger
than 1\%. From MC simulations we find that these criteria result in a
6.6\%, 5.1\%, and 3.9\% loss of prompt \Jpsi, non-prompt \Jpsi, and
\PgUa\ events, respectively, given two reconstructed tracks associated
with the double muon trigger.

\section{Signal Extraction}
\label{sec:signal-extraction}
\subsection{\texorpdfstring{\Jpsi}{J/psi} Analysis}
\label{sec:jpsi}

\subsubsection{Inclusive \texorpdfstring{\Jpsi}{J/psi}}
\label{sec:inclJpsi}

\begin{figure}[htbp]
  \begin{center}
    \includegraphics[width=0.5\linewidth]{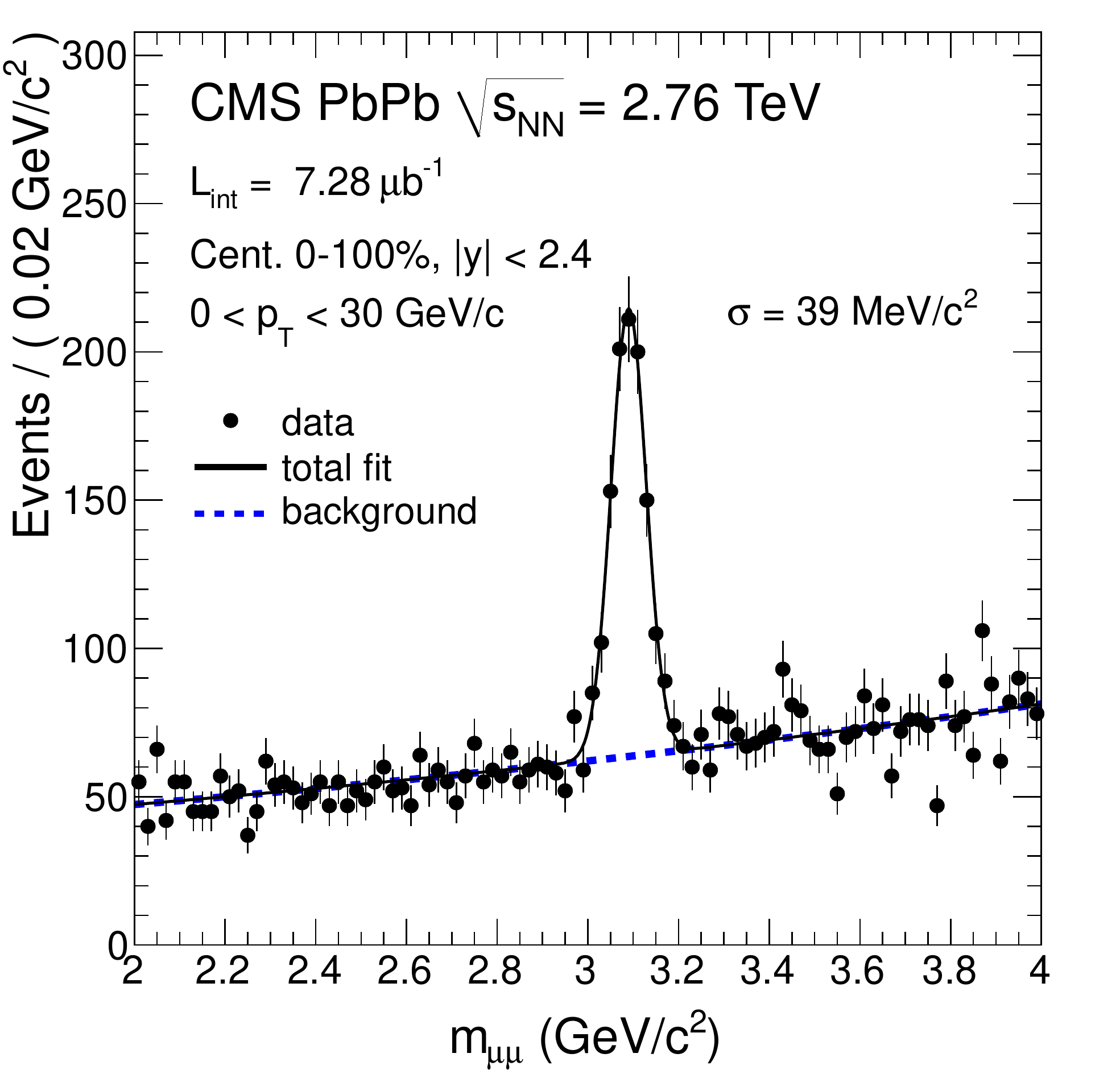}
    \caption{Invariant-mass spectrum of \mumu pairs (black circles)
      with $|y|<2.4$ and \mbox{$0<\pt<30\GeVc$} integrated over
      centrality. The fit to the data with the functions discussed in
      the text is shown as the black line. The dashed blue line shows
      the fitted background contribution.}
    \label{fig:jpsi_invmass}
  \end{center}
\end{figure}

The \mumu pair invariant-mass $m_{\Pgm\Pgm}$ spectrum is shown in
\fig{fig:jpsi_invmass} in the region $2< m_{\Pgm\Pgm}<4\GeVcc$ for muon
pairs with $0 < \pt < 30\GeVc$ and rapidity $|y|<2.4$, after applying
the single-muon quality requirements. No minimum pair-\pt requirement
is applied explicitly. However, the CMS acceptance for \mumu pairs in
this mass range requires a minimum \pt that is strongly $y$-dependent
and is $\approx\!6.5\GeVc$ at $y=0$. The black curve in
\fig{fig:jpsi_invmass} represents an unbinned maximum likelihood fit
to the \mumu pair spectrum, with the signal described by the sum of a
Gaussian and a Crystal Ball function, with common mean $m_0$ and width
$\sigma$, and the background described by an exponential. The Crystal
Ball function $f_\text{CB}(m)$ combines a Gaussian core and a
power-law tail with an exponent $n$ to account for energy loss due to
final-state photon radiation,
\begin{equation}
  \label{eq:cb}
    f_{\text CB}(m) = \left\{
      \begin{array}{ll}
        \frac{N}{\sqrt{2\pi}\sigma}\exp\Bigl(-\frac{(m-m_0)^2}{2\sigma^2}\Bigr),& \quad \text{for}~\frac{m-m_0}{\sigma}>-\alpha;\\
        \frac{N}{\sqrt{2\pi}\sigma}\Bigl(\!\frac{n}{\lvert\alpha\rvert}\!\Bigr)^n\exp\Bigl(-\frac{\lvert\alpha\rvert^2}{2}\Bigr)\Bigl(\frac{n}{\lvert\alpha\rvert}-\lvert\alpha\lvert-\frac{m-m_0}{\sigma}\Bigr)^{-n},& \quad \text{for}~\frac{m-m_0}{\sigma}\leq-\alpha.
      \end{array}\right.
\end{equation}
The parameter $\alpha$ defines the transition between the Gaussian and
the power-law functions. The fit function has eight free parameters;
in addition to the five parameters used in \eq{eq:cb}, one parameter
is the fraction of the Gaussian contribution to the total signal yield
(typically $\approx\!0.47$) and two parameters define the
normalization and the slope of the exponential background. The
fitted mean value, $m_0 = (3.090\pm0.002)\GeVcc$, is 0.2\% below the
PDG value of $m_{\Jpsi} = 3.097\GeVcc$~\cite{Nakamura:2010zzi} because
of slight momentum scale biases in the data reconstruction; the width
is $\sigma = (39\pm2)\MeVcc$, consistent with MC expectations. The
number of inclusive \Jpsi mesons obtained by the fit is $734\pm 54$.

The analysis is performed in bins of the \Jpsi meson \pt and $y$, as
well as in bins of event centrality. Integrating over all centrality
(0--100\%) and \pt ($6.5<\pt<30\GeVc$) the rapidity bins are
\begin{equation*}
    |y|<1.2,~1.2<|y|<1.6,~\text{and}~1.6<|y|<2.4.
\end{equation*}
For the two forward bins, the CMS acceptance extends to lower \pt, so
results are also presented for the bins
\begin{equation*}
    1.2<|y|<1.6~\text{and}~5.5<\pt<30\GeVc,~\text{as well as}~1.6<|y|<2.4~\text{and}~3<\pt<30\GeVc.
\end{equation*}
These values allow a better comparison with the low-\pt measurements
of the ALICE experiment, which has acceptance for \Jpsi with
$\pt>0\GeVc$ for the rapidity intervals $|y|<0.9$ and $2.4<y<4.0$, in
the electron and muon decay channels,
respectively~\cite{Alessandro:2006yt}.

Integrating over all centrality (0--100\%) and rapidity ($|y|<2.4$)
the \pt bins are
\begin{equation*}
    6.5<\pt<10\GeVc~\text{and}~10<\pt<30\GeVc.
\end{equation*}

Integrating over the \pt range $6.5<\pt<30\GeVc$ and rapidity
$|y|<2.4$, the centrality bins are: 0--10\%, 10--20\%, 20--30\%,
30--40\%, 40--50\%, and 50--100\%.

The unbinned maximum likelihood fit with the sum of Crystal Ball and
Gaussian functions is performed in each of these bins. Because of the
small sample size, the parameters of the signal shape are determined
for each rapidity and \pt interval, integrated over centrality, as the
dominant effect on the mass shape is the \pt- and rapidity-dependent
mass resolution. As a function of rapidity, the width of the Crystal
Ball function varies from 24\MeVcc ($|y|<1.2$) to 51\MeVcc
($1.6<|y|<2.4$), for the \pt range $6.5<\pt<30\GeVc$. As a function of
\pt, the width changes from 39\MeVcc ($6.5<\pt<10\GeVc$) to 23\MeVcc
($10<\pt<30\GeVc$), when integrated over rapidity. The values are then
fixed for the finer centrality bins. The background shape is allowed
to vary in each bin. The raw yields of inclusive \Jpsi are listed in
\tab{tab:inclyields} of Appendix~\ref{app:datatables}.

\subsubsection{Prompt and Non-prompt \texorpdfstring{\Jpsi}{J/psi}}
\label{sec:promptJpsi}

The identification of \Jpsi mesons coming from \cPqb-hadron decays
relies on the measurement of a secondary \mumu vertex displaced from
the primary collision vertex. The displacement vector between the
\mumu vertex and the primary vertex $\vec{r}$ is measured in the plane
transverse to the beam direction. The most probable transverse
\cPqb-hadron decay length in the laboratory
frame~\cite{Buskulic:1993vi,Buskulic:1993viErratum} is calculated as
\begin{equation}
  \label{eq:lxy}
    L_{xy} = \frac{\hat{u}^TS^{-1}\vec{r}}{\hat{u}^TS^{-1}\hat{u}}\,,
\end{equation}
where $\hat{u}$ is the unit vector in the direction of the \Jpsi meson
$\vec{\pt}$ and $S^{-1}$ is the inverse of the sum of the primary and
secondary vertex covariance matrices. From $L_{xy}$ the pseudo-proper
decay length $\ell_{\Jpsi} = L_{xy}\, m_{\Jpsi}/\pt$ is computed as an
estimate of the \cPqb-hadron decay length. The pseudo-proper decay
length is measured with a resolution of $\sim\!35\mum$.

To measure the fraction of non-prompt \Jpsi, the invariant-mass
spectrum of \mumu pairs and their $\ell_{\Jpsi}$ distribution are
fitted simultaneously using a two-dimensional unbinned
maximum-likelihood fit in bins of \pt, rapidity, and centrality with
the fraction of non-prompt \Jpsi as a free parameter. The fitting
procedure is similar to the one used in the pp analysis at \sqrts =
7\TeV~\cite{Khachatryan:2010yr}. The differences are: (i) the
parametrisation of the $\ell_{\Jpsi}$ resolution function and (ii) the
MC template used for the true $\ell_{\Jpsi}$ distribution of generated
non-prompt \Jpsi for which both muons have been
reconstructed. Regarding (i), the reconstructed $\ell_{\Jpsi}$
distribution of simulated prompt \Jpsi is better parametrised with a
resolution function that is the sum of four Gaussians (the \pp
analysis at 7\TeV used the sum of three Gaussians). Four of the eight
fit parameters are fixed to the MC fit result and only the common
mean, two widths, and one relative fraction are left free in the fits
to the data. Regarding (ii), the $\ell_{\Jpsi}$ distribution of
non-prompt \Jpsi differs from that of the \pp analysis because of the
different heavy-ion tracking algorithm. In order to cope with the much
higher detector occupancy, the \PbPb tracking algorithm is done in one
iteration and requires a pixel triplet seed to point to the
reconstructed primary vertex within 1\,mm. Furthermore, the algorithm
includes a filter at the last step that requires the track to point
back to the primary vertex within six times the primary vertex
resolution. This reduces the reconstruction efficiency for \Jpsi with
large values of $\ell_{\Jpsi}$, \ie it causes a difference in the
prompt and non-prompt \Jpsi reconstruction efficiencies that increases
with the \Jpsi meson \pt.

\begin{figure*}[htbp]
  \begin{center}
    \includegraphics[width=0.45\linewidth]{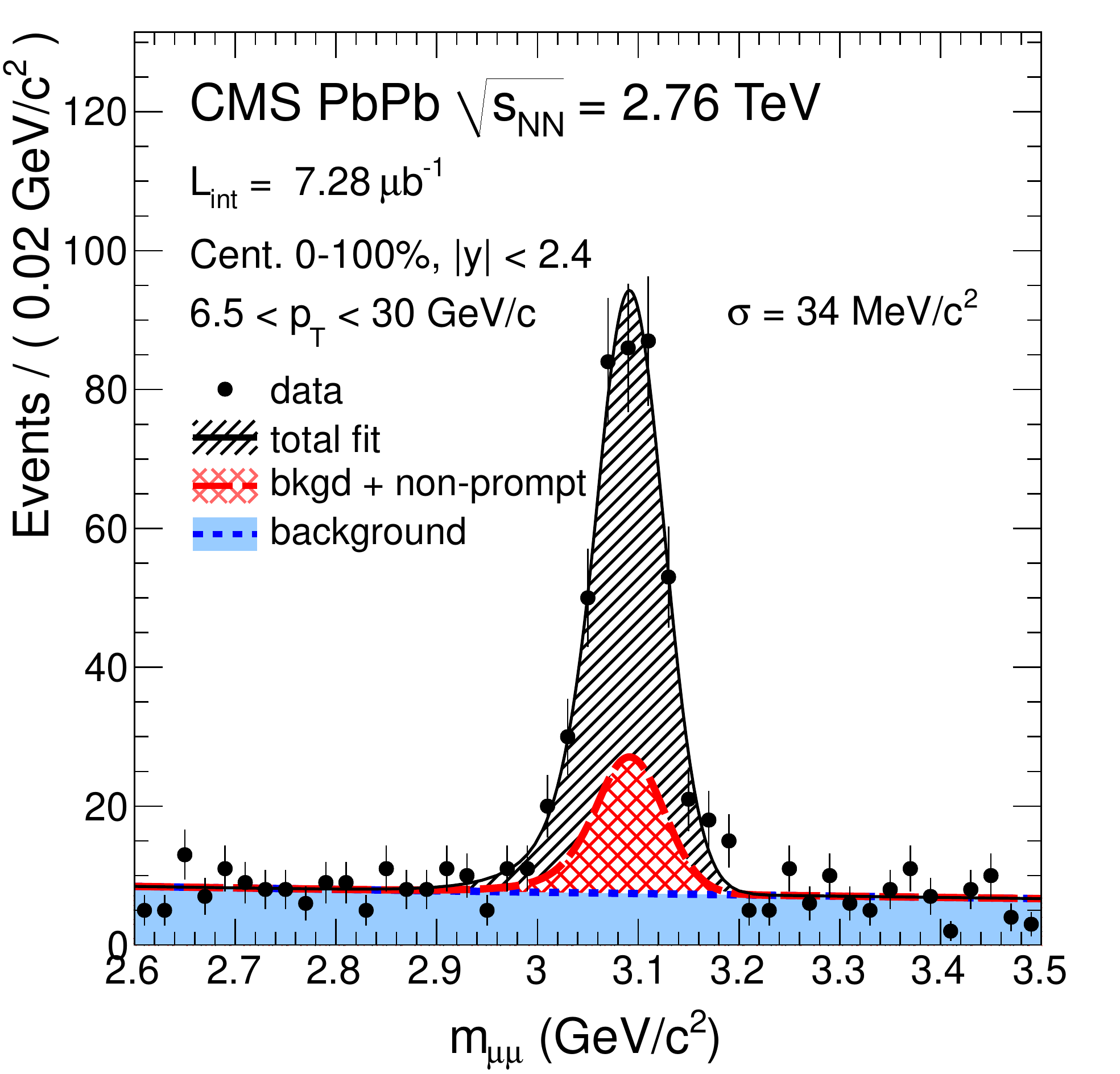}
    \includegraphics[width=0.45\linewidth]{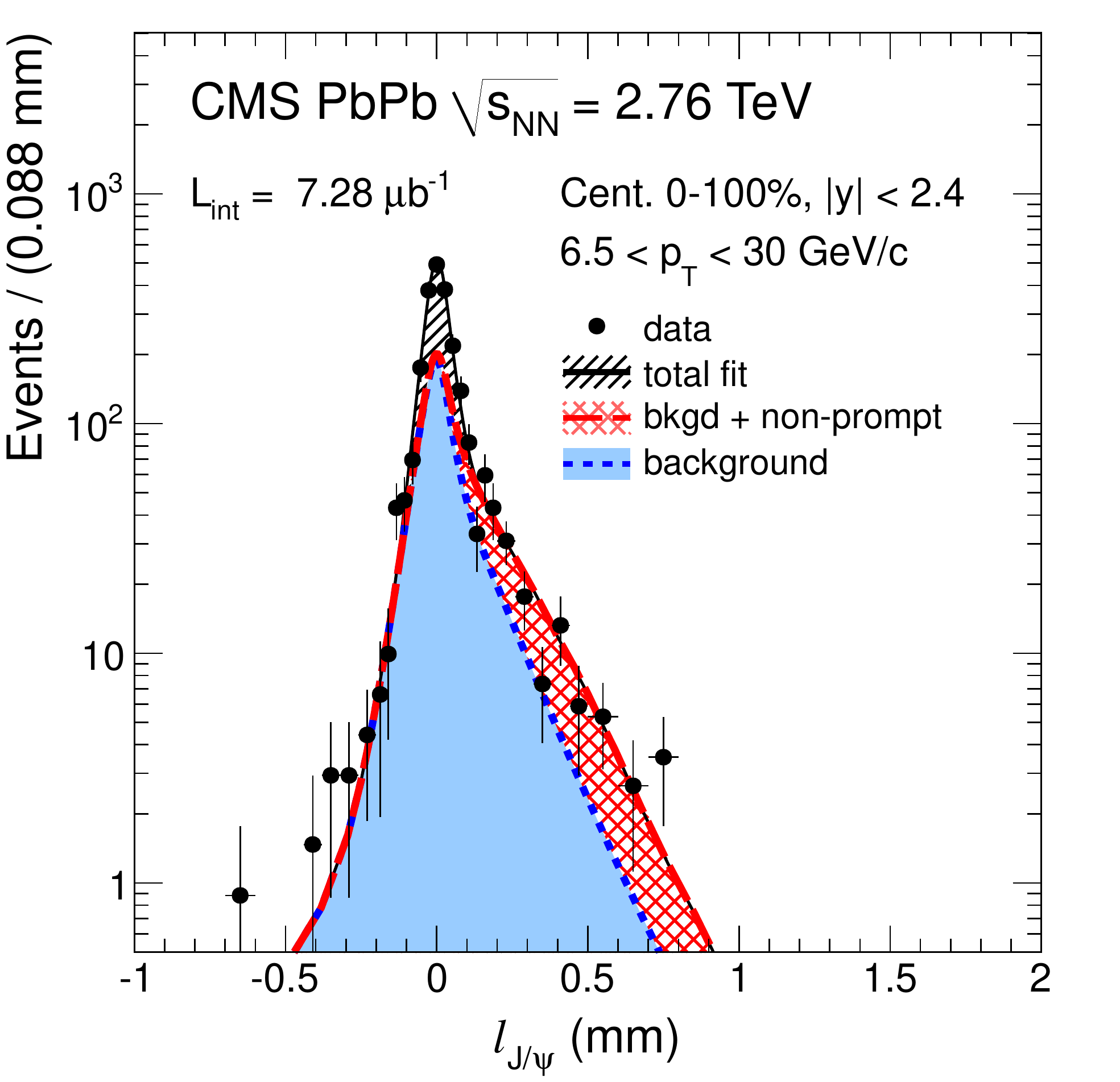}\\
    \includegraphics[width=0.45\linewidth]{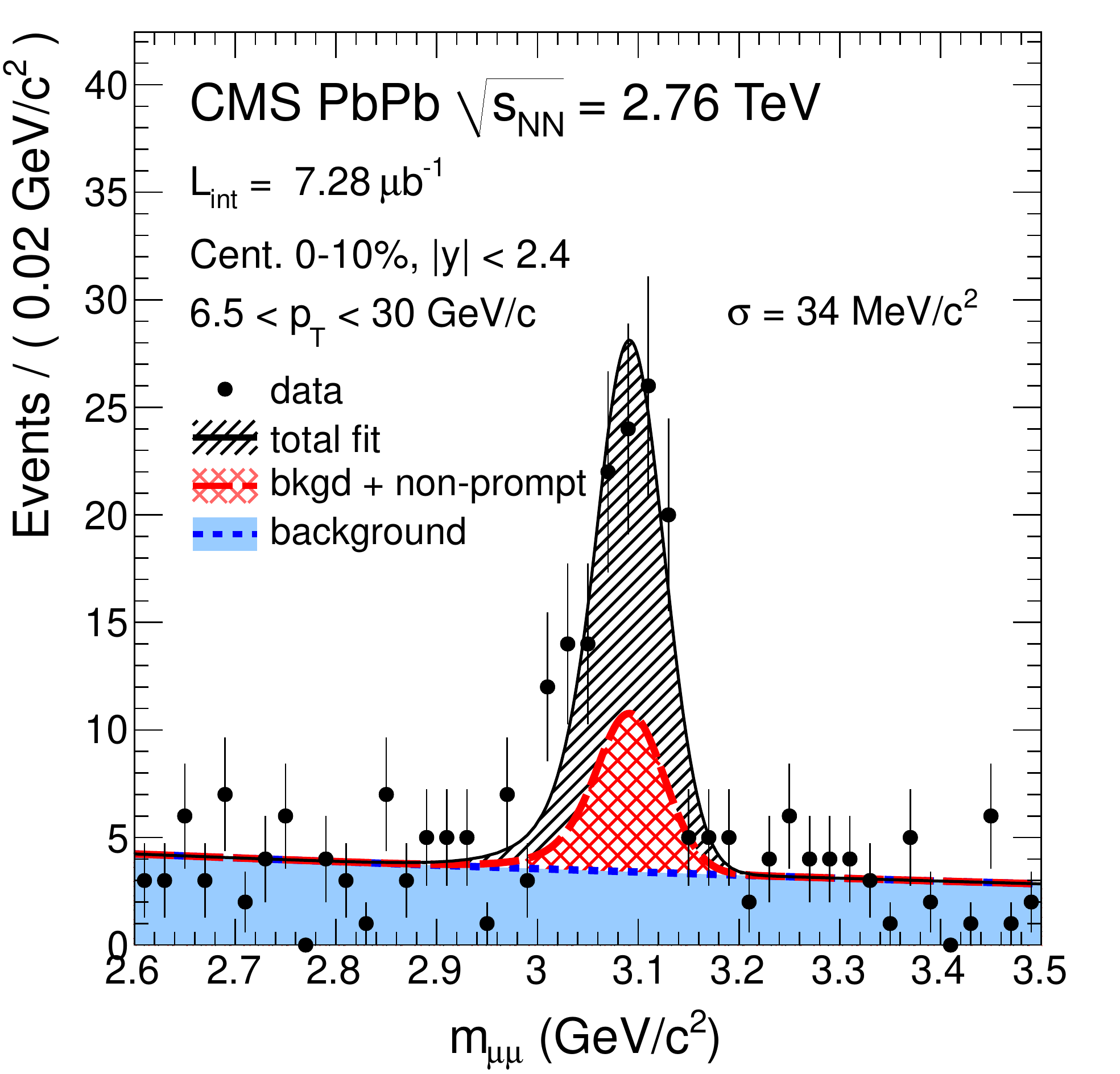}
    \includegraphics[width=0.45\linewidth]{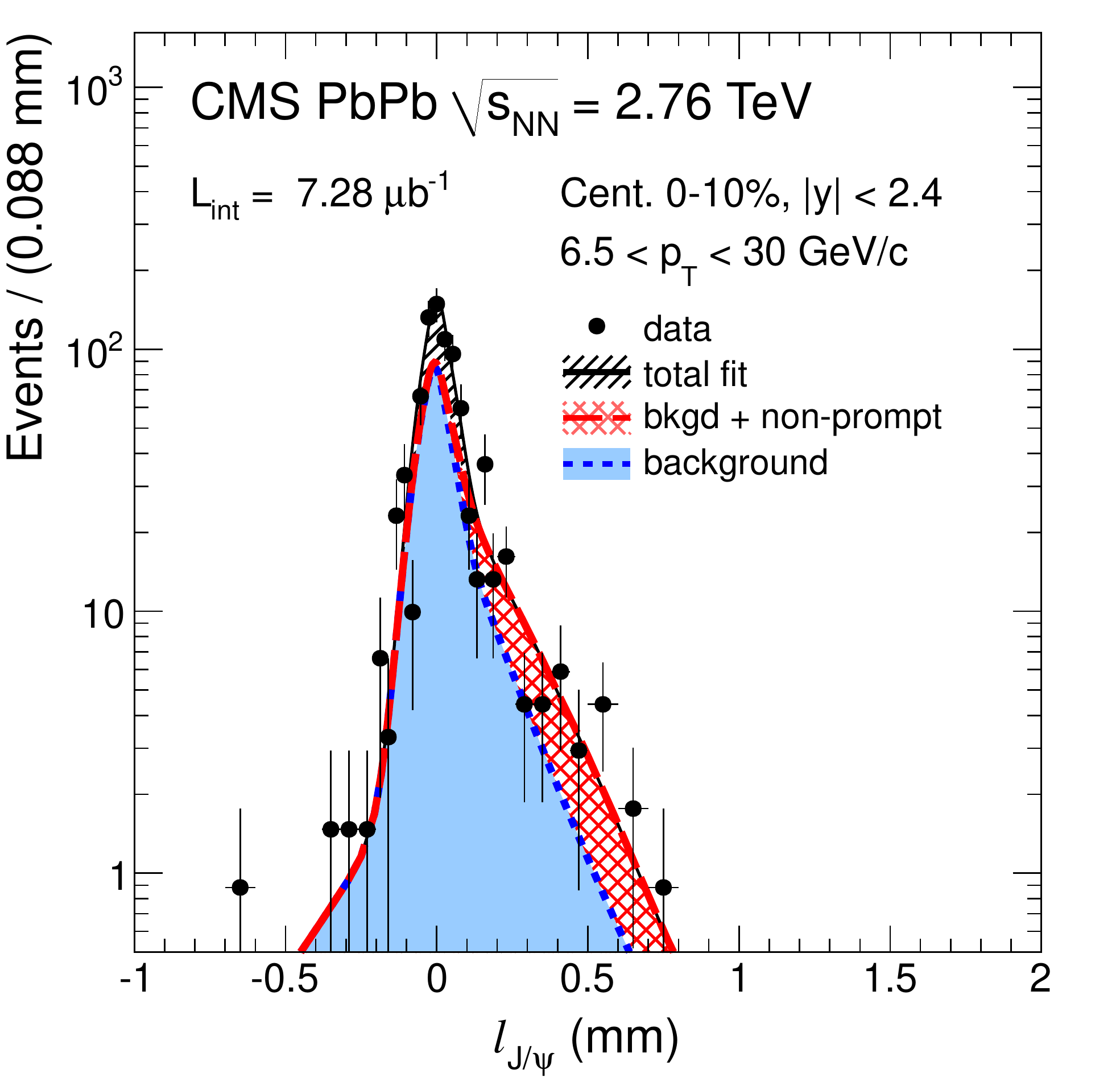}
    \caption{Invariant-mass spectra (left) and pseudo-proper decay
      length distributions (right) of \mumu pairs integrated over
      centrality (top) and for the 0--10\% centrality bin
      (bottom). The spectra are integrated over the rapidity range
      $|y|<2.4$ and the \pt range $6.5<\pt<30\GeVc$. The projections
      of the two-dimensional fit onto the respective axes are overlaid
      as solid black lines. The dashed red lines show the fitted
      contribution of non-prompt \Jpsi. The fitted background
      contributions are shown as dotted blue lines.}
    \label{fig:jpsi_2dmassfits}
  \end{center}
\end{figure*}

The prompt \Jpsi result is presented (in
Section~\ref{sec:promptResults}) in the centrality bins 0--10\%,
10--20\%, 20--30\%, 30--40\%, 40--50\%, and 50--100\%, while the
non-prompt \Jpsi result, given the smaller sample, is presented (in
Section~\ref{sec:nonpromptResults}) in only two centrality bins,
0--20\% and 20--100\%. Examples of $m_{\mumu}$ and $\ell_{\Jpsi}$
distributions are shown in \fig{fig:jpsi_2dmassfits}, including the
one for the 0--10\% centrality bin, which is one of the worst in terms
of signal over background ratio. The two-dimensional fit results are
shown as projections onto the mass and $\ell_{\Jpsi}$ axes. Integrated
over centrality, the numbers of prompt and non-prompt \Jpsi mesons
with $|y|<2.4$ and $6.5<\pt<30\GeVc$ are $307\pm22$ and $90\pm13$,
respectively.

In order to determine the systematic uncertainty on the yield
extraction, the signal and background shapes are varied: for the
signal mass shape, in addition to the default sum of the Crystal Ball
and Gaussian functions, a single Gaussian and a single Crystal Ball
function are tried. Alternatively, the $\alpha$ and $n$ parameters of
the Crystal Ball function are fixed individually for each \pt and
rapidity bin to the values found in the centrality integrated
bin. This is in contrast to the default procedure in which the values
for each rapidity bin are fixed to the values found in the bin
integrated over centrality and all \pt. For the background mass shape,
a straight line is tried as an alternative. A crosscheck using a
simple counting of the yield in the signal region after the
subtraction of the same-sign spectrum leads to consistent results. The
uncertainty on the fraction of non-prompt \Jpsi due to the
parametrisation of the $\ell_{\Jpsi}$ distribution is estimated by
varying the number of free parameters in the resolution function while
the other parameters are fixed to their MC values. The systematic
uncertainty is taken as the RMS of the yields obtained from the
different variations of the fit function. The systematic uncertainties
vary between 0.5\% and 5.7\% for the prompt \Jpsi yield, while the
non-prompt \Jpsi yield has uncertainties up to the extreme case of
14\% in the most forward rapidity ($1.6<|y|<2.4$) and lowest \pt
($3<\pt<30\GeVc$) bin.

\subsection{\texorpdfstring{\PgUa}{Upsilon(1S)} Analysis}
\label{sec:upsilon}

\begin{figure*}[htbp]
  \begin{center}
    \includegraphics[width=0.45\textwidth]{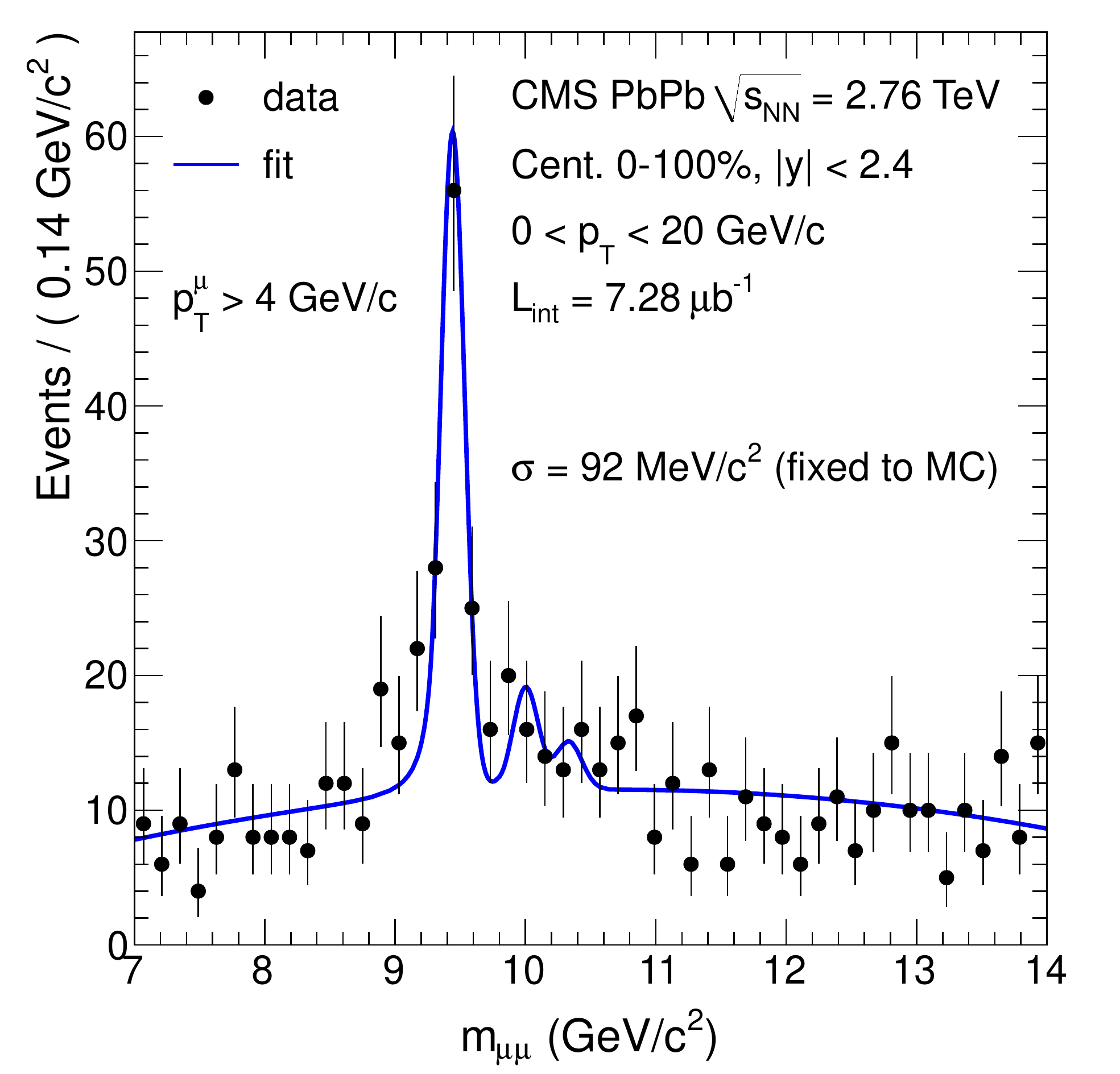}
    \includegraphics[width=0.45\textwidth]{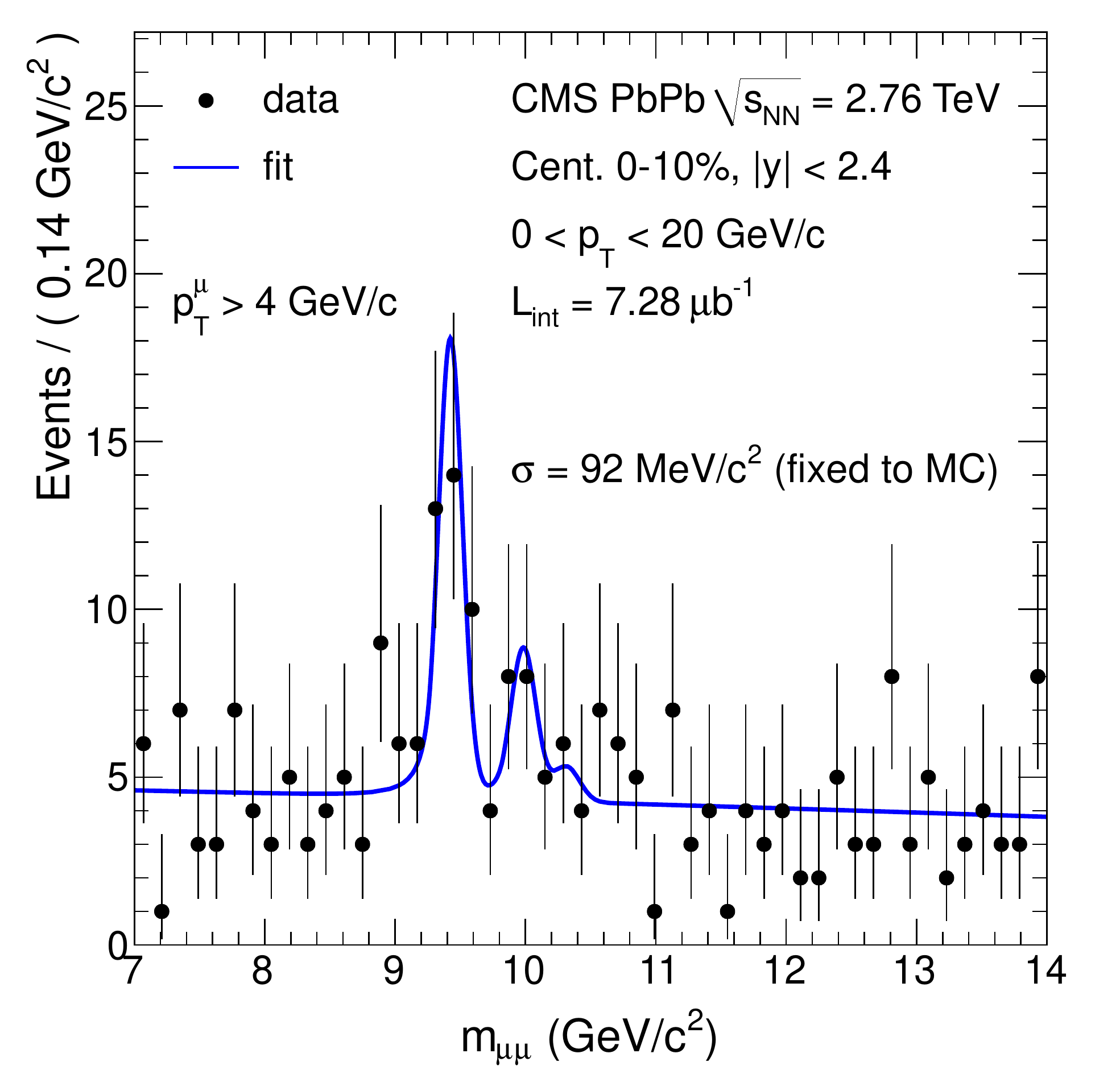}
    \caption{Invariant-mass spectrum of \mumu pairs (black circles)
      with $\pt<20\GeVc$ and $|y|<2.4$, for muons above 4\GeVc,
      integrated over centrality (left) and for the 0--10\% centrality
      bin (right).}
    \label{fig:ups_invmass}
  \end{center}
\end{figure*}

To extract the \PgUa\ yield, an extended unbinned maximum-likelihood
fit to the \mumu invariant mass spectrum between 7 and 14\GeVcc is
performed, integrated over \pt, rapidity, and centrality, as shown in
the left panel of \fig{fig:ups_invmass}. The measured mass line shape
of each \PgU\ state is parametrised by a Crystal Ball function. Since
the three \PgU\ resonances partially overlap in the measured dimuon
mass spectrum, they are fitted simultaneously. Therefore, the
probability distribution function describing the signal consists of
three Crystal Ball functions. In addition to the three \PgUn\ yields,
the \PgUa\ mass is the only parameter left free, to accommodate a
possible bias in the momentum scale calibration. The mass ratios
between the states are fixed to their world average
values~\cite{Nakamura:2010zzi}, and the mass resolution is forced to
scale linearly with the resonance mass. The \PgUa\ resolution is fixed
to the value found in the simulation, $92 \MeVcc$. This value is
consistent with what is measured when leaving this parameter free in a
fit to the data, $(122 \pm 30)\MeVcc$. The low-side tail parameters in
the Crystal Ball function are also fixed to the values obtained from
simulation. Finally, a second-order polynomial is chosen to describe
the background in the mass range 7--14\GeVcc. From this fit, before
accounting for acceptance and efficiencies, the measured \PgUa\ raw
yield is $86 \pm 12$. The observed suppression of the excited states
was discussed in~\cite{Chatrchyan:2011pe}. The fitted mean value is
$m_0 = (9.441\pm0.016)\GeVcc$, which, for the same reason as for the
\Jpsi, is slightly below the PDG value $m_{\PgUa} =
9.460\GeVcc$~\cite{Nakamura:2010zzi}.

The data are binned in \pt and rapidity of the \mumu pairs, as well as
in bins of the event centrality (0--10\%, 10--20\%, and
20--100\%). The bins in rapidity are $|y|<1.2$ and $1.2<|y|<2.4$. In
contrast to the \Jpsi case, CMS has acceptance for \PgU\ down to
\mbox{$\pt = 0\GeVc$} over the full rapidity range. The \pt bins in
this analysis are $0<\pt<6.5\GeVc$, $6.5<\pt<10\GeVc$, and
$10<\pt<20\GeVc$. There are only two events with a \mumu pair in the
\PgU\ mass region and $\pt > 20\GeVc$. The invariant-mass distribution
for the centrality bin 0--10\% is illustrated in the right panel of
\fig{fig:ups_invmass}. The raw yields of \PgUa are tabulated
in \tab{tab:upsilonyields} of Appendix~\ref{app:datatables}.

The systematic uncertainties are computed by varying the line shape in
the following ways: (i) the Crystal Ball function tail parameters are
varied randomly according to their covariance matrix and within
conservative values covering imperfect knowledge of the amount of
detector material and final-state radiation in the underlying process;
(ii) the width is varied by $\pm 5\MeVcc$, a value motivated by the
current understanding of the detector performance (\eg, the dimuon
mass resolution, accurately measured at the \Jpsi mass, is identical
in \pp and \PbPb collisions); (iii) the background shape is changed
from quadratic to linear, and the mass range of the fit is varied from
6--15 to 8--12\GeVcc; the observed RMS of the results in each category
is taken as the systematic uncertainty. The quadratic sum of these
three systematic uncertainties is dominated by the variation of the
resolution of the mass fit, and is of the order of 10\%, reaching 13\%
for the 0--10\% centrality bin. As was the case for the \Jpsi
selection, a simple counting of the yield in the signal region after
the subtraction of the same-sign spectrum leads to consistent results.

\section{Acceptance and Efficiency}
\label{sec:acceff}
\subsection{Acceptance}
\label{sec:acc}

The dimuon acceptance, $A$, is defined as the fraction of \mumu pairs
for which both muons are declared detectable in the CMS detector with
respect to all muon pairs produced in $|y|<2.4$,
\begin{equation}
  \label{eq:acc}
    A(\pt,y;\lambda_{\theta}) = \frac{N^{\Pgm\Pgm}_{\text{detectable}}(\pt,y;\lambda_{\theta})}{N^{\Pgm\Pgm}_{\text{generated}}(\pt,y;\lambda_{\theta})},
\end{equation}
where:
\begin{itemize}
\item N$^{\Pgm\Pgm}_{\text{detectable}}$ is the number of generated
  events in a given quarkonium (\pt, $y$) bin in the MC simulation,
  for which both muons are detectable according to the selections
  defined in Eqs.~\eqref{eq:singleMuonAcc}
  and~\eqref{eq:singleMuonAccUps};
\item N$^{\Pgm\Pgm}_{\text{generated}}$ is the number of all \mumu pairs
  generated within the considered (\pt, $y$) bin.
\end{itemize}
The acceptance depends on the \pt and $y$ of the \mumu pair, and the
polarization parameter $\lambda_{\theta}$. Different polarizations of
the \Jpsi and \PgUa\ will cause different single-muon angular
distributions in the laboratory frame and, hence, different
probabilities for the muons to fall inside the CMS detector
acceptance. Since the quarkonium polarization has not been measured in
heavy-ion or \pp collisions at \sqrtsnn = 2.76\TeV, the prompt \Jpsi
and \PgUa\ results are quoted for the unpolarized scenario only. For
non-prompt \Jpsi the results are reported for the polarization
predicted by \EVTGEN. The impact of the polarization on the
acceptance is studied for the most extreme polarization scenarios in
the Collins--Soper and helicity frames. For fully longitudinal
(transverse) polarized \Jpsi in the Collins--Soper frame, the effect
is found to be at most $-20\%$ (6\%). In the helicity frame, the
effects are at most 40\% and $-20\%$ for the two scenarios. For \PgUa\
the polarization effects range between $-20\%$ for longitudinal
polarization in the Collins--Soper frame to 40\% for transverse
polarization in the helicity frame.

\begin{figure*}[htbp]
  \centering
  \includegraphics[width=0.45\linewidth]{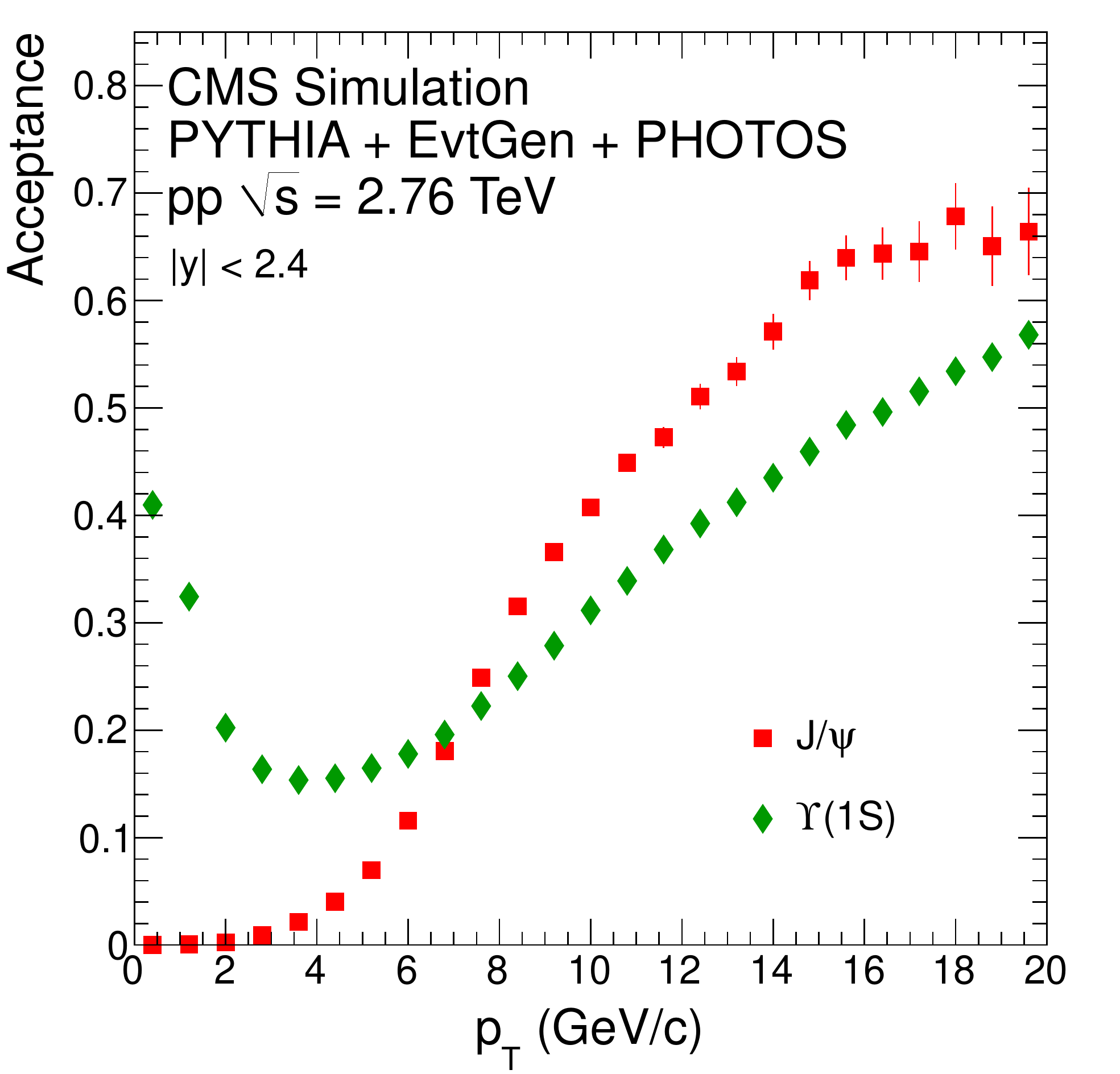}
  \includegraphics[width=0.45\linewidth]{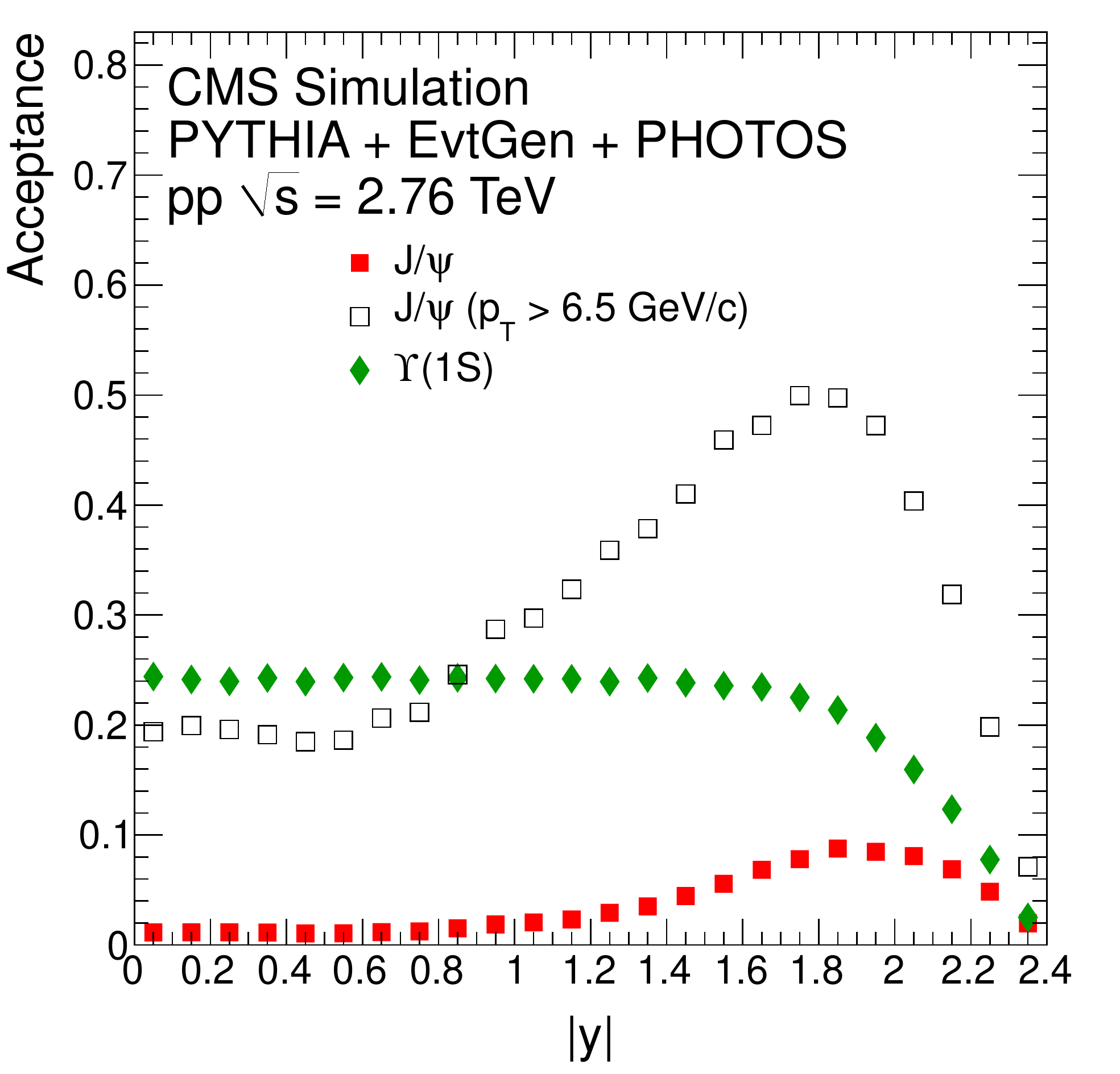}
  \caption{Dimuon acceptance as a function of \pt (left) and $|y|$
    (right) for \Jpsi (red squares) and \PgUa\ (green diamonds). Also
    shown in the right panel is the acceptance for \Jpsi with
    $\pt>6.5\GeVc$ (open black squares). The error bars represent the
    statistical uncertainties only.}
  \label{fig:acceptance}
\end{figure*}

The acceptance is calculated using the MC sample described in
Section~\ref{sec:event-sel}. The \pt and rapidity dependencies of the
\Jpsi and \PgUa\ acceptances are shown in \fig{fig:acceptance}.

Since the acceptance is a function of both \pt and $y$, uncertainties
in the predicted distributions for these variables can lead to a
systematic uncertainty in the average acceptance over a \pt or $y$
bin.  To estimate these uncertainties, the shapes of the generated MC
\pt and $|y|$ distributions are varied by applying a weight that
increases linearly from 0.7 to 1.3 over the range $0<|y|<2.4$ and
$0<\pt<30\GeVc$ (20\GeVc) for \Jpsi (\PgUa).  The RMS of the resulting
changes in the acceptance for each \pt and $y$ bin are summed in
quadrature to compute the overall systematic uncertainty from this
source.  The largest relative systematic uncertainties obtained are
4.2\%, 3.2\%, and 2.8\% for the prompt \Jpsi, non-prompt \Jpsi, and
\PgUa\ acceptances, respectively.

\subsection{Efficiency}
\label{sec:eff}

The trigger, reconstruction, and selection efficiencies of \mumu pairs
are evaluated using simulated MC signal events embedded in simulated
\PbPb events, as described in Section~\ref{sec:event-sel}. The overall
efficiency is calculated, in each analysis bin, as the fraction of
generated events (passing the single muon phase space cuts) where both
muons are reconstructed, fulfil the quality selection criteria and
pass the trigger requirements. In the embedded sample, the signal over
background ratio is by construction higher than in data, so the
background contribution underneath the resonance peak is negligible
and the signal is extracted by simply counting the \mumu pairs in the
quarkonium mass region. The counting method is crosschecked by using
exactly the same fitting procedure as if the MC events were collision
data. Only muons in the kinematic region defined by
Eqs.~\eqref{eq:singleMuonAcc} and~\eqref{eq:singleMuonAccUps} are
considered.

\begin{figure}[htbp]
  \centering
  \includegraphics[width=0.45\linewidth]{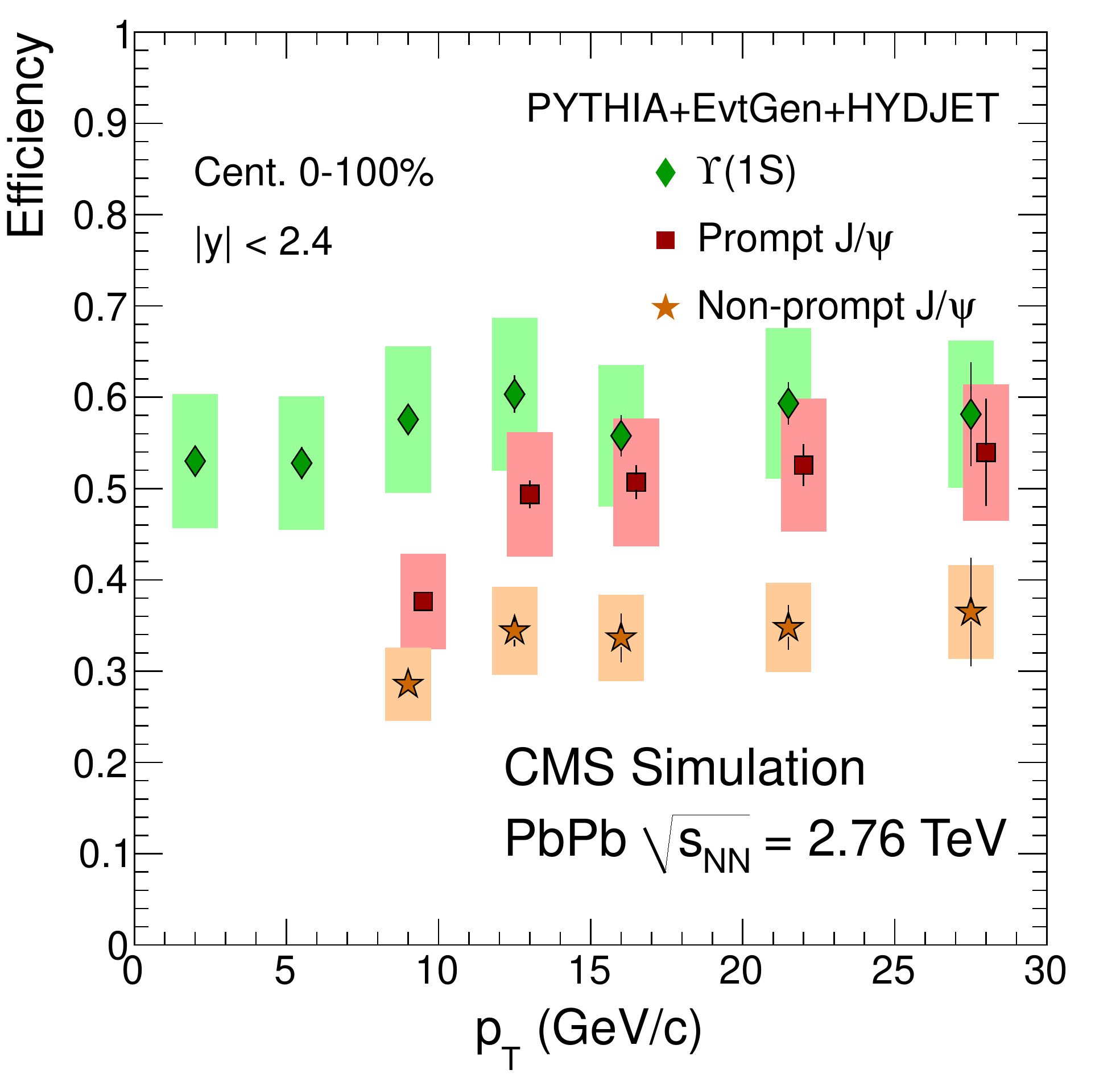}
  \includegraphics[width=0.45\linewidth]{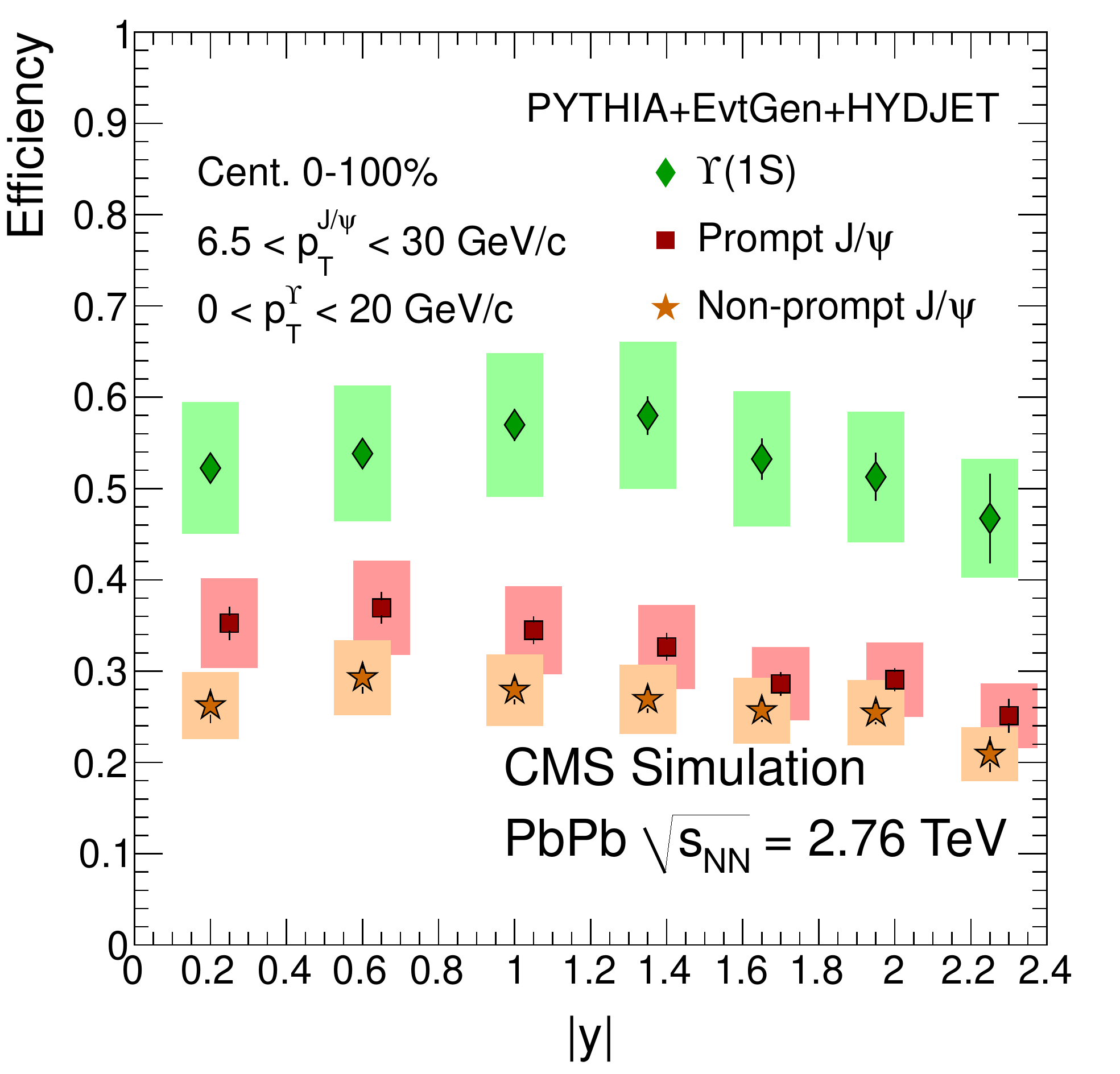}\\
  \includegraphics[width=0.45\linewidth]{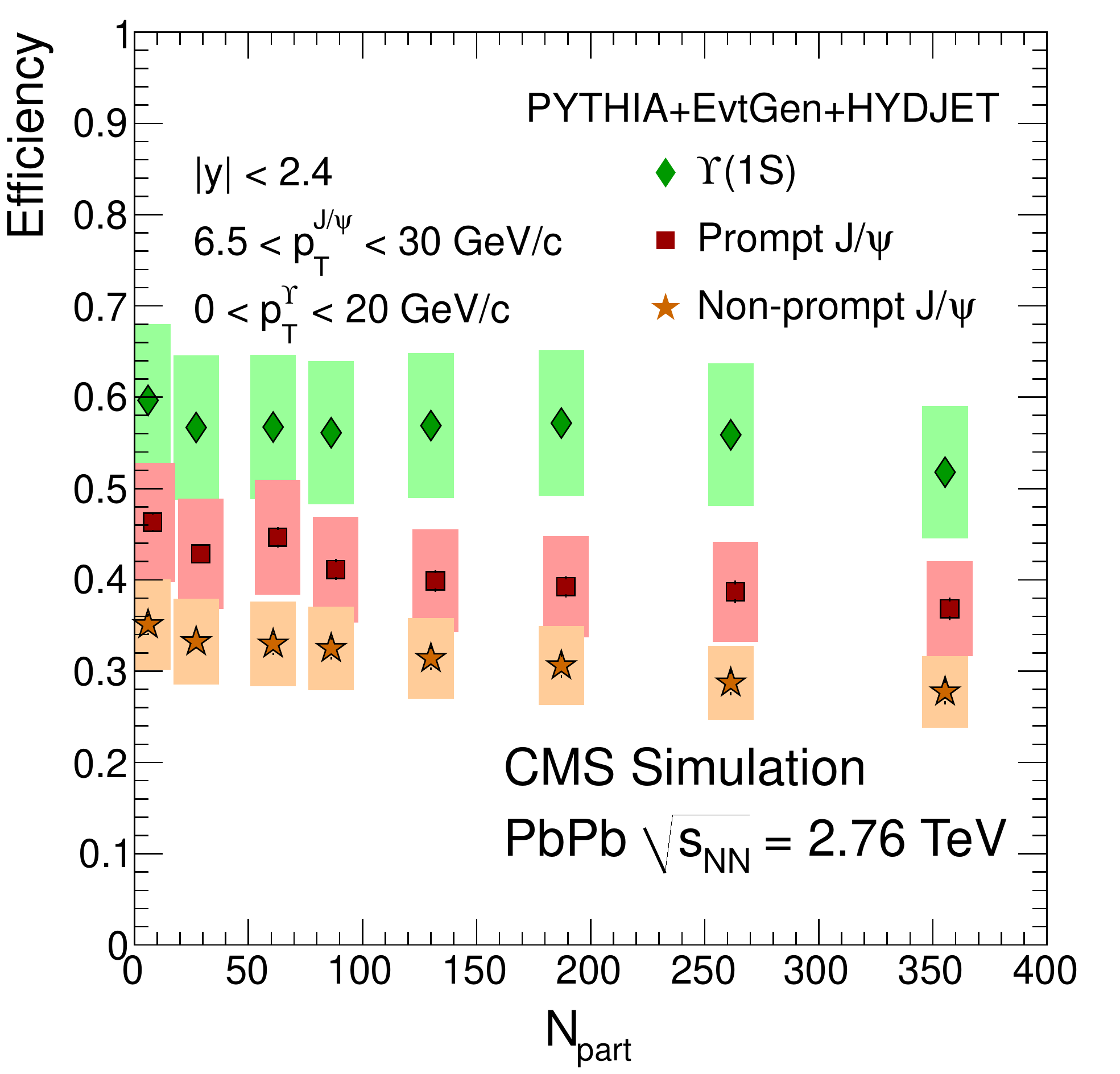}
  \caption{Combined trigger, reconstruction, and selection
    efficiencies as a function of quarkonium \pt and $|y|$, and event
    centrality, for each signal: red squares and orange stars for
    prompt and non-prompt \Jpsi, respectively, and green diamonds for
    \PgUa. For better visibility, the prompt \Jpsi points are shifted
    by $\Delta \pt = 0.5\GeVc$, $\Delta y=0.05$, and $\Delta \npart =
    2$. Statistical (systematic) uncertainties are shown as bars
    (boxes). The systematic uncertainties are the quadratic sum of the
    uncertainty on the kinematic distributions and the MC validation
    uncertainty.}
  \label{fig:eff}
\end{figure}

In \fig{fig:eff}, the efficiencies are shown as a function of the
\mumu pair \pt, $y$, and the event centrality, for each signal: red
squares for prompt \Jpsi, orange stars for non-prompt \Jpsi, and green
diamonds for \PgUa. As discussed in Section~\ref{sec:promptJpsi}, the
efficiency of non-prompt \Jpsi is lower than that of prompt \Jpsi,
reaching about 35\% for $\pt > 12\GeVc$. The prompt \Jpsi efficiency
increases with \pt until reaching a plateau slightly above 50\% at \pt
of about 12\GeVc, while the \PgUa\ efficiency is $\sim\!55\%$,
independent of \pt. The efficiencies decrease slowly as a function of
centrality because of the increasing occupancy in the silicon tracker;
the relative difference between peripheral and central collisions is
17\% for \Jpsi and 10\% for \PgUa. The integrated efficiency values
are 38.3\%, 29.2\%, and 54.5\% for the prompt \Jpsi, non-prompt \Jpsi
(both with $6.5<\pt<30\GeVc$, $|y|<2.4$, and 0--100\% centrality), and
\PgUa\ (with $0<\pt<20\GeVc$, $|y|<2.4$, and 0--100\% centrality),
respectively.

The systematic uncertainty on the final corrections due to the
kinematic distributions is estimated by a $\pm30\%$ variation of the
slopes of the generated \pt and rapidity shapes, similar to the
acceptance variation described in the previous section. The systematic
uncertainties are in the ranges 1.8--3.4\%, 2.2--4.2\%, and 1.4--2.7\%
for prompt \Jpsi, non-prompt \Jpsi, and \PgUa, respectively, including
the statistical precision of the MC samples.

The individual components of the MC efficiency are crosschecked using
muons from \Jpsi decays in simulated and collision data with a
technique called \emph{tag-and-probe}, similar to the one used for the
corresponding \pp measurement~\cite{Khachatryan:2010yr}. In this
method, high quality muons (the \emph{tags}) are combined with muons
that are selected without applying the selections whose efficiency is
to be measured (the \emph{probes}). Probe muons that fulfil these
selections are then categorized as \emph{passing probes}, the others
as \emph{failing probes}. A simultaneous fit of the two resulting
invariant mass spectra (passing and failing) provides the efficiency
of the probed selection. Because of correlations in the efficiency of
matching silicon-tracker tracks to standalone muons, the total
efficiency does not fully factorize into the individual components
probed by this method. Therefore, the reconstruction and trigger
efficiencies for \mumu pairs are directly obtained from the MC
simulation, rather than as a product of the partial components.

The fits are performed for \emph{tag-probe} pairs with a \pt above
6.5\GeVc as this is the region measured over the full rapidity range,
with and without applying the probed selection on one of the muons:
\begin{enumerate}
\item The trigger efficiency is estimated by measuring the fraction of
  global muons (used as probes) associated to the double-muon trigger
  in an event sample selected by tag-muons associated to a single-muon
  trigger. A Crystal Ball function is used to describe the \Jpsi
  peak. The $\pt^{\mu}$ and $\eta^{\mu}$ dependencies of the trigger
  efficiency are compatible between data and MC. For \Jpsi with
  $\pt>6.5\GeVc$, the $\pt^{\mu}$ and $\eta^{\mu}$ integrated trigger
  efficiency is 95.9\% in MC and $(95.1\pm0.9)\%$ in data.
\item Standalone muons passing the quality selections required in this
  analysis are used to evaluate the efficiency of the silicon tracker
  reconstruction, which includes losses induced by the matching
  between the silicon-tracker track and the muon detector track, and
  by the imposed quality selection criteria (both on the global track
  and on its silicon-tracker segment).  For this efficiency
  measurement, the signal is fitted with a Gaussian function and the
  background with a second-order polynomial. A Gaussian, rather than a
  Crystal Ball function, is used because of the poor momentum
  resolution of the standalone muons. No $\pt>6.5$\GeVc requirement
  was used, since the poorer momentum resolution of standalone muons
  would have biased the measurement. The single-muon efficiencies
  measured in MC and data of 84.9\% and $(83.7^{+5.7}_{-5.3})\%$,
  respectively, are in good agreement.
\end{enumerate}
The systematic uncertainty of the muon pair efficiency, 13.7\%, is
determined by comparing the \emph{tag-and-probe} efficiencies
evaluated in \PbPb data and MC samples, and is dominated by the
statistical uncertainties of the measurements. The standalone muon
reconstruction efficiency (99\% in the plateau) cannot be probed with
silicon-tracker tracks because of the large charged particle
multiplicity in PbPb collisions. Since this part of the reconstruction
is identical to that used for pp data, a systematic uncertainty of
1\%, reported in Ref.~\cite{muon-pog}, is assumed.

\section{The pp Baseline Measurement}
\label{sec:ppRef}
A \pp run at \sqrts = 2.76\TeV was taken in March 2011. The integrated
luminosity was 231\nbinv, with an associated uncertainty of 6\%. For
hard-scattering processes, the integrated luminosity of the \pp sample
is comparable to that of the \PbPb sample (\mbox{$7.28\mubinv \cdot
  208^2 \approx 315 \nbinv$}).

Given the higher instantaneous luminosity, the \Lone trigger required
slightly higher quality muons in the \pp run than in the \PbPb
run. The offline event selection is the same as in the \PbPb analysis,
only slightly relaxed for the HF coincidence requirement: instead of
three towers, only one tower with at least 3\GeV deposited is required
in the \pp case. The same reconstruction algorithm, \ie the one
optimized for the heavy-ion environment, is used for both \pp and
\PbPb data. The products of the trigger, reconstruction, and selection
efficiencies determined in \pp MC simulations are 42.5\%, 34.5\%, and
55.1\% for the prompt \Jpsi, non-prompt \Jpsi (both with
$6.5<\pt<30\GeVc$, $|y|<2.4$), and \PgUa\ (with $0<\pt<20\GeVc$,
$|y|<2.4$), respectively.

The accuracy of the MC simulation in describing the trigger efficiency
is crosschecked with the \emph{tag-and-probe} method in the same way
as for the \PbPb analysis discussed in Section~\ref{sec:eff}. For
muons from decays of \Jpsi with $\pt>6.5$\GeVc, the $\pt^{\mu}$ and
$\eta^{\mu}$ integrated trigger efficiencies are ($92.5 \pm 0.6$)\% in
data and ($94.3 \pm 0.2$)\% in MC. In the same phase-space, the
tracking and muon selection efficiency is ($82.5\pm2.4$)\% in data and
($84.6\pm1.0$)\% in MC. For the standalone muon reconstruction
efficiency a systematic uncertainty of 1\% is assigned, as reported in
Ref.~\cite{muon-pog}. As in the \PbPb case, the systematic uncertainty
of the muon pair efficiency in \pp collisions, 13.7\%, is determined
by comparing the \emph{tag-and-probe} efficiencies evaluated in data
and MC samples, and is dominated by the statistical uncertainties of
the measurements.

\begin{figure}[htbp]
  \begin{center}
      \includegraphics[width=0.45\textwidth]{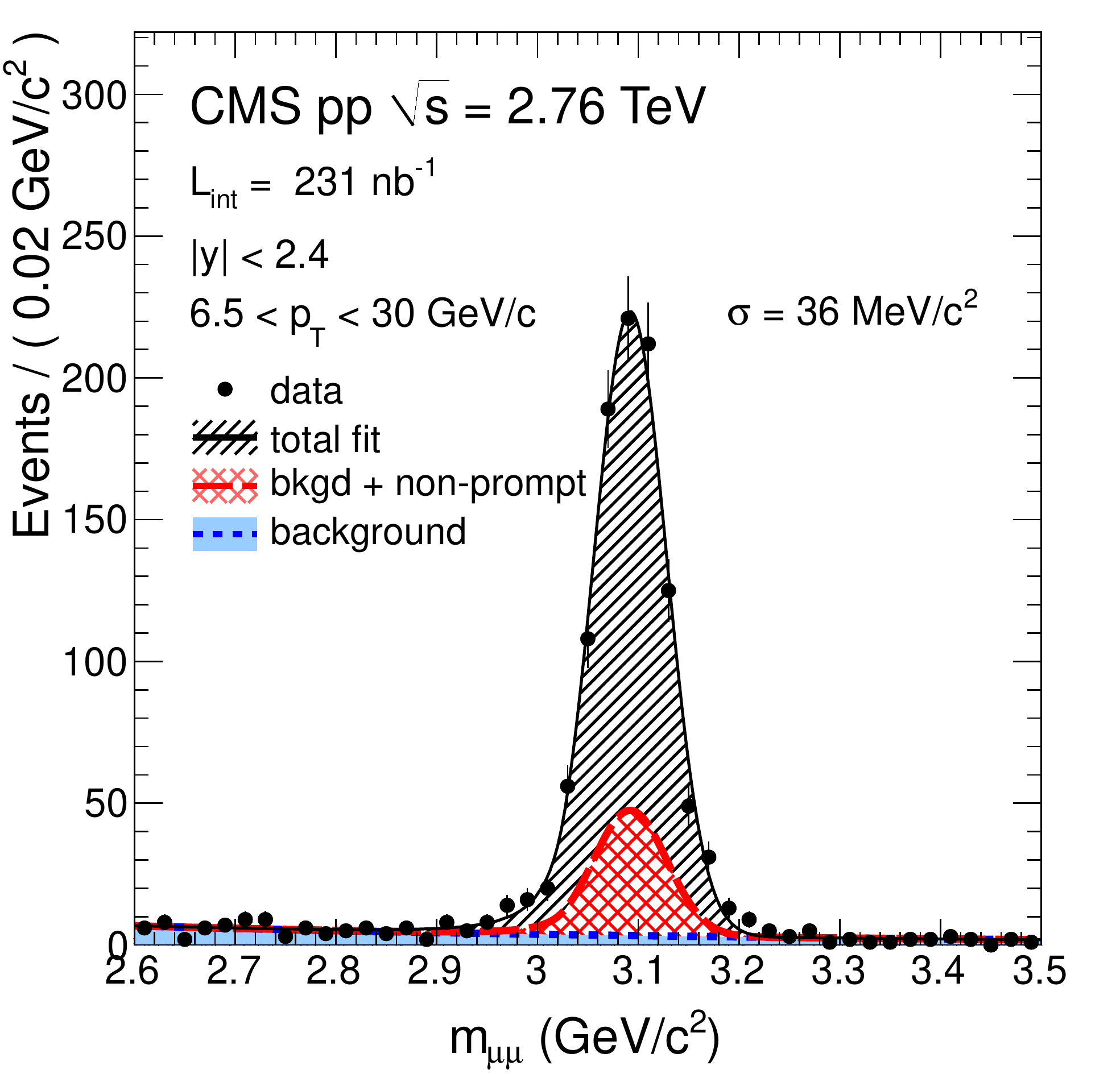}
    \includegraphics[width=0.45\textwidth]{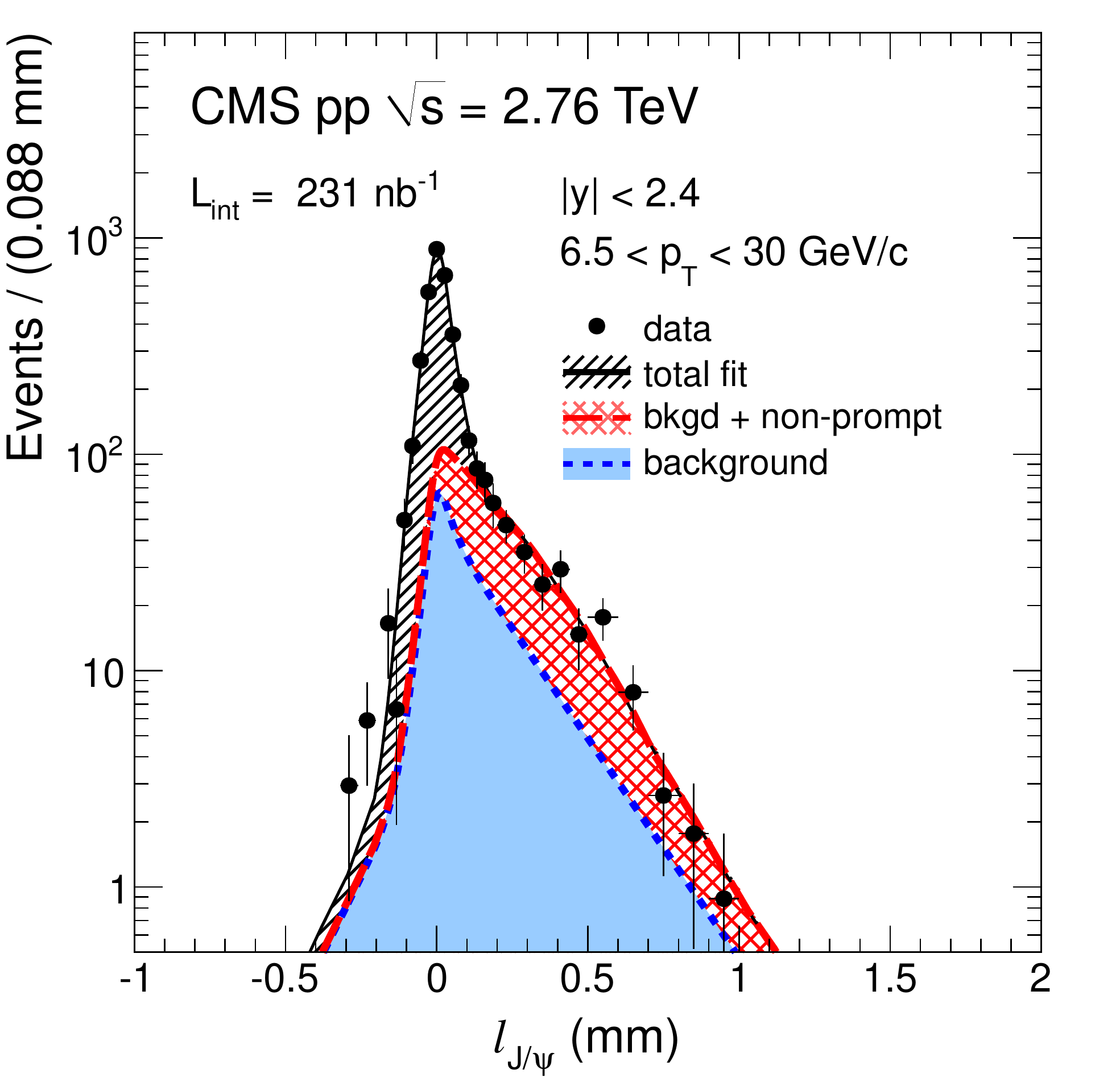}\\
    \caption{Non-prompt \Jpsi signal extraction for \pp collisions at
      $\sqrts = 2.76\TeV$: dimuon invariant mass fit (left) and
      pseudo-proper decay length fit (right).}
    \label{fig:jpsi_pp}
  \end{center}
\end{figure}

The quarkonium signals in \pp collisions are extracted following the
same methods as in \PbPb collisions, described in
Sections~\ref{sec:jpsi} and~\ref{sec:upsilon}, apart from the
non-prompt \Jpsi signal extraction: the four Gaussians of the lifetime
resolution are fixed to the MC values because of the lack of events in
the dimuon mass sidebands. The systematic uncertainty on the signal
extraction in \pp is 10\% for \PgUa\ and varies, depending on \pt and
rapidity, between 0.4 and 6.2\% for prompt \Jpsi and between 5 and
20\% for non-prompt \Jpsi. The fit results for the prompt and
non-prompt \Jpsi yield extraction are shown in \fig{fig:jpsi_pp} for
$|y|<2.4$ and $6.5<\pt<30\GeVc$. The numbers of prompt and non-prompt
\Jpsi mesons in this rapidity and \pt range are $820\pm34$ and
$206\pm20$, respectively.

\begin{figure}[hbtp]
  \begin{center}
  \includegraphics[width=0.5\linewidth]{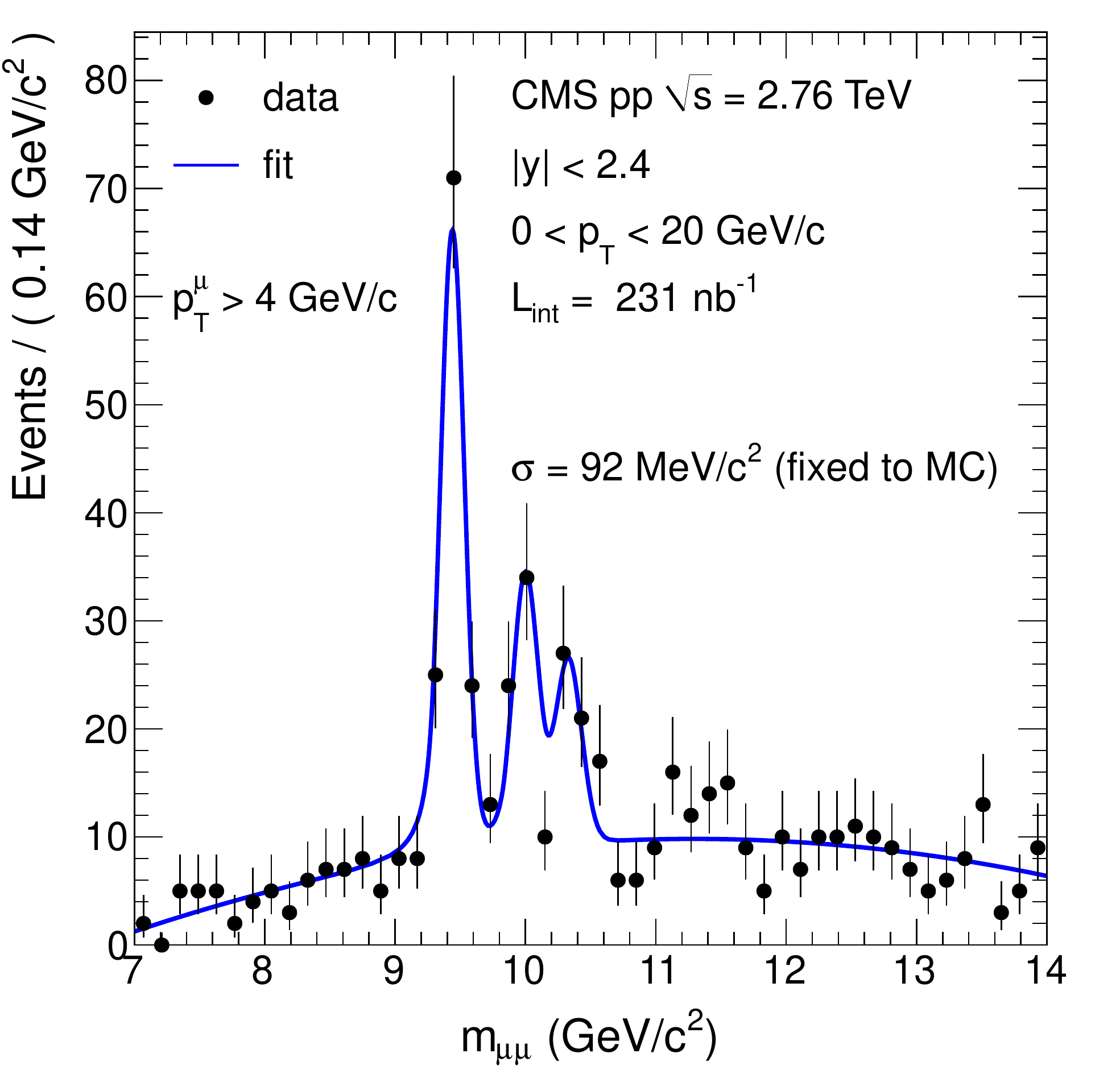}
  \caption{The \pp dimuon invariant-mass distribution in the range
    $\pt<20 \GeVc$ for $|y|<2.4$ and the result of the fit to the
    $\PgU$ resonances.}
  \label{fig:ups_pp}
  \end{center}
\end{figure}

The invariant-mass spectrum of \mumu pairs in the $\PgU$ region from
\pp collisions is shown in \fig{fig:ups_pp}. The same procedure as the
one described for the \PbPb analysis is used. The number of $\PgUa$
mesons with $|y|<2.4$ and $0<\pt<20\GeVc$ is $101\pm12$. The fit
result of the excited states is discussed in~\cite{Chatrchyan:2011pe}.

The differential cross section results include the systematic
uncertainties of the reconstruction efficiency and acceptance,
estimated in the same way as for the \PbPb analysis. The systematic
uncertainties on the efficiencies are 1.6--3\%, 1.4--2\%, and
0.4--0.9\% for prompt \Jpsi, non-prompt \Jpsi, and \PgUa,
respectively. The uncertainty on the acceptance is identical in the
\pp and \PbPb analyses.

For the measurement of the nuclear modification factors, in which the
ratio of \PbPb to \pp results is computed, most of the reconstruction
systematic uncertainties cancel out because the same algorithm is
used. However, the following factors must be accounted for:
\begin{enumerate}
\item The luminosity uncertainty. This is a global systematic
  uncertainty of 6\% that allows all measured nuclear modification
  factors to change by a common scale-factor. Since the \PbPb yield is
  normalized by the number of minimum-bias events, which has a
  negligible uncertainty, no systematic uncertainty on the \PbPb
  luminosity has to be considered.
\item The uncertainty on \taa. For results integrated over centrality,
  this is a global systematic uncertainty of 5.7\%, based on the
  Glauber model employed. For results as a function of centrality, the
  uncertainty varies between a minimum of 4.3\% in the most central
  bin and a maximum of 15\% in the most peripheral
  bin~\cite{Chatrchyan:2011sx}.
\item The systematic uncertainty associated with the trigger
  efficiency. The ratios between the \emph{tag-and-probe} efficiencies
  obtained in \pp and \PbPb are the same in data and MC events, within
  the statistical accuracy of the data (1\% for the single-muon
  efficiency).  Twice this value (2\%) is assigned as the uncertainty
  on the difference of the trigger efficiencies of \mumu pairs in
  \PbPb and \pp collisions.
\item The tracking efficiency uncertainty due to different charged
  particle multiplicities in \pp and \PbPb collisions.  The ratios
  between the \emph{tag-and-probe} efficiencies obtained in \pp and
  central \PbPb events are the same in data and MC events, within the
  statistical accuracy of the data (6.8\% for the single-muon
  efficiency).  This value is propagated as the tracking systematic
  uncertainty in all the ratios of \PbPb to \pp data.
\end{enumerate}

\section{Results}
\label{sec:results}
The double-differential quarkonium cross sections in \PbPb collisions
are reported in the form
\begin{equation}
  \label{eq:xsection}
    \frac{1}{\taa} \cdot
    \frac{\mathrm{d}^2N}{\mathrm{d}y\,\mathrm{d}\pt} = \frac{1}{\taa\, N_{\text{MB}}} \cdot \frac{1}{\Delta y\, \Delta \pt} \cdot \frac{N_{\QQbar}}{A\, \varepsilon},
\end{equation}
while in \pp collisions they are calculated as
\begin{equation}
  \label{eq:xsection_pp}
    \frac{d^2\sigma}{\mathrm{d}y\,\mathrm{d}\pt} = \frac{1}{\lumi_{\pp}} \cdot \frac{1}{\Delta y\, \Delta \pt} \cdot \frac{N_{\QQbar}}{A\, \varepsilon},
\end{equation}
where:
\begin{itemize}
\item $N_{\QQbar}$ is the number of measured prompt \Jpsi, non-prompt
  \Jpsi, or \PgUa\ in the \mumu decay channel;
\item $N_{\text{MB}}$ is the number of minimum-bias events sampled by
  the event selection; when binned in centrality, only the fraction of
  minimum-bias events in that centrality bin is considered;
\item $A$ is the geometric acceptance, which depends on the \pt and $y$
  of the quarkonium state;
\item $\varepsilon$ is the combined trigger and reconstruction
  efficiency, which depends on the \pt and $y$ of the quarkonium state
  and on the centrality of the collision;
\item $\Delta y$ and $\Delta \pt$ are the bin widths in rapidity and
  \pt, respectively;
\item \taa is the nuclear overlap function, which depends on the
  collision centrality;
\item $\lumi_{\pp} = (231\pm14)\nbinv$ is the integrated luminosity of
  the \pp data set.
\end{itemize}

Following \eq{eq:xsection}, the uncorrected yields of inclusive,
prompt and non-prompt \Jpsi, and \PgUa, measured in \PbPb collisions
are corrected for acceptance and efficiency (reported in
Figs.~\ref{fig:acceptance} and \ref{fig:eff}), and converted into
yields divided by the nuclear overlap function \taa. These quantities
can be directly compared to cross sections in \pp collisions measured
from the raw yields according to \eq{eq:xsection_pp}. The rapidity and
centrality-dependent results are presented integrated over \pt. All
results are presented for the unpolarized scenario and are tabulated
in Tables~\ref{tab:inclxsec}--\ref{tab:upsilonxsectpp} of
Appendix~\ref{app:datatables}.

The systematic uncertainties detailed in the previous sections are
summarized in Tables~\ref{tab:syst_PbPb} and~\ref{tab:syst_pp}. The
relative uncertainties for all terms appearing in
Eqs.~(\ref{eq:xsection}) and~(\ref{eq:xsection_pp}) are added in
quadrature, leading to a total of 15--21\% on the corrected
yields. For results plotted as a function of \pt or rapidity, the
systematic uncertainty on \taa enters as a global uncertainty on the
scale and is not included in the systematic uncertainties of the
yields. As a function of centrality, the uncertainty on \taa varies
point-to-point and is included in the systematic uncertainties of the
yields.

\begin{table}[htbp]
  \begin{center}
    \caption{Point-to-point systematic uncertainties on
      the prompt \Jpsi, non-prompt \Jpsi, and \PgUa\ yields measured in
      \PbPb collisions.}
    \label{tab:syst_PbPb}
    \begin{tabular}{lr@{--}lr@{--}lr@{--}l}
      \hline
      ~ & \multicolumn{2}{c}{prompt \Jpsi (\%)} & \multicolumn{2}{c}{non-prompt \Jpsi (\%)} & \multicolumn{2}{c}{\PgUa\ (\%)}\\\hline
      Yield extraction & 0.5&5.7 & 1.5&14.0 & 8.7&13.4\\
      Efficiency & 1.8&3.4 & 2.2&4.2 & 1.4&2.7\\
      Acceptance & 0.9&4.2 & 2.0&3.2 & 1.5&2.8\\
      MC Validation & \multicolumn{2}{l}{13.7} & \multicolumn{2}{l}{13.7} & \multicolumn{2}{l}{13.7}\\
      Stand-alone $\mu$ reco. & \multicolumn{2}{l}{1.0} & \multicolumn{2}{l}{1.0} & \multicolumn{2}{l}{1.0}\\
      \taa & 4.3&15.0 & 4.6&8.6 & 4.3&8.6\\\hline
      Total & 15&21 & 15&21 & 18&20\\\hline
    \end{tabular}
  \end{center}
\end{table}

\begin{table}[htbp]
  \begin{center}
    \caption{Point-to-point systematic uncertainties on
      the prompt \Jpsi, non-prompt \Jpsi, and \PgUa\ yields measured in
      \pp collisions.}
    \label{tab:syst_pp}
    \begin{tabular}{lr@{--}lr@{--}lr@{--}l}
      \hline
      ~ & \multicolumn{2}{c}{prompt \Jpsi (\%)} & \multicolumn{2}{c}{non-prompt \Jpsi (\%)} & \multicolumn{2}{c}{\PgUa\ (\%)}\\\hline
      Yield extraction & 0.8&5.3 & 5.3&16.8 & \multicolumn{2}{l}{10.0}\\
      Efficiency & 1.6&3.0 & 1.4&2.0 & 0.4&0.9\\
      Acceptance & 0.9&4.2 & 2.0&3.2 & 1.5&2.8\\
      MC Validation & \multicolumn{2}{l}{13.7} & \multicolumn{2}{l}{13.7} & \multicolumn{2}{l}{13.7}\\
      Stand-alone $\mu$ reco. & \multicolumn{2}{l}{1.0} & \multicolumn{2}{l}{1.0} & \multicolumn{2}{l}{1.0}\\\hline
      Total & 14&16 & 15&22 & 17&18\\\hline
    \end{tabular}
  \end{center}
\end{table}

The nuclear modification factor,
\begin{equation}
    \raa = \frac{\lumi_{\pp}}{\taa N_{\text{MB}}}\frac{N_{\PbPb} (\QQbar)}{N_{\pp} (\QQbar)}\cdot \frac{\varepsilon_{\pp}}{\varepsilon_{\PbPb}}\,,
\end{equation}
is calculated from the raw yields $N_{\PbPb} (\QQbar)$ and $N_{\pp}
(\QQbar)$, correcting only for the multiplicity-dependent fraction of
the efficiency ($\frac{\varepsilon_{\pp}}{\varepsilon_{\PbPb}}
\sim\!1.16$ for the most central bin); the \pt and rapidity
dependencies of the efficiency cancel in the ratio. These results are
also tabulated in Appendix~\ref{app:datatables}. It should be noted
that the \raa would be sensitive to changes of the \Jpsi polarization
between \pp and \PbPb collisions, an interesting physics effect on its
own~\cite{Faccioli:2012kp}.

In all figures showing results, statistical uncertainties are
represented by error bars and systematic uncertainties by
boxes. Results as a function of rapidity are averaged over the
positive and negative rapidity regions.

\subsection{Inclusive and Prompt \texorpdfstring{\Jpsi}{J/psi}}
\label{sec:promptResults}

\begin{figure}[htbp]
  \begin{center}
    \includegraphics[width=0.45\textwidth]{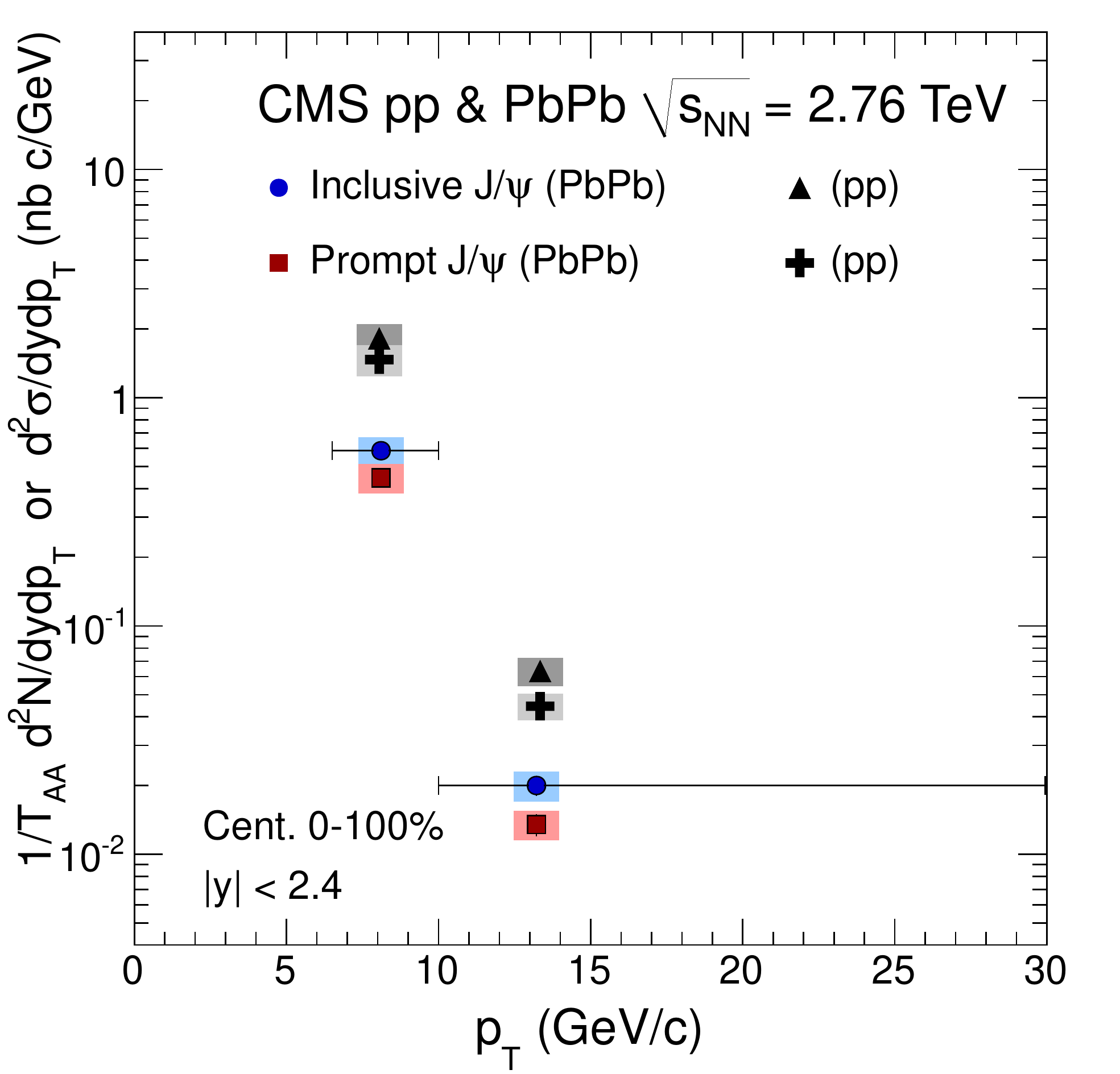}\hspace{1em}
    \includegraphics[width=0.45\textwidth]{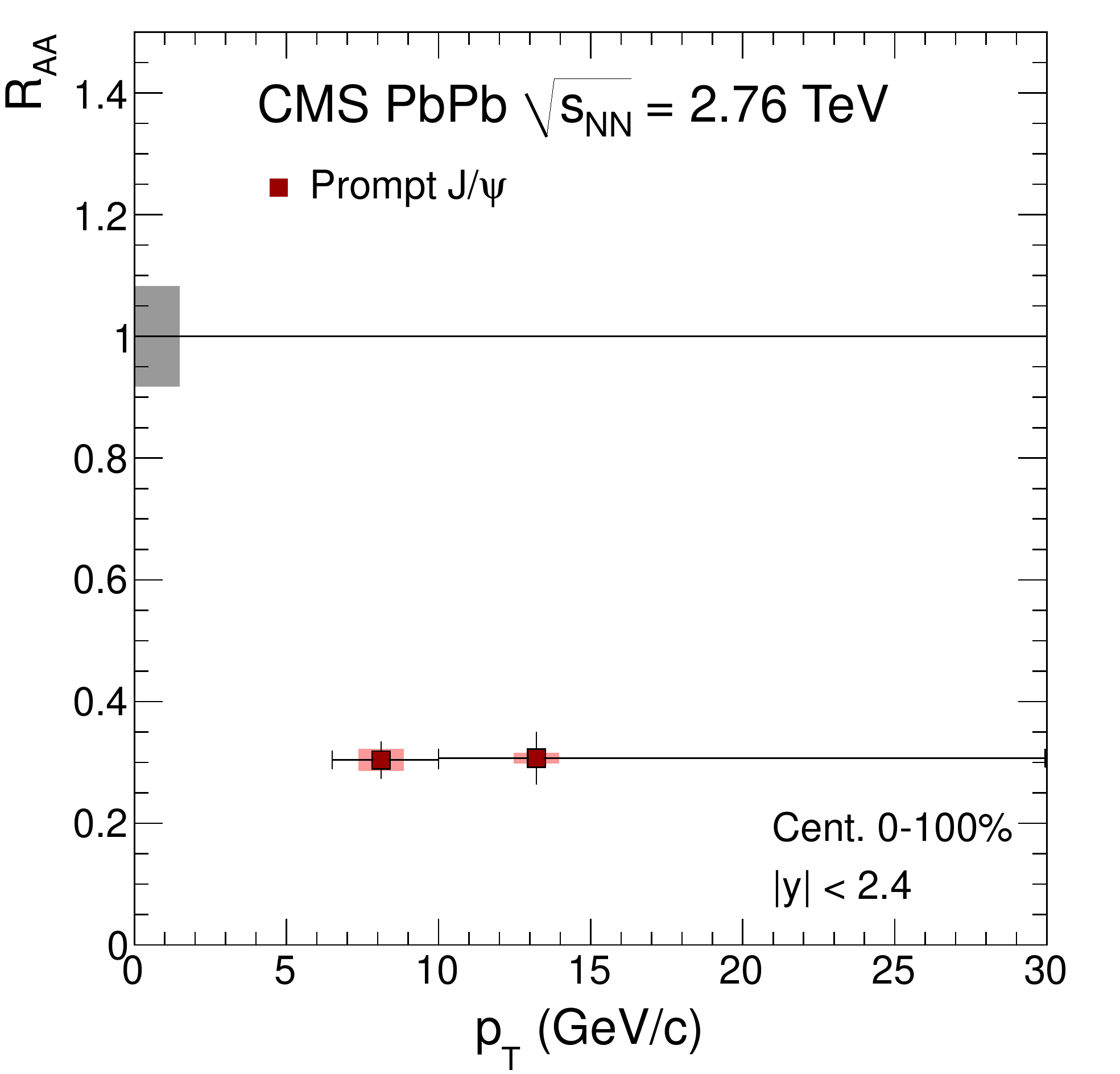}
    \caption{Left: yield of inclusive \Jpsi (blue circles) and prompt
      \Jpsi (red squares) divided by \taa as a function of \pt. The
      results are compared to the cross sections of inclusive \Jpsi
      (black triangles) and prompt \Jpsi (black crosses) measured in
      \pp. The global scale uncertainties on the \PbPb data due to
      \taa (5.7\%) and the \pp integrated luminosity (6.0\%) are not
      shown. Right: nuclear modification factor \raa of prompt \Jpsi
      as a function of \pt. A global uncertainty of 8.3\%, from \taa
      and the integrated luminosity of the \pp data sample, is shown
      as a grey box at $\raa=1$. Points are plotted at their measured
      average \pt. Statistical (systematic) uncertainties are shown as
      bars (boxes). Horizontal bars indicate the bin width.}
    \label{fig:inclJpsi_pt}
  \end{center}
\end{figure}

\begin{figure}[htbp]
  \begin{center}
    \includegraphics[width=0.45\textwidth]{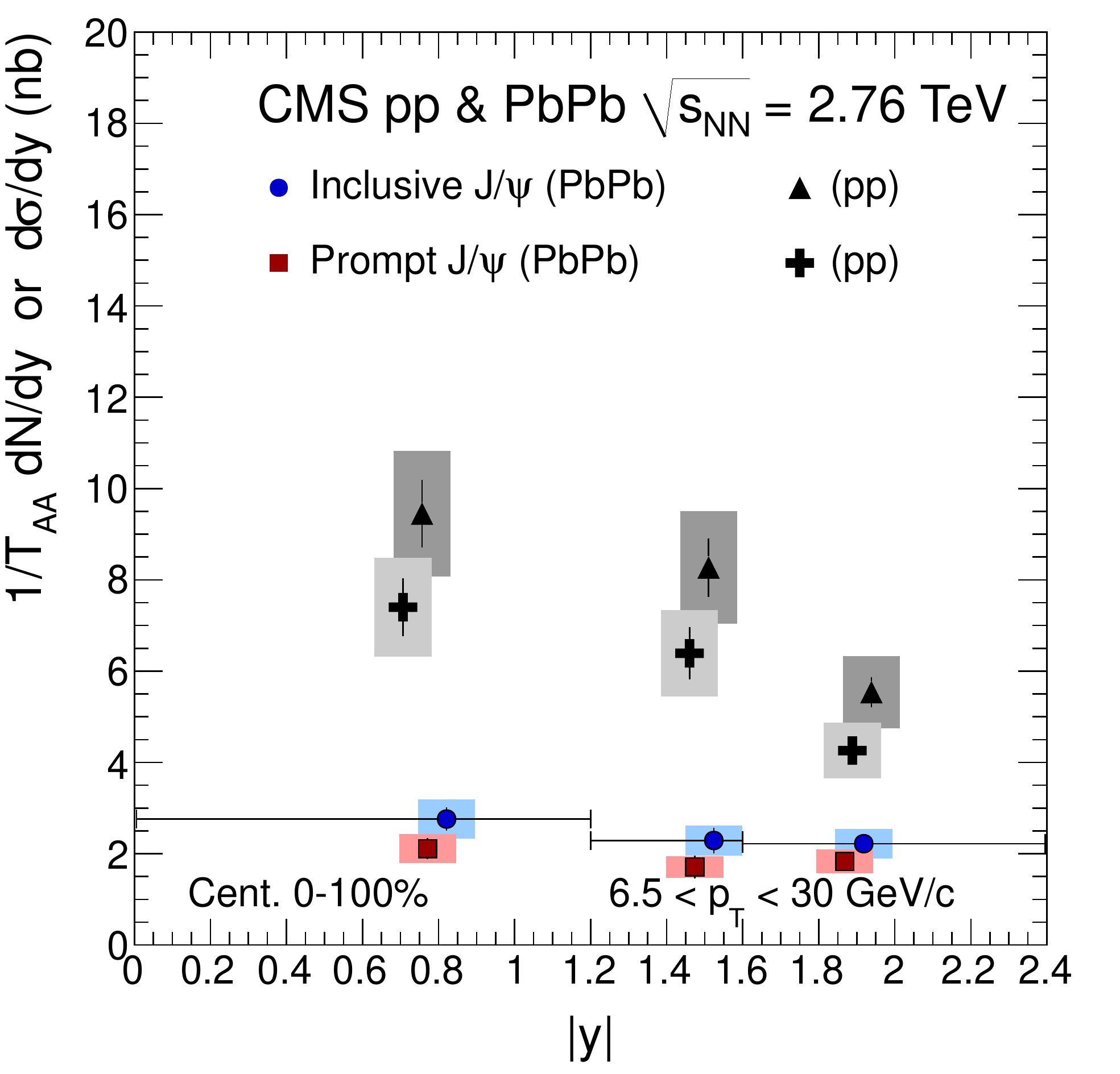}\hspace{1em}
    \includegraphics[width=0.45\textwidth]{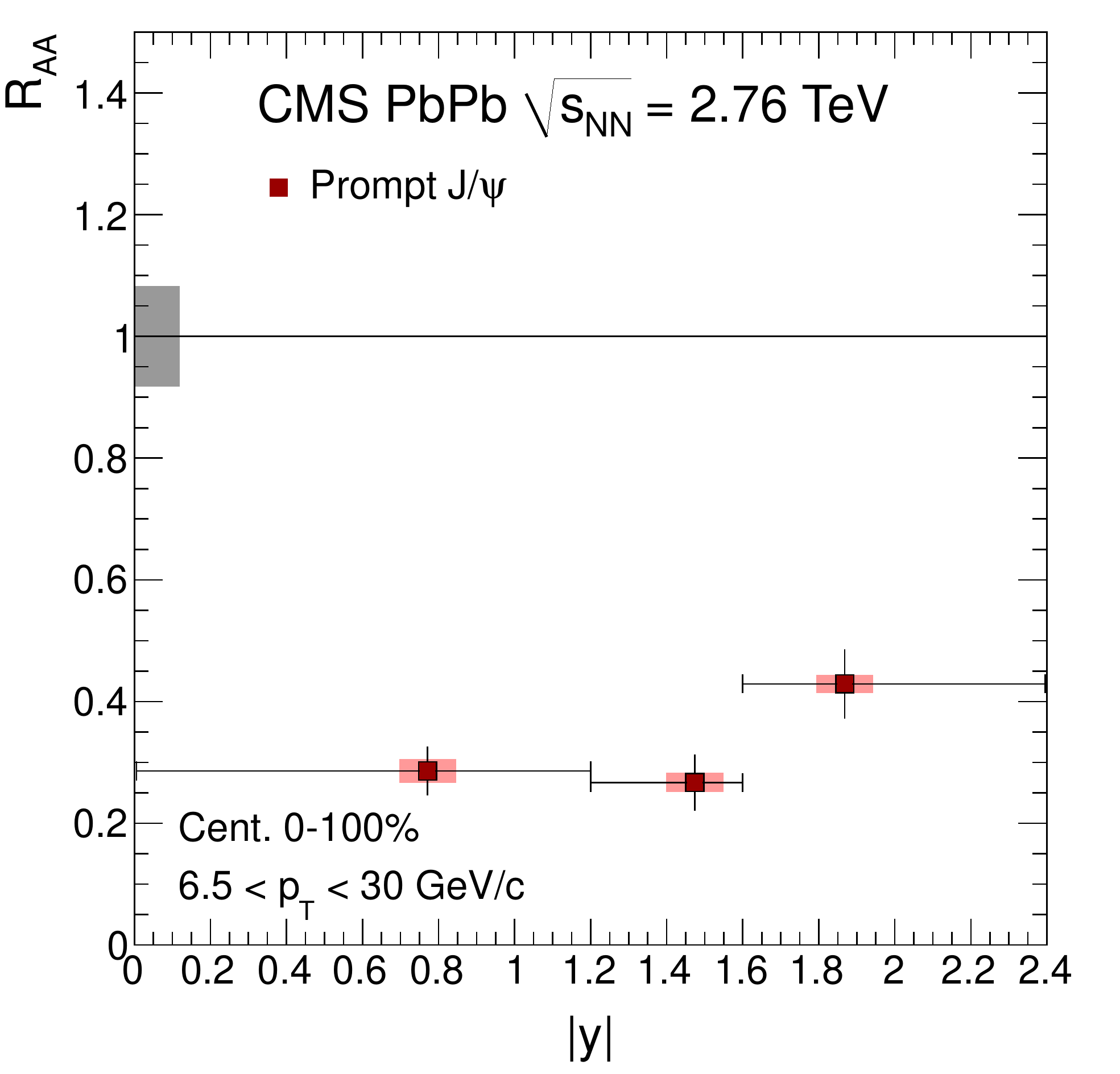}
    \caption{Left: yield of inclusive \Jpsi (blue circles) and prompt
      \Jpsi (red squares) divided by \taa as a function of
      rapidity. The results are compared to the cross sections of
      inclusive \Jpsi (black triangles) and prompt \Jpsi (black
      crosses) measured in \pp. The inclusive \Jpsi points are shifted
      by $\Delta y = 0.05$ for better visibility. The global scale
      uncertainties on the \PbPb data due to \taa (5.7\%) and the \pp
      luminosity (6.0\%) are not shown. Right: nuclear modification
      factor \raa of prompt \Jpsi as a function of rapidity. A global
      uncertainty of 8.3\%, from \taa and the integrated luminosity of
      the \pp data sample, is shown as a grey box at $\raa=1$. Points
      are plotted at their measured average $|y|$. Statistical
      (systematic) uncertainties are shown as bars (boxes). Horizontal
      bars indicate the bin width.}
    \label{fig:inclJpsi_eta}
  \end{center}
\end{figure}

The inclusive and prompt \Jpsi differential yields in \PbPb
collisions, divided by \taa, are shown in the left panel of
\fig{fig:inclJpsi_pt} as a function of \pt, for $|y|<2.4$ and
integrated over centrality. The corresponding \pp cross sections are
also shown. The suppression of the prompt \Jpsi yield by a factor of
$\sim\!3$ with respect to \pp is easier to appreciate through the \raa
observable, shown in the right panel of \fig{fig:inclJpsi_pt}. The
\raa measurements do not exhibit a \pt dependence over the measured
\pt range, while there is an indication of less suppression in the
most forward rapidity bin ($1.6 < |y| < 2.4$) in comparison to the
mid-rapidity bin, as shown in \fig{fig:inclJpsi_eta}. At forward
rapidity, in addition to $6.5<\pt<30\GeVc$ the nuclear modification
factor is measured for lower \pt (down to 3\GeVc) without observing a
significant change, as can be seen in \tab{tab:promptxsec}.


\begin{figure}[htbp]
  \begin{center}
    \includegraphics[width=0.45\textwidth]{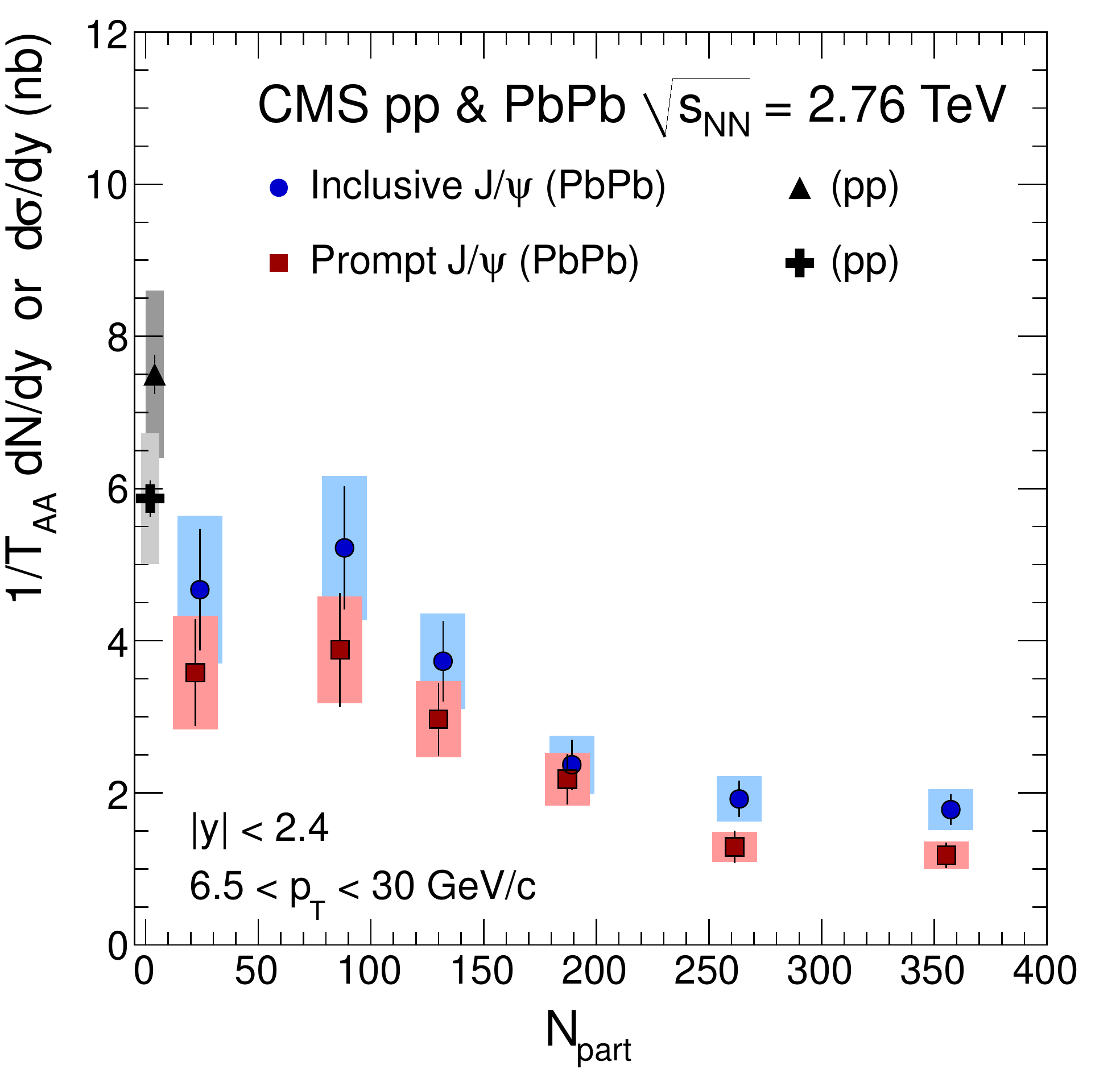}\hspace{1em}
    \includegraphics[width=0.45\textwidth]{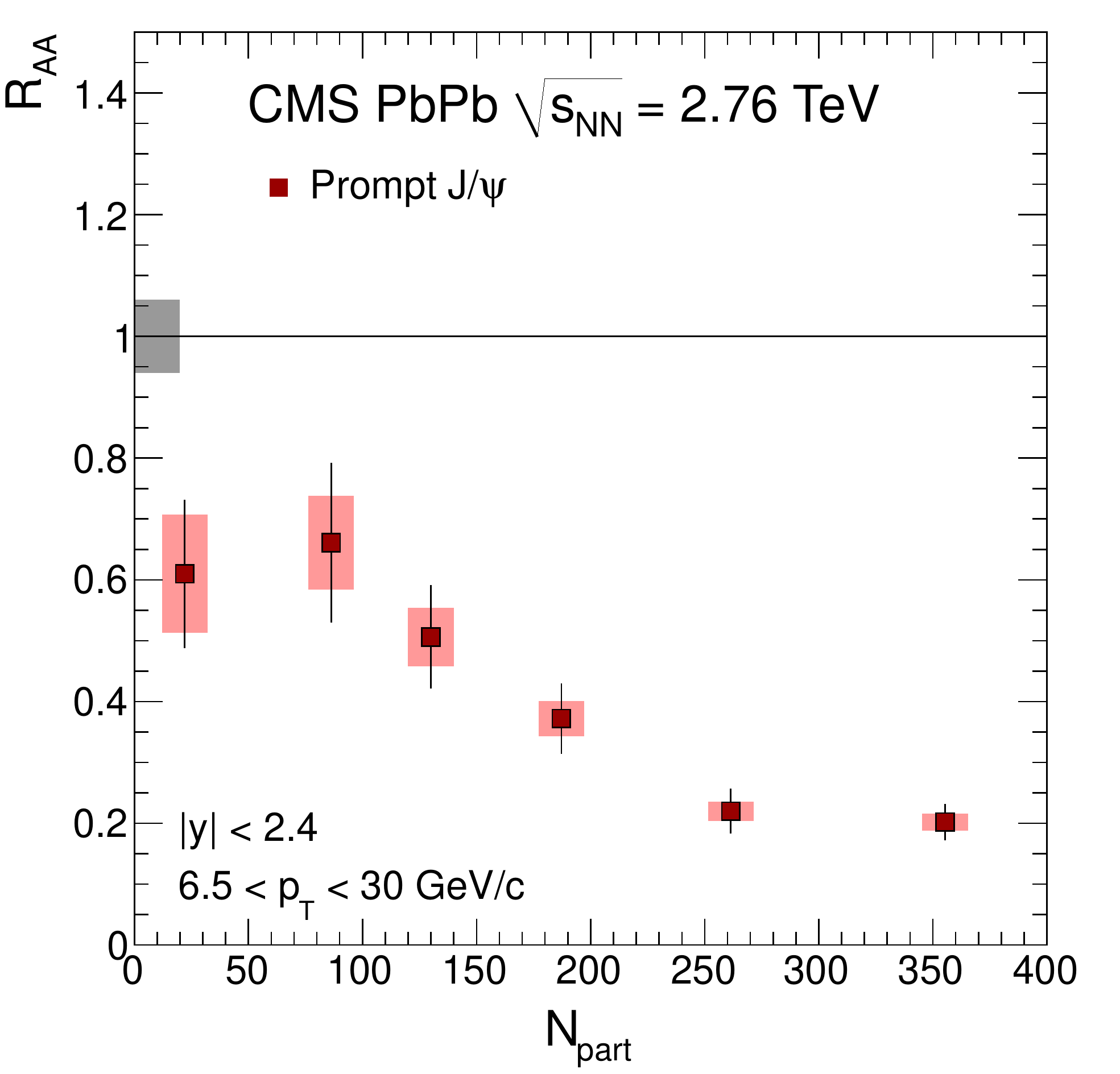}
    \caption{Left: yield of inclusive \Jpsi (blue circles) and prompt
      \Jpsi (red squares) divided by \taa as a function of \npart. The
      results are compared to the cross sections of inclusive \Jpsi
      (black triangle) and prompt \Jpsi (black cross) measured in
      \pp. The inclusive \Jpsi points are shifted by $\Delta \npart =
      2$ for better visibility. Right: nuclear modification factor
      \raa of prompt \Jpsi as a function of \npart. A global
      uncertainty of 6\%, from the integrated luminosity of the \pp
      data sample, is shown as a grey box at $\raa=1$. Statistical
      (systematic) uncertainties are shown as bars (boxes).}
    \label{fig:inclJpsi_cent}
  \end{center}
\end{figure}

The inclusive \Jpsi yield in \PbPb collisions divided by \taa,
integrated over the \pt range 6.5--30\GeVc and $|y|<2.4$, is shown in
the left panel of \fig{fig:inclJpsi_cent} as a function of
\npart. Also included is the prompt \Jpsi yield, which exhibits the
same centrality dependence as the inclusive \Jpsi: from the 50--100\%
centrality bin ($\langle\npart\rangle=22.1$) to the 10\% most central
collisions ($\langle\npart\rangle=355.4$) the yield divided by \taa
falls by a factor of $\sim\!2.6$. The results are compared to the
cross sections measured in \pp, showing that prompt \Jpsi are already
suppressed in peripheral \PbPb collisions. The \raa of prompt \Jpsi as
a function of \npart is shown in the right panel of
\fig{fig:inclJpsi_cent}: a suppression of $\sim\!5$ is observed in the
10\% most central \PbPb collisions with respect to \pp. This
suppression is reduced in more peripheral collisions, reaching a
factor of $\sim\!1.6$ in the 50--100\% centrality bin.

\subsection{Non-prompt \texorpdfstring{\Jpsi}{J/psi}}
\label{sec:nonpromptResults}

\begin{figure}[htbp]
  \begin{center}
    \includegraphics[width=0.6\textwidth]{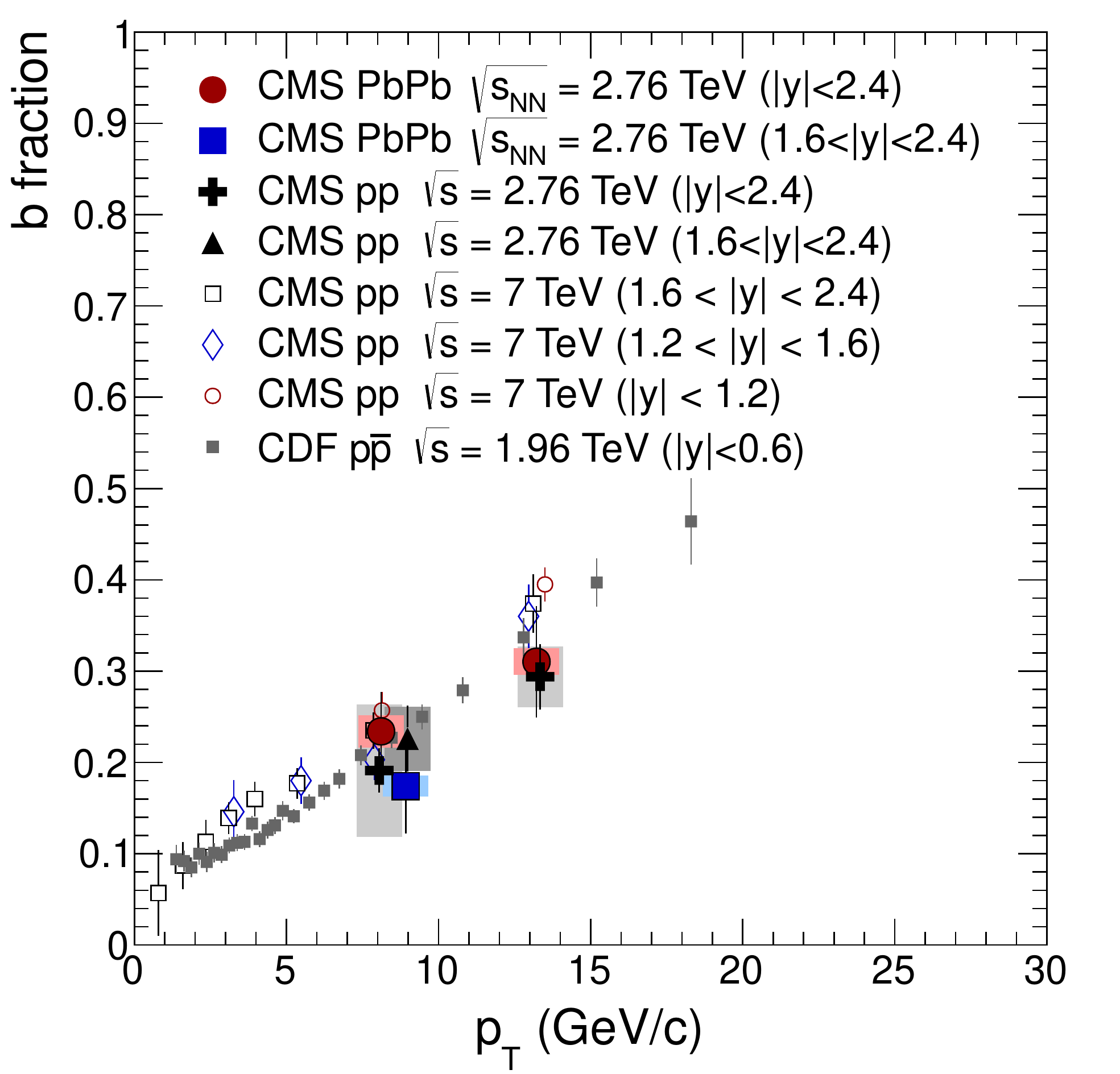}
    \caption{\cPqb\ fraction of \Jpsi production in \pp and \PbPb
      collisions at \sqrtsnn = 2.76~\TeV as a function of \pt for the
      rapidity bins $|y|<2.4$ and $1.6<|y|<2.4$, compared to \cPqb\
      fractions measured by CDF in \ppbar collisions at \sqrts =
      1.96~\TeV~\cite{Acosta:2004yw} and by CMS in \pp collisions at
      \sqrts = 7~\TeV~\cite{Khachatryan:2010yr}. Points are plotted at
      their measured average \pt. Statistical (systematic)
      uncertainties are shown as bars (boxes).}
    \label{fig:nonpromptJpsi_pt}
  \end{center}
\end{figure}

The uncorrected fraction of non-prompt \Jpsi is obtained from the
two-dimensional fit to the invariant mass and $\ell_{\Jpsi}$ spectra
discussed in Section~\ref{sec:promptJpsi}. To obtain the corrected
\cPqb\ fraction, which is the ratio of non-prompt to inclusive \Jpsi,
the raw fraction is corrected for the different reconstruction
efficiencies and acceptances for prompt and non-prompt \Jpsi. The
\cPqb\ fraction in \pp and in \PbPb (integrated over centrality) at
\sqrtsnn = 2.76\TeV is presented in \fig{fig:nonpromptJpsi_pt} as a
function of \pt, for several rapidity bins, together with results from
CDF~\cite{Acosta:2004yw} and CMS~\cite{Khachatryan:2010yr} at other
collision energies. There is good agreement, within uncertainties,
between the earlier results and the present measurements.

\begin{figure}[htbp]
  \begin{center}
    \includegraphics[width=0.45\textwidth]{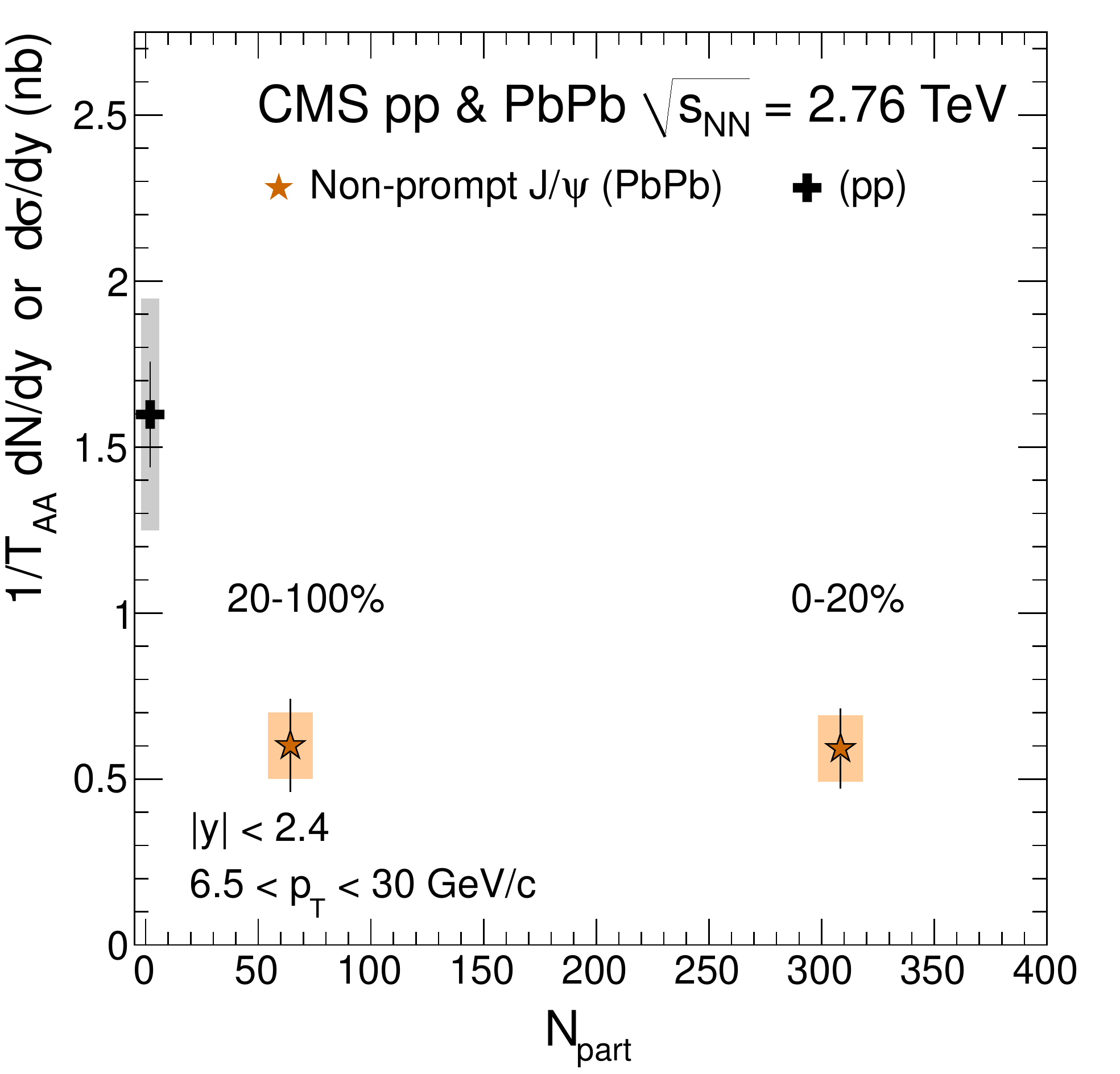}\hspace{1em}
    \includegraphics[width=0.45\textwidth]{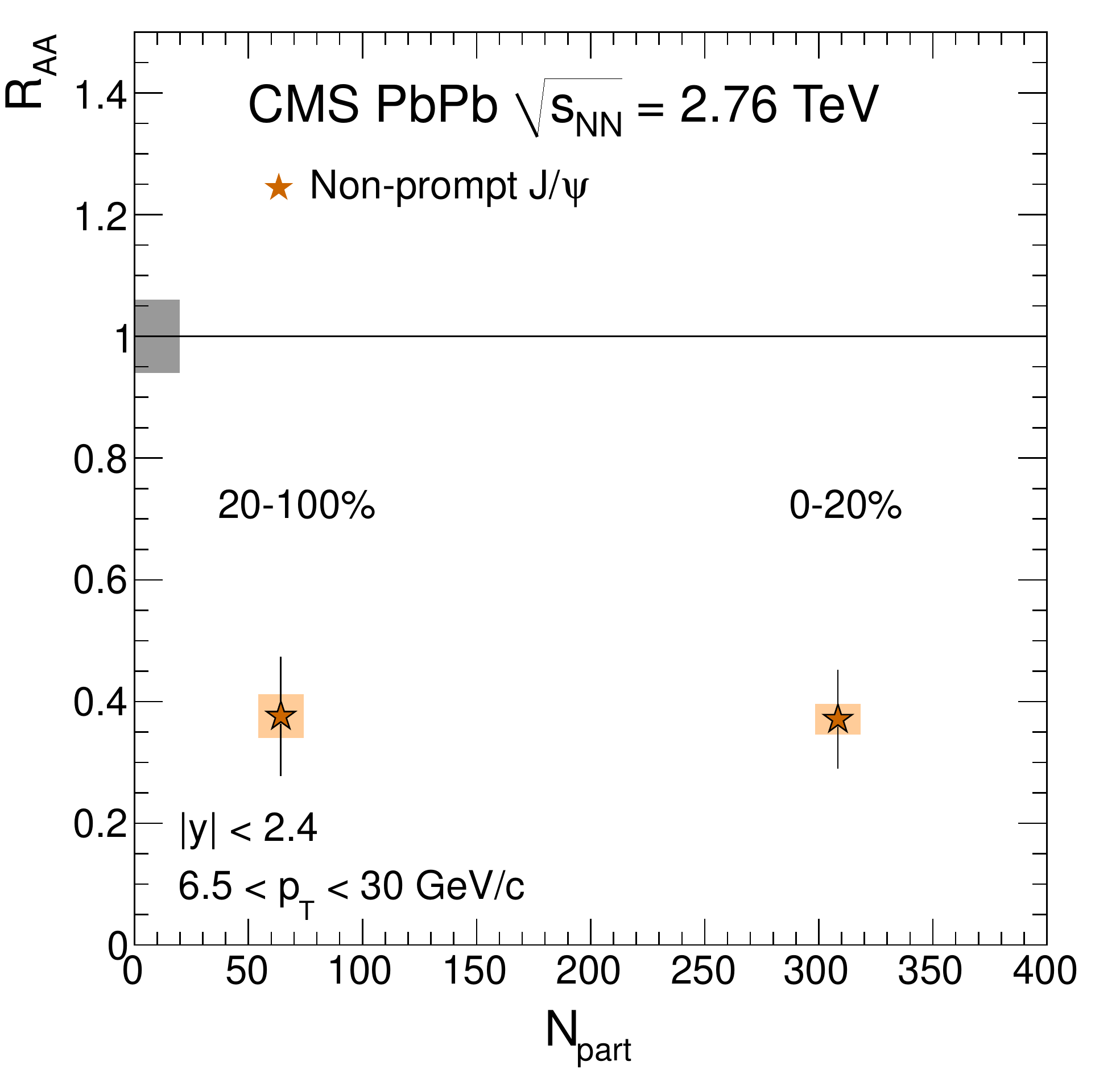}
    \caption{Left: non-prompt \Jpsi yield divided by \taa (orange
      stars) as a function of \npart compared to the non-prompt \Jpsi
      cross section measured in \pp (black cross). Right: nuclear
      modification factor \raa of non-prompt \Jpsi as a function of
      \npart. A global uncertainty of 6\%, from the integrated
      luminosity of the \pp data sample, is shown as a grey box at
      $\raa=1$. Statistical (systematic) uncertainties are shown as
      bars (boxes).}
    \label{fig:nonpromptJpsi_cent}
  \end{center}
\end{figure}

The non-prompt \Jpsi yield in \PbPb collisions divided by \taa,
integrated over the \pt range 6.5--30\GeVc and $|y|<2.4$, is shown in
the left panel of \fig{fig:nonpromptJpsi_cent} as a function of
\npart, together with the \pp cross section. Non-prompt \Jpsi are
suppressed by a factor of $\sim\!2.6$ with respect to \pp collisions,
as can be seen in the right panel of \fig{fig:nonpromptJpsi_cent}. The
suppression does not exhibit a centrality dependence, but the most
peripheral centrality bin (20--100\%, $\langle\npart\rangle=64.2$) is
very broad. Hard processes, such as quarkonium and \cPqb-hadron
production, are produced following a scaling with the number of
nucleon-nucleon collisions, thus most events in such a large bin occur
towards its most central edge.

\subsection{\texorpdfstring{\PgUa}{Upsilon(1S)}}
\label{sec:upsResults}

\begin{figure}[htbp]
  \begin{center}
    \includegraphics[width=0.45\textwidth]{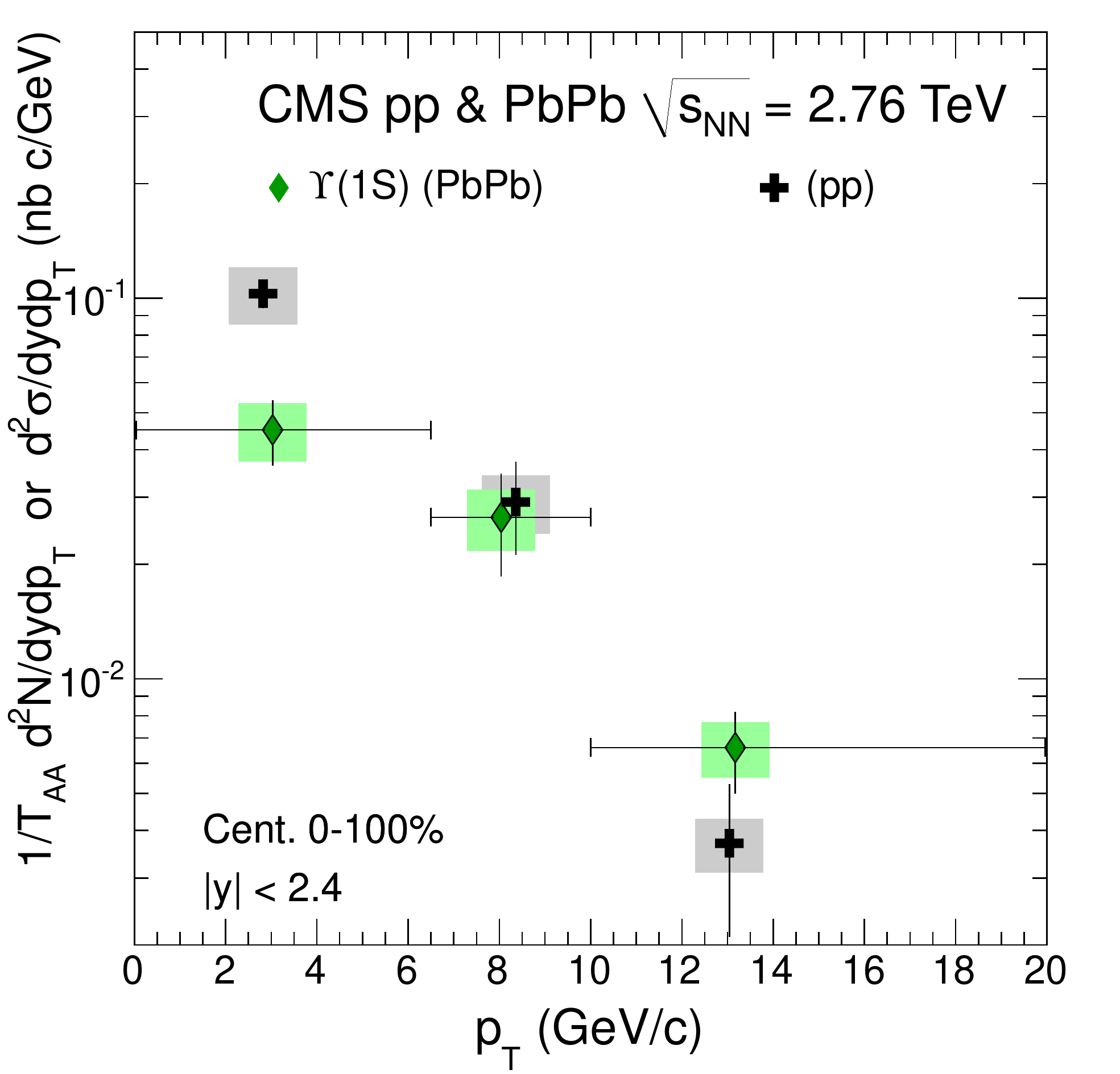}\hspace{1em}
    \includegraphics[width=0.45\textwidth]{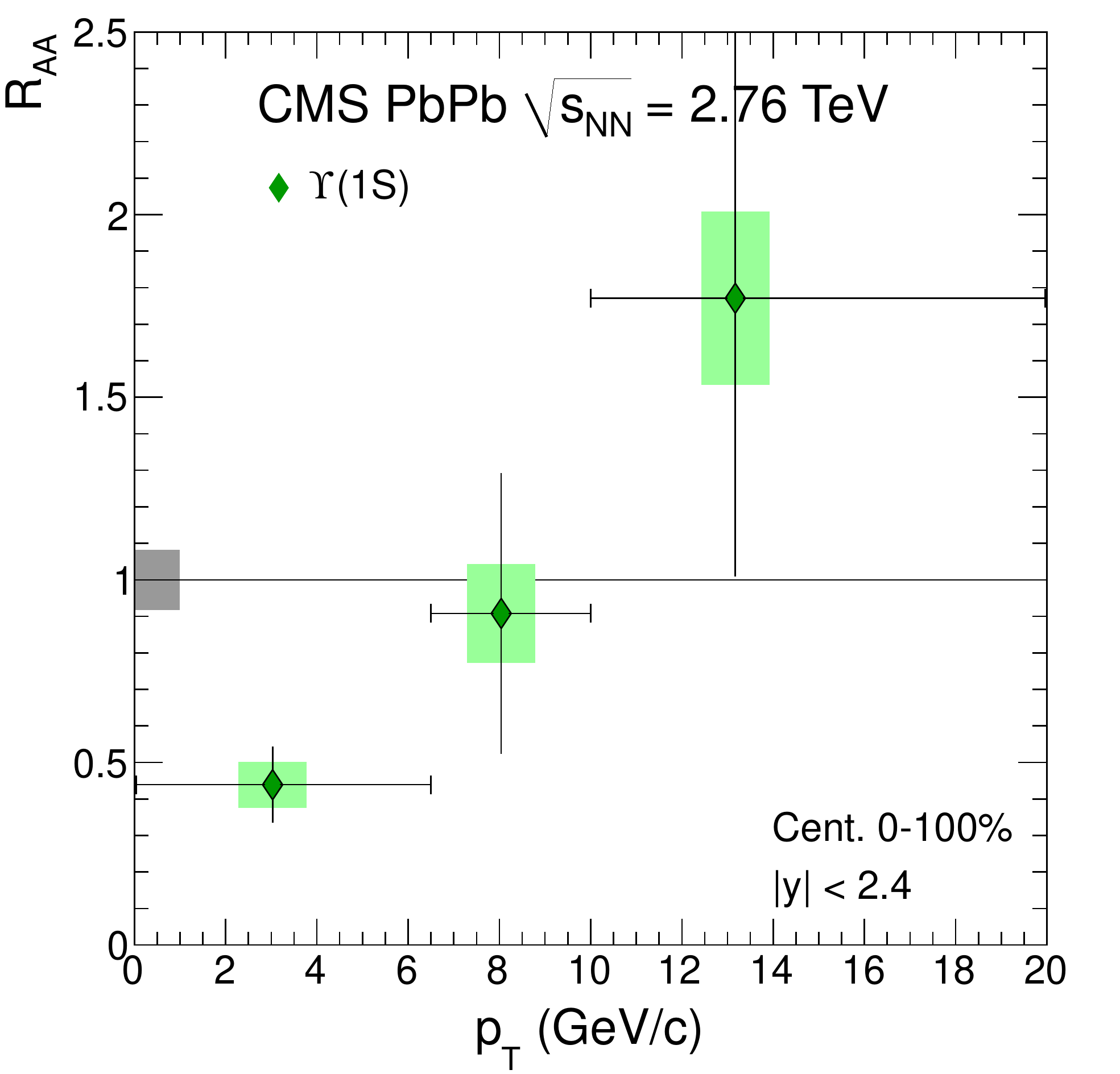}
    \caption{Left: \PgUa\ yield divided by \taa in \PbPb collisions
      (green diamonds) as a function of \pt. The result is compared to
      the cross section measured in \pp collisions (black crosses).
      The global scale uncertainties on the \PbPb data due to \taa
      (5.7\%) and the \pp integrated luminosity (6.0\%) are not
      shown. Right: nuclear modification factor \raa of \PgUa\ as a
      function of \pt. A global uncertainty of 8.3\%, from \taa and
      the integrated luminosity of the \pp data sample, is shown as a
      grey box at $\raa=1$. Points are plotted at their measured
      average \pt. Statistical (systematic) uncertainties are shown as
      bars (boxes). Horizontal bars indicate the bin width.}
    \label{fig:upsilon_pt}
  \end{center}
\end{figure}

\begin{figure}[htbp]
  \begin{center}
    \includegraphics[width=0.45\textwidth]{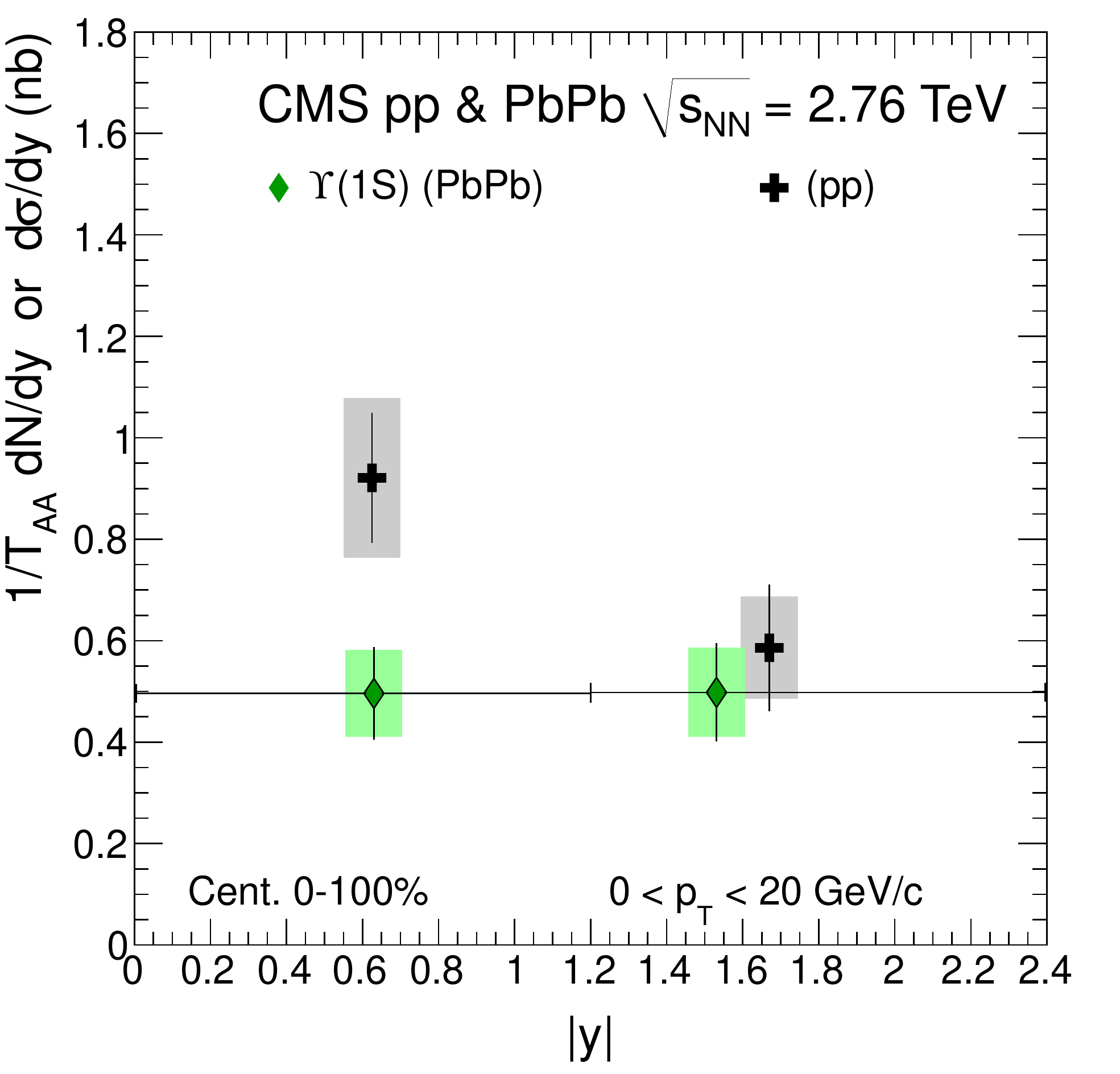}\hspace{1em}
    \includegraphics[width=0.45\textwidth]{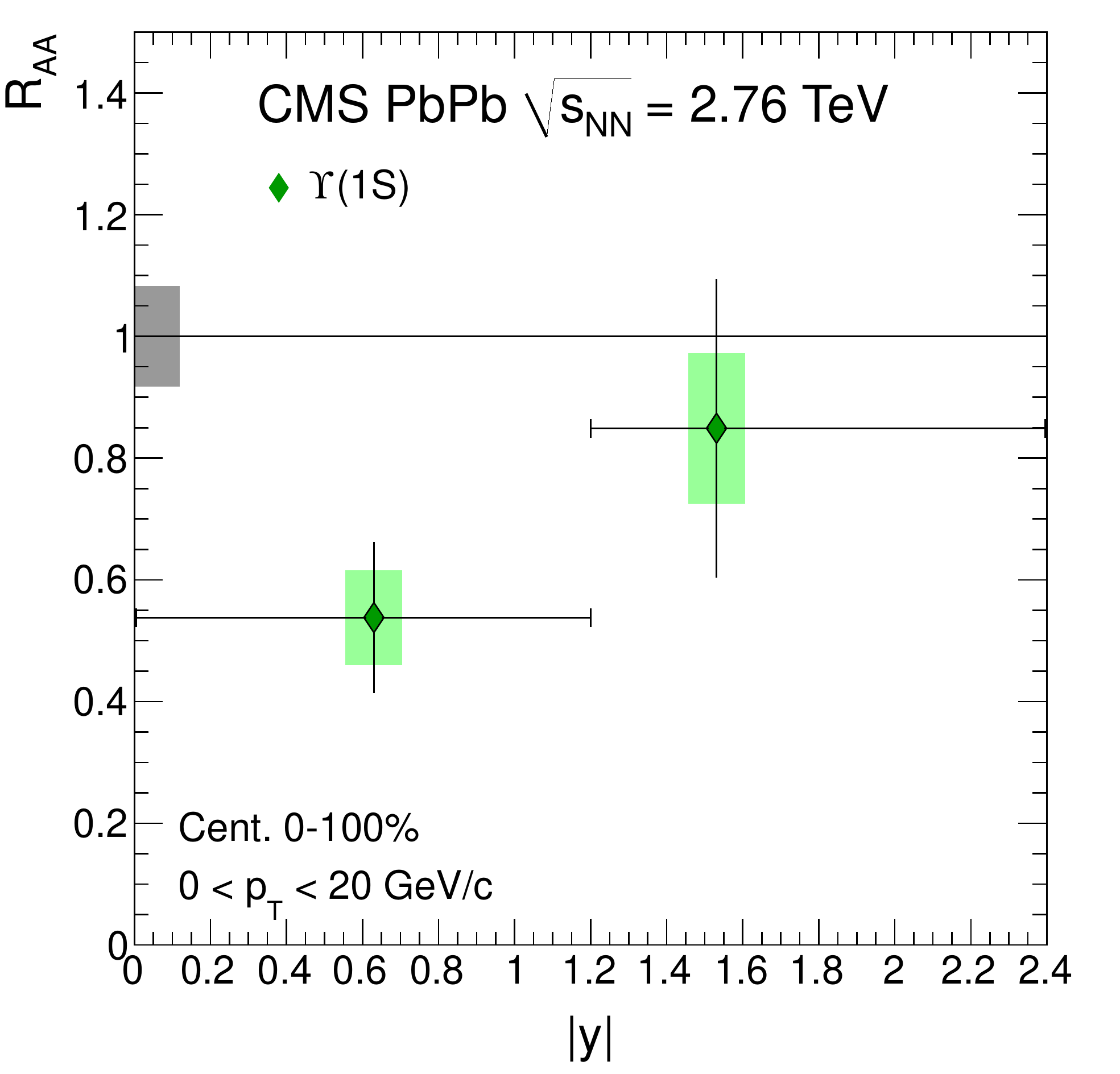}
    \caption{Left: \PgUa\ yield divided by \taa in \PbPb collisions
      (green diamonds) as a function of rapidity. The result is
      compared to the cross section measured in \pp collisions (black
      crosses). The global scale uncertainties on the \PbPb data due
      to \taa (5.7\%) and the \pp integrated luminosity (6.0\%) are
      not shown. Right: nuclear modification factor \raa of \PgUa\ as a
      function of rapidity. A global uncertainty of 8.3\%, from \taa
      and the integrated luminosity of the \pp data sample, is shown
      as a grey box at $\raa=1$. Points are plotted at their measured
      average $|y|$. Statistical (systematic) uncertainties are shown
      as bars (boxes). Horizontal bars indicate the bin width.}
    \label{fig:upsilon_eta}
  \end{center}
\end{figure}

\begin{figure}[htbp]
  \begin{center}
    \includegraphics[width=0.45\textwidth]{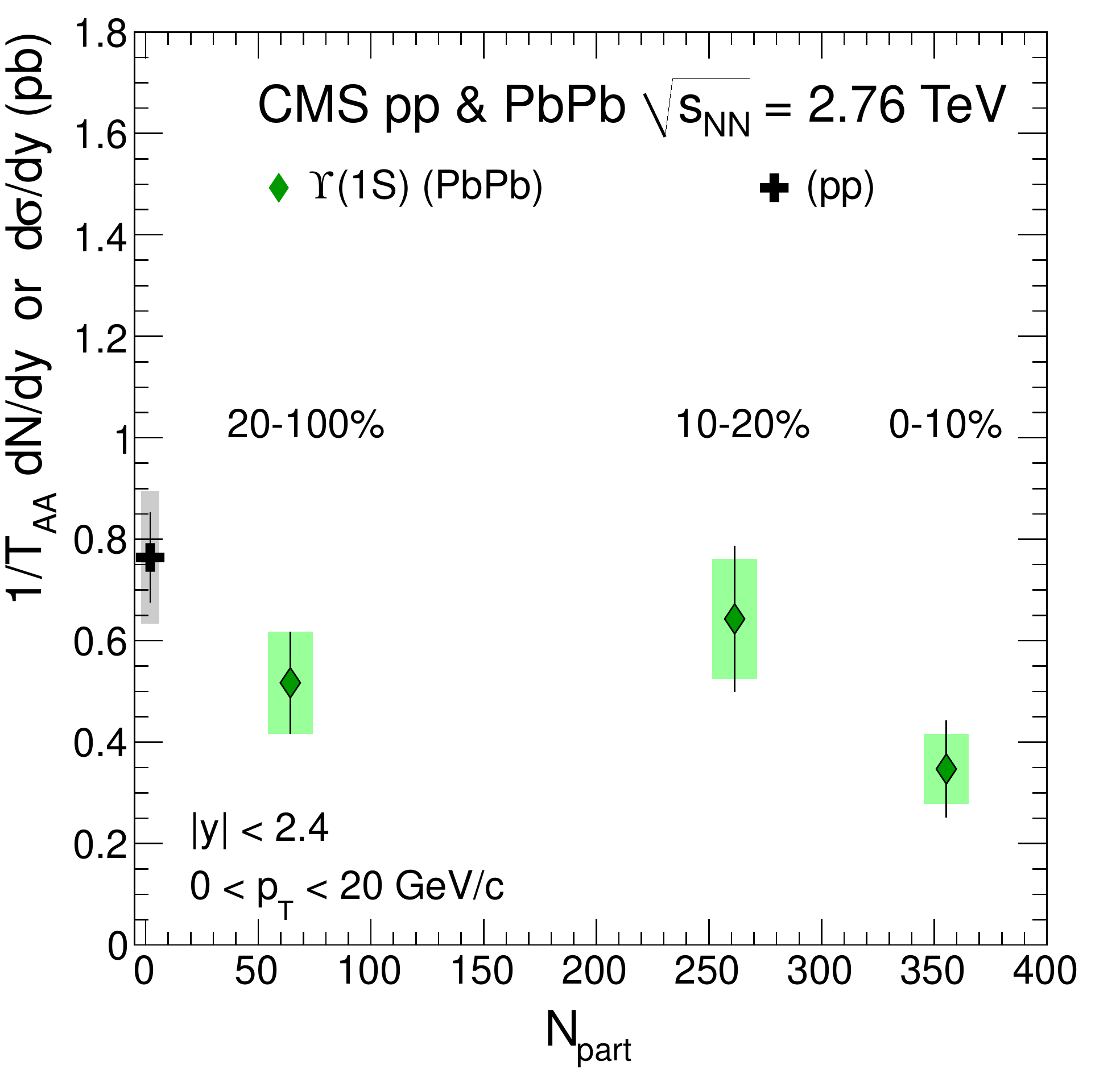}\hspace{1em}
    \includegraphics[width=0.45\textwidth]{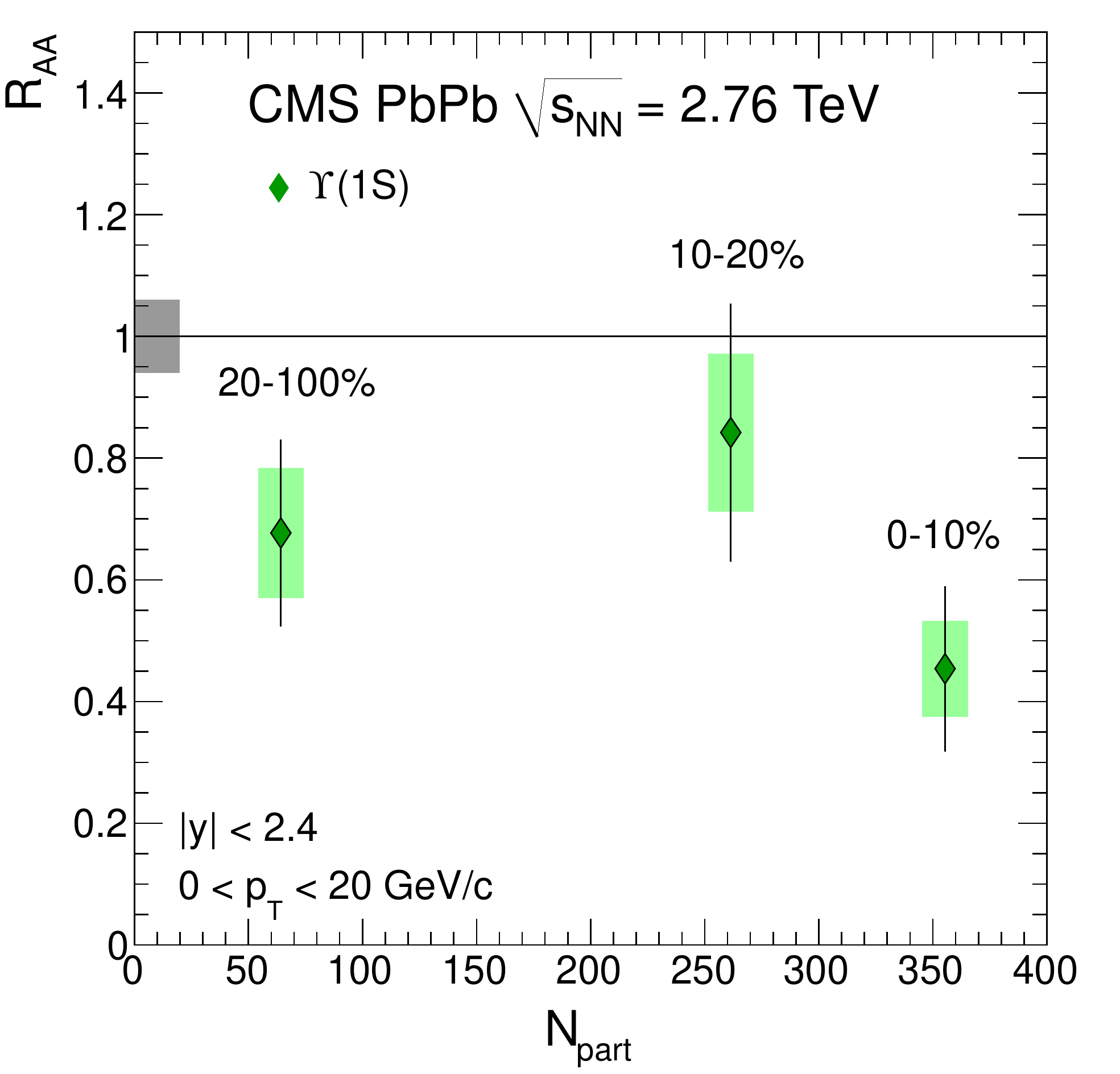}
    \caption{Left: \PgUa\ yield divided by \taa (green diamonds) as a
      function of \npart compared to the \PgUa\ cross section measured
      in \pp (black cross). Right: nuclear modification factor \raa of
      \PgUa\ as a function of \npart. A global uncertainty of 6\%, from
      the integrated luminosity of the \pp data sample, is shown as a
      grey box at $\raa=1$. Statistical (systematic) uncertainties are
      shown as bars (boxes).}
    \label{fig:upsilon_cent}
  \end{center}
\end{figure}

In \fig{fig:upsilon_pt}, the \PgUa\ yield divided by \taa in \PbPb
collisions and its cross section in \pp collisions are shown as a
function of \pt; the \raa of \PgUa\ is displayed in the right panel of
\fig{fig:upsilon_pt}. The \pt dependence shows a significant
suppression, by a factor of $\sim\!2.3$ at low \pt, that disappears
for $\pt > 6.5\GeVc$. The rapidity dependence indicates a slightly
smaller suppression at forward rapidity, as shown in
\fig{fig:upsilon_eta}. However, the statistical uncertainties are too
large to draw strong conclusions on any \pt or rapidity
dependence. The \PgUa\ yield in \PbPb collisions divided by \taa and
the \PgUa\ \raa are presented as a function of \npart in the left and
right panels of \fig{fig:upsilon_cent}, respectively. Within
uncertainties, no centrality dependence of the \PgUa\ suppression is
observed.


\section{Discussion}
\label{sec:discussion}
This paper has presented the first measurements of the prompt and
non-prompt \Jpsi, as well as the \PgUa\ mesons, via their decays into
\mumu pairs in \PbPb and \pp collisions at \sqrtsnn = 2.76\TeV. The
results are based on data recorded with the CMS detector from the
first LHC \PbPb run in 2010, and from a \pp run during March 2011 at
\sqrts = 2.76\TeV.

The prompt \Jpsi cross section shows a factor of two suppression in
central \PbPb collisions with respect to peripheral collisions for
\Jpsi with $6.5<\pt<30\GeVc$. With respect to \pp, a nuclear
modification factor of $\raa =
0.20\pm0.03\,(\text{stat.})\pm0.01\,(\text{syst.})$ has been measured
in the 10\% most central collisions. Prompt \Jpsi produced in
peripheral collisions are already suppressed with respect to \pp:
$\raa = 0.61\pm0.12\,(\text{stat.})\pm0.10\,(\text{syst.})$ in the
50--100\% centrality bin. While no \pt dependence is observed in the
measured \pt range, within uncertainties, less suppression is observed
at forward rapidity ($\raa =
0.43\pm0.06\,(\text{stat.})\pm0.01\,(\text{syst.})$) than at
mid-rapidity ($\raa =
0.29\pm0.04\,(\text{stat.})\pm0.02\,(\text{syst.})$).

A comparison of the \raa centrality dependence to results measured for
$\pt < 5\GeVc$ by PHENIX~\cite{Adare:2006ns} in \AuAu collisions at
\sqrtsnn = 200\GeV shows a similar suppression, despite the different
collision energies and kinematic ranges.  Integrated over centrality,
CMS has measured an inclusive \Jpsi nuclear modification factor of
$\raa = 0.41\pm0.05\,(\text{stat.})\pm0.02\,(\text{syst.})$ in the
most forward rapidity bin ($1.6<|y|<2.4$) in the \pt range
$3<\pt<30\GeVc$. This result is consistent with the ALICE measurement
of an inclusive \Jpsi \raa of $\sim\!0.5$ at rapidity $2.5<y<3.25$ for
$\pt>3\GeVc$~\cite{Abelev:2012rv}.

A strong suppression of non-prompt \Jpsi mesons is observed in \PbPb
collisions when compared to \pp collisions. This is the first
unambiguous measurement of \cPqb-hadron suppression in heavy-ion
collisions, which is likely connected to in-medium energy loss of
\cPqb\ quarks. The average \pt of the non-prompt \Jpsi in the measured
kinematic range is $\sim\!10\GeVc$. Based on simulations of
\cPqb-hadron decays, this translates into an average \cPqb-hadron \pt
of $\sim\!13\GeVc$. The suppression of non-prompt \Jpsi is of a
comparable magnitude to the charged hadron \raa measured by
ALICE~\cite{Aamodt:2010jd}, which reflects the in-medium energy loss
of light quarks. The non-prompt \Jpsi yield, though strongly
suppressed ($\raa = 0.37\pm0.08\,(\text{stat.})\pm0.02
(\text{syst.})$) in the 20\% most central collisions, shows no strong
centrality dependence, within uncertainties, when compared to a broad
peripheral region (20--100\%). Furthermore, this suppression of
non-prompt \Jpsi is comparable in size to that observed for high-\pt
single electrons from semileptonic heavy-flavour decays at
RHIC~\cite{Adare:2006nq,Abelev:2006db,Abelev:2006dbErratum} in which
charm and bottom decays were not separated.

The \PgUa\ yield divided by \taa as a function of \pt, rapidity, and
centrality has been measured in \PbPb collisions.  No strong
centrality dependence is observed within the uncertainties. The
nuclear modification factor integrated over centrality is $\raa =
0.63\pm0.11\,(\text{stat.})\pm0.09\,(\text{syst.})$. This suppression
is observed predominantly at low \pt. Using \ppbar collisions at
\sqrts = 1.8\TeV, CDF measured the fraction of directly produced \PgUa\
as $(50.9\pm8.2\,(\text{stat.})\pm9.0\,(\text{syst.}))\%$ for \PgUa\
with $\pt>8\GeVc$~\cite{Affolder:1999wm}. Therefore, the \PgUa\
suppression presented in this paper could be indirectly caused by the
suppression of excited \PgU\ states, as indicated by earlier results
from CMS~\cite{Chatrchyan:2011pe}.

\section{Summary}
\label{sec:conclusion}
In summary, CMS has presented the first measurements of prompt \Jpsi,
non-prompt \Jpsi, and \PgUa\ suppression in \PbPb collisions at
\sqrtsnn = 2.76\TeV. Prompt \Jpsi are found to be suppressed, with a
strong centrality dependence.
By measuring non-prompt \Jpsi, CMS has directly observed the
suppression of \cPqb\ hadrons for the first time. The measurement of
\PgUa\ suppression, together with the suppression of the \PgUbc\
states~\cite{Chatrchyan:2011pe}, marks the first steps of detailed
bottomonium studies in heavy-ion collisions.

\section*{Acknowledgements}
\label{sec:ack}
\hyphenation{Bundes-ministerium Forschungs-gemeinschaft
Forschungs-zentren} We wish to congratulate our colleagues in the CERN
accelerator departments for the excellent performance of the LHC
machine. We thank the technical and administrative staff at CERN and
other CMS institutes. This work was supported by the Austrian Federal
Ministry of Science and Research; the Belgium Fonds de la Recherche
Scientifique, and Fonds voor Wetenschappelijk Onderzoek; the Brazilian
Funding Agencies (CNPq, CAPES, FAPERJ, and FAPESP); the Bulgarian
Ministry of Education and Science; CERN; the Chinese Academy of
Sciences, Ministry of Science and Technology, and National Natural
Science Foundation of China; the Colombian Funding Agency
(COLCIENCIAS); the Croatian Ministry of Science, Education and Sport;
the Research Promotion Foundation, Cyprus; the Estonian Academy of
Sciences and NICPB; the Academy of Finland, Finnish Ministry of
Education and Culture, and Helsinki Institute of Physics; the Institut
National de Physique Nucl\'eaire et de Physique des Particules~/~CNRS,
and Commissariat \`a l'\'Energie Atomique et aux \'Energies
Alternatives~/~CEA, France; the Bundesministerium f\"ur Bildung und
Forschung, Deutsche Forschungsgemeinschaft, and Helmholtz-Gemeinschaft
Deutscher Forschungszentren, Germany; the General Secretariat for
Research and Technology, Greece; the National Scientific Research
Foundation, and National Office for Research and Technology, Hungary;
the Department of Atomic Energy and the Department of Science and
Technology, India; the Institute for Studies in Theoretical Physics
and Mathematics, Iran; the Science Foundation, Ireland; the Istituto
Nazionale di Fisica Nucleare, Italy; the Korean Ministry of Education,
Science and Technology and the World Class University program of NRF,
Korea; the Lithuanian Academy of Sciences; the Mexican Funding
Agencies (CINVESTAV, CONACYT, SEP, and UASLP-FAI); the Ministry of
Science and Innovation, New Zealand; the Pakistan Atomic Energy
Commission; the Ministry of Science and Higher Education and the
National Science Centre, Poland; the Funda\c{c}\~ao para a Ci\^encia e
a Tecnologia, Portugal; JINR (Armenia, Belarus, Georgia, Ukraine,
Uzbekistan); the Ministry of Education and Science of the Russian
Federation, the Federal Agency of Atomic Energy of the Russian
Federation, Russian Academy of Sciences, and the Russian Foundation
for Basic Research; the Ministry of Science and Technological
Development of Serbia; the Ministerio de Ciencia e Innovaci\'on, and
Programa Consolider-Ingenio 2010, Spain; the Swiss Funding Agencies
(ETH Board, ETH Zurich, PSI, SNF, UniZH, Canton Zurich, and SER); the
National Science Council, Taipei; the Scientific and Technical
Research Council of Turkey, and Turkish Atomic Energy Authority; the
Science and Technology Facilities Council, UK; the US Department of
Energy, and the US National Science Foundation.

Individuals have received support from the Marie-Curie programme and
the European Research Council (European Union); the Leventis
Foundation; the A. P. Sloan Foundation; the Alexander von Humboldt
Foundation; the Belgian Federal Science Policy Office; the Fonds pour
la Formation \`a la Recherche dans l'Industrie et dans l'Agriculture
(FRIA-Belgium); the Agentschap voor Innovatie door Wetenschap en
Technologie (IWT-Belgium); the Council of Science and Industrial
Research, India; and the HOMING PLUS programme of Foundation for
Polish Science, cofinanced from European Union, Regional Development
Fund.

\bibliography{auto_generated}   

\providecommand{\href}[2]{#2}\begingroup\raggedright\begin{thebibliography}{10}%
\makeatletter
\providecommand{\hrefCMSnoop }[0]{\@secondoftwo}%
\makeatother
\providecommand{\doi}{\texttt{doi:}\begingroup \urlstyle{tt}\Url}

\bibitem{Karsch:2003jg}
\hrefCMSnoop {} {F.~Karsch and E.~Laermann, ``Thermodynamics and in-medium
  hadron properties from lattice {QCD}'',} in \textit{ Quark-Gluon Plasma III},
  R.~C. Hwa and X.-N. Wang, eds.
\newblock World Scientific Publishing Co. Pte. Ltd., 2004.
\newblock
\href{http://www.arXiv.org/abs/hep-lat/0305025}{\texttt{
  arXiv:hep-lat/0305025}}.
\newblock

\bibitem{Shuryak:1977ut}
\hrefCMSnoop {} {E.~V. Shuryak, ``Theory of Hadronic Plasma'',} \textit{ Sov.
  Phys. JETP} \textbf{ 47} (1978)
212.

\bibitem{Arsene:2004fa}
\hrefCMSnoop {} {{ BRAHMS} Collaboration, ``{Quark-gluon plasma and color glass
  condensate at RHIC? The perspective from the BRAHMS experiment}'',} \textit{
  Nucl. Phys. A} \textbf{ 757} (2005) 1,
  \href{http://dx.doi.org/10.1016/j.nuclphysa.2005.02.130}{\doi{10.1016/j.nuclphysa.2005.02.130}},
  \href{http://www.arXiv.org/abs/nucl-ex/0410020}{\texttt{
  arXiv:nucl-ex/0410020}}.

\bibitem{Back:2004je}
\hrefCMSnoop {} {{ PHOBOS} Collaboration, ``{The PHOBOS perspective on
  discoveries at RHIC}'',} \textit{ Nucl. Phys. A} \textbf{ 757} (2005) 28,
  \href{http://dx.doi.org/10.1016/j.nuclphysa.2005.03.084}{\doi{10.1016/j.nuclphysa.2005.03.084}},
  \href{http://www.arXiv.org/abs/nucl-ex/0410022}{\texttt{
  arXiv:nucl-ex/0410022}}.

\bibitem{Adcox:2004mh}
\hrefCMSnoop {} {{ PHENIX} Collaboration, ``{Formation of dense partonic matter
  in relativistic nucleus-nucleus collisions at RHIC: Experimental evaluation
  by the PHENIX Collaboration}'',} \textit{ Nucl. Phys. A} \textbf{ 757} (2005)
  184,
  \href{http://dx.doi.org/10.1016/j.nuclphysa.2005.03.086}{\doi{10.1016/j.nuclphysa.2005.03.086}},
  \href{http://www.arXiv.org/abs/nucl-ex/0410003}{\texttt{
  arXiv:nucl-ex/0410003}}.

\bibitem{Adams:2005dq}
\hrefCMSnoop {} {{ STAR} Collaboration, ``{Experimental and theoretical
  challenges in the search for the quark--gluon plasma: The STAR
  Collaboration's critical assessment of the evidence from RHIC collisions}'',}
  \textit{ Nucl. Phys. A} \textbf{ 757} (2005) 102,
  \href{http://dx.doi.org/10.1016/j.nuclphysa.2005.03.085}{\doi{10.1016/j.nuclphysa.2005.03.085}},
  \href{http://www.arXiv.org/abs/nucl-ex/0501009}{\texttt{
  arXiv:nucl-ex/0501009}}.

\bibitem{Matsui:1986dk}
\hrefCMSnoop {} {T.~Matsui and H.~Satz, ``\Jpsi suppression by quark-gluon
  plasma formation'',} \textit{ Phys. Lett. B} \textbf{ 178} (1986) 416,
\href{http://dx.doi.org/10.1016/0370-2693(86)91404-8}{\doi{10.1016/0370-2693(86)91404-8}}.

\bibitem{Mocsy:2007jz}
\hrefCMSnoop {} {{\'A}.~M{\'o}csy and P.~Petreczky, ``{Color screening melts
  quarkonium}'',} \textit{ Phys. Rev. Lett.} \textbf{ 99} (2007) 211602,
  \href{http://dx.doi.org/10.1103/PhysRevLett.99.211602}{\doi{10.1103/PhysRevLett.99.211602}},
  \href{http://www.arXiv.org/abs/0706.2183}{\texttt{ arXiv:0706.2183}}.

\bibitem{Vogt:2010aa}
\hrefCMSnoop {} {R.~Vogt, ``{Cold Nuclear Matter Effects on \Jpsi and \PgU\
  Production at energies available at the CERN Large Hadron Collider (LHC)}'',}
  \textit{ Phys. Rev. C} \textbf{ 81} (2010) 044903,
  \href{http://dx.doi.org/10.1103/PhysRevC.81.044903}{\doi{10.1103/PhysRevC.81.044903}},
  \href{http://www.arXiv.org/abs/1003.3497}{\texttt{ arXiv:1003.3497}}.

\bibitem{Zhao:2011cv}
\hrefCMSnoop {} {X.~Zhao and R.~Rapp, ``{Medium modifications and production of
  charmonia at LHC}'',} \textit{ Nucl. Phys. A} \textbf{ 859} (2011) 114,
  \href{http://dx.doi.org/10.1016/j.nuclphysa.2011.05.001}{\doi{10.1016/j.nuclphysa.2011.05.001}},
  \href{http://www.arXiv.org/abs/1102.2194}{\texttt{ arXiv:1102.2194}}.

\bibitem{Zhao:2010nk}
\hrefCMSnoop {} {X.~Zhao and R.~Rapp, ``{Charmonium in medium: From correlators
  to experiment}'',} \textit{ Phys. Rev. C} \textbf{ 82} (2010) 064905,
  \href{http://dx.doi.org/10.1103/PhysRevC.82.064905}{\doi{10.1103/PhysRevC.82.064905}},
  \href{http://www.arXiv.org/abs/1008.5328}{\texttt{ arXiv:1008.5328}}.

\bibitem{Andronic:2006ky}
A.~Andronic\hrefCMSnoop {} { {et~al.}, ``{Statistical hadronization of heavy
  quarks in ultra-relativistic nucleus--nucleus collisions}'',} \textit{ Nucl.
  Phys. A} \textbf{ 789} (2007) 334,
  \href{http://dx.doi.org/10.1016/j.nuclphysa.2007.02.013}{\doi{10.1016/j.nuclphysa.2007.02.013}},
  \href{http://www.arXiv.org/abs/nucl-th/0611023}{\texttt{
  arXiv:nucl-th/0611023}}.

\bibitem{Capella:2007jv}
A.~Capella\hrefCMSnoop {} { {et~al.}, ``{Charmonium dissociation and
  recombination at RHIC and LHC}'',} \textit{ Eur. Phys. J. C} \textbf{ 58}
  (2008) 437,
  \href{http://dx.doi.org/10.1140/epjc/s10052-008-0772-6}{\doi{10.1140/epjc/s10052-008-0772-6}},
  \href{http://www.arXiv.org/abs/0712.4331}{\texttt{ arXiv:0712.4331}}.

\bibitem{Thews:2005vj}
\hrefCMSnoop {} {R.~L. Thews and M.~L. Mangano, ``{Momentum spectra of
  charmonium produced in a quark-gluon plasma}'',} \textit{ Phys. Rev. C}
  \textbf{ 73} (2006) 014904,
  \href{http://dx.doi.org/10.1103/PhysRevC.73.014904}{\doi{10.1103/PhysRevC.73.014904}},
  \href{http://www.arXiv.org/abs/nucl-th/0505055}{\texttt{
  arXiv:nucl-th/0505055}}.

\bibitem{Yan:2006ve}
\hrefCMSnoop {} {L.~Yan, P.~Zhuang, and N.~Xu, ``{\Jpsi production in
  quark-gluon plasma}'',} \textit{ Phys. Rev. Lett.} \textbf{ 97} (2006)
  232301,
  \href{http://dx.doi.org/10.1103/PhysRevLett.97.232301}{\doi{10.1103/PhysRevLett.97.232301}},
  \href{http://www.arXiv.org/abs/nucl-th/0608010}{\texttt{
  arXiv:nucl-th/0608010}}.

\bibitem{Grandchamp:2005yw}
L.~Grandchamp\hrefCMSnoop {} { {et~al.}, ``{Bottomonium production at \sqrtsnn
  = 200\GeV and \sqrtsnn = 5.5\TeV}'',} \textit{ Phys. Rev. C} \textbf{ 73}
  (2006) 064906,
  \href{http://dx.doi.org/10.1103/PhysRevC.73.064906}{\doi{10.1103/PhysRevC.73.064906}},
  \href{http://www.arXiv.org/abs/hep-ph/0507314}{\texttt{
  arXiv:hep-ph/0507314}}.

\bibitem{Baglin:1994ui}
\hrefCMSnoop {} {{ NA38} Collaboration, ``{$\psi'$ and \Jpsi production in pW,
  pU and SU interactions at 200\GeV/nucleon}'',} \textit{ Phys. Lett. B}
  \textbf{ 345} (1995) 617,
\href{http://dx.doi.org/10.1016/0370-2693(94)01614-I}{\doi{10.1016/0370-2693(94)01614-I}}.

\bibitem{Alessandro:2004ap}
\hrefCMSnoop {} {{ NA50} Collaboration, ``A new measurement of \Jpsi
  suppression in \PbPb collisions at 158\GeV per nucleon'',} \textit{ Eur.
  Phys. J. C} \textbf{ 39} (2005) 335,
  \href{http://dx.doi.org/10.1140/epjc/s2004-02107-9}{\doi{10.1140/epjc/s2004-02107-9}},
\href{http://www.arXiv.org/abs/hep-ex/0412036}{\texttt{ arXiv:hep-ex/0412036}}.

\bibitem{Alessandro:2006ju}
\hrefCMSnoop {} {{ NA50} Collaboration, ``{$\psi'$ production in \PbPb
  collisions at 158\GeV/nucleon}'',} \textit{ Eur. Phys. J. C} \textbf{ 49}
  (2007) 559,
  \href{http://dx.doi.org/10.1140/epjc/s10052-006-0153-y}{\doi{10.1140/epjc/s10052-006-0153-y}},
\href{http://www.arXiv.org/abs/nucl-ex/0612013}{\texttt{
  arXiv:nucl-ex/0612013}}.

\bibitem{Arnaldi:2007zz}
\hrefCMSnoop {} {{ NA60} Collaboration, ``{\Jpsi production in InIn collisions
  at 158\GeV/nucleon}'',} \textit{ Phys. Rev. Lett.} \textbf{ 99} (2007)
  132302,
\href{http://dx.doi.org/10.1103/PhysRevLett.99.132302}{\doi{10.1103/PhysRevLett.99.132302}}.

\bibitem{Adare:2006ns}
\hrefCMSnoop {} {{ PHENIX} Collaboration, ``\Jpsi production versus centrality,
  transverse momentum, and rapidity in \AuAu collisions at \sqrtsnn =
  200\GeV'',} \textit{ Phys. Rev. Lett.} \textbf{ 98} (2007) 232301,
  \href{http://dx.doi.org/10.1103/PhysRevLett.98.232301}{\doi{10.1103/PhysRevLett.98.232301}},
\href{http://www.arXiv.org/abs/nucl-ex/0611020}{\texttt{
  arXiv:nucl-ex/0611020}}.

\bibitem{Abelev:2010am}
\hrefCMSnoop {} {{ STAR} Collaboration, ``{\PgU\ cross section in \pp
  collisions at \sqrts = 200 GeV}'',} \textit{ Phys. Rev. D} \textbf{ 82}
  (2010) 012004,
  \href{http://dx.doi.org/10.1103/PhysRevD.82.012004}{\doi{10.1103/PhysRevD.82.012004}},
  \href{http://www.arXiv.org/abs/1001.2745}{\texttt{ arXiv:1001.2745}}.

\bibitem{Aad:2010px}
\hrefCMSnoop {} {{ ATLAS} Collaboration, ``{Measurement of the centrality
  dependence of \Jpsi yields and observation of \Z production in \PbPb
  collisions with the ATLAS detector at the LHC}'',} \textit{ Phys. Lett. B}
  \textbf{ 697} (2011) 294,
  \href{http://dx.doi.org/10.1016/j.physletb.2011.02.006}{\doi{10.1016/j.physletb.2011.02.006}},
  \href{http://www.arXiv.org/abs/1012.5419}{\texttt{ arXiv:1012.5419}}.

\bibitem{Abelev:2012rv}
\hrefCMSnoop {} {{ ALICE} Collaboration, ``\Jpsi production at low transverse
  momentum in \PbPb collisions at \sqrtsnn = 2.76\TeV'',}
  \href{http://www.arXiv.org/abs/1202.1383}{\texttt{ arXiv:1202.1383}}.
submitted to Phys. Rev. Lett.

\bibitem{Chatrchyan:2011pe}
\hrefCMSnoop {} {{ CMS} Collaboration, ``Indications of suppression of excited
  \PgU\ states in \PbPb collisions at \sqrtsnn = 2.76\TeV'',} \textit{ Phys.
  Rev. Lett.} \textbf{ 107} (2011) 052302,
  \href{http://dx.doi.org/10.1103/PhysRevLett.107.052302}{\doi{10.1103/PhysRevLett.107.052302}},
\href{http://www.arXiv.org/abs/1105.4894}{\texttt{ arXiv:1105.4894}}.

\bibitem{Aaij:2011jh}
\hrefCMSnoop {} {{ LHCb} Collaboration, ``Measurement of \Jpsi production in
  \pp collisions at \sqrts = 7\TeV'',} \textit{ Eur. Phys. J. C} \textbf{ 71}
  (2011) 1645,
  \href{http://dx.doi.org/10.1140/epjc/s10052-011-1645-y}{\doi{10.1140/epjc/s10052-011-1645-y}},
  \href{http://www.arXiv.org/abs/1103.0423}{\texttt{ arXiv:1103.0423}}.

\bibitem{Khachatryan:2010yr}
\hrefCMSnoop {} {{ CMS} Collaboration, ``Prompt and non-prompt \Jpsi production
  in \pp collisions at \sqrts = 7\TeV'',} \textit{ Eur. Phys. J. C} \textbf{
  71} (2011) 1575,
  \href{http://dx.doi.org/10.1140/epjc/s10052-011-1575-8}{\doi{10.1140/epjc/s10052-011-1575-8}},
\href{http://www.arXiv.org/abs/1011.4193}{\texttt{ arXiv:1011.4193}}.

\bibitem{Aad:2011sp}
\hrefCMSnoop {} {{ ATLAS} Collaboration, ``Measurement of the differential
  cross-sections of inclusive, prompt and non-prompt \Jpsi production in \pp
  collisions at \sqrts = 7\TeV'',} \textit{ Nucl. Phys. B} \textbf{ 850} (2011)
  387,
  \href{http://dx.doi.org/10.1016/j.nuclphysb.2011.05.015}{\doi{10.1016/j.nuclphysb.2011.05.015}},
  \href{http://www.arXiv.org/abs/1104.3038}{\texttt{ arXiv:1104.3038}}.

\bibitem{Adare:2006nq}
\hrefCMSnoop {} {{ PHENIX} Collaboration, ``Energy loss and flow of heavy
  quarks in \AuAu collisions at \sqrtsnn = 200\GeV'',} \textit{ Phys. Rev.
  Lett.} \textbf{ 98} (2007) 172301,
  \href{http://dx.doi.org/10.1103/PhysRevLett.98.172301}{\doi{10.1103/PhysRevLett.98.172301}},
\href{http://www.arXiv.org/abs/nucl-ex/0611018}{\texttt{
  arXiv:nucl-ex/0611018}}.

\bibitem{Abelev:2006db}
\hrefCMSnoop {} {{ STAR} Collaboration, ``{Transverse momentum and centrality
  dependence of high-\pt nonphotonic electron suppression in \AuAu collisions
  at \sqrtsnn = 200\GeV}'',} \textit{ Phys. Rev. Lett.} \textbf{ 98} (2007)
  192301,
  \href{http://dx.doi.org/10.1103/PhysRevLett.98.192301}{\doi{10.1103/PhysRevLett.98.192301}},
  \href{http://www.arXiv.org/abs/nucl-ex/0607012v2}{\texttt{
  arXiv:nucl-ex/0607012v2}}.

\bibitem{Abelev:2006dbErratum}
\hrefCMSnoop {} {{ STAR} Collaboration, ``{Erratum: Transverse momentum and
  centrality dependence of high-\pt nonphotonic electron suppression in \AuAu
  collisions at \sqrtsnn = 200\GeV}'',} \textit{ Phys. Rev. Lett.} \textbf{
  106} (2011) 159902(E),
  \href{http://dx.doi.org/10.1103/PhysRevLett.106.159902}{\doi{10.1103/PhysRevLett.106.159902}},
  \href{http://www.arXiv.org/abs/nucl-ex/0607012v3}{\texttt{
  arXiv:nucl-ex/0607012v3}}.

\bibitem{Dokshitzer:2001zm}
\hrefCMSnoop {} {Y.~L. Dokshitzer and D.~Kharzeev, ``{Heavy quark colorimetry
  of QCD matter}'',} \textit{ Phys. Lett. B} \textbf{ 519} (2001) 199,
  \href{http://dx.doi.org/10.1016/S0370-2693(01)01130-3}{\doi{10.1016/S0370-2693(01)01130-3}},
  \href{http://www.arXiv.org/abs/hep-ph/0106202}{\texttt{
  arXiv:hep-ph/0106202}}.

\bibitem{Armesto:2005iq}
N.~Armesto\hrefCMSnoop {} { {et~al.}, ``{Testing the color charge and mass
  dependence of parton energy loss with heavy-to-light ratios at BNL RHIC and
  CERN LHC}'',} \textit{ Phys. Rev. D} \textbf{ 71} (2005) 054027,
  \href{http://dx.doi.org/10.1103/PhysRevD.71.054027}{\doi{10.1103/PhysRevD.71.054027}},
  \href{http://www.arXiv.org/abs/hep-ph/0501225}{\texttt{
  arXiv:hep-ph/0501225}}.

\bibitem{Peigne:2008nd}
\hrefCMSnoop {} {S.~Peigne and A.~Peshier, ``{Collisional energy loss of a fast
  heavy quark in a quark-gluon plasma}'',} \textit{ Phys. Rev. D} \textbf{ 77}
  (2008) 114017,
  \href{http://dx.doi.org/10.1103/PhysRevD.77.114017}{\doi{10.1103/PhysRevD.77.114017}},
  \href{http://www.arXiv.org/abs/0802.4364}{\texttt{ arXiv:0802.4364}}.

\bibitem{Adolphi:2008zzk}
\hrefCMSnoop {} {{ CMS} Collaboration, ``{The CMS experiment at the CERN
  LHC}'',} \textit{ JINST} \textbf{ 3} (2008) S08004,
  \href{http://dx.doi.org/10.1088/1748-0221/3/08/S08004}{\doi{10.1088/1748-0221/3/08/S08004}}.

\bibitem{TRK-10-004}
\href {http://cdsweb.cern.ch/record/1279137/files/TRK-10-004-pas.pdf} {{ CMS}
  Collaboration, ``Measurement of momentum scale and resolution using low-mass
  resonances and cosmic ray muons'',} {CMS Physics Analysis Summary}
  TRK-2010/004, (2010).

\bibitem{Chatrchyan:2011sx}
\hrefCMSnoop {} {{ CMS} Collaboration, ``{Observation and studies of jet
  quenching in \PbPb collisions at \sqrtsnn = 2.76\TeV}'',} \textit{ Phys. Rev.
  C} \textbf{ 84} (2011) 024906,
  \href{http://dx.doi.org/10.1103/PhysRevC.84.024906}{\doi{10.1103/PhysRevC.84.024906}},
  \href{http://www.arXiv.org/abs/1102.1957}{\texttt{ arXiv:1102.1957}}.

\bibitem{Chatrchyan:2011pb}
\hrefCMSnoop {} {{ CMS} Collaboration, ``{Dependence on pseudorapidity and
  centrality of charged hadron production in \PbPb collisions at \sqrtsnn =
  2.76\TeV}'',} \textit{ JHEP} \textbf{ 08} (2011) 141,
  \href{http://dx.doi.org/10.1007/JHEP08(2011)141}{\doi{10.1007/JHEP08(2011)141}},
\href{http://www.arXiv.org/abs/1107.4800}{\texttt{ arXiv:1107.4800}}.

\bibitem{Miller:2007ri}
M.~L. Miller\hrefCMSnoop {} { {et~al.}, ``{Glauber modeling in high-energy
  nuclear collisions}'',} \textit{ Ann. Rev. Nucl. Part. Sci.} \textbf{ 57}
  (2007) 205,
  \href{http://dx.doi.org/10.1146/annurev.nucl.57.090506.123020}{\doi{10.1146/annurev.nucl.57.090506.123020}},
  \href{http://www.arXiv.org/abs/nucl-ex/0701025}{\texttt{
  arXiv:nucl-ex/0701025}}.

\bibitem{Sjostrand:2006za}
\hrefCMSnoop {} {T.~Sj{\"o}strand, S.~Mrenna, and P.~Z. Skands, ``{PYTHIA 6.4
  physics and manual}'',} \textit{ JHEP} \textbf{ 05} (2006) 026,
  \href{http://dx.doi.org/10.1088/1126-6708/2006/05/026}{\doi{10.1088/1126-6708/2006/05/026}},
  \href{http://www.arXiv.org/abs/hep-ph/0603175}{\texttt{
  arXiv:hep-ph/0603175}}.

\bibitem{Bargiotti:2007zz}
\href {http://cdsweb.cern.ch/record/1042611/files/lhcb-2007-042.pdf} {{M.
  Bargiotti and V. Vagnoni}, ``{Heavy quarkonia sector in PYTHIA 6.324: Tuning,
  validation and perspectives at LHCb}'',} {LHCb Note} {LHCb-2007-042}, (2007).

\bibitem{Acosta:2004yw}
\hrefCMSnoop {} {{ CDF} Collaboration, ``{Measurement of the \Jpsi meson and
  \cPqb-hadron production cross sections in \ppbar collisions at \sqrts =
  1960\GeV}'',} \textit{ Phys. Rev. D} \textbf{ 71} (2005) 032001,
  \href{http://dx.doi.org/10.1103/PhysRevD.71.032001}{\doi{10.1103/PhysRevD.71.032001}},
\href{http://www.arXiv.org/abs/hep-ex/0412071}{\texttt{ arXiv:hep-ex/0412071}}.

\bibitem{Lange:2001uf}
\hrefCMSnoop {} {D.~J. Lange, ``{The EvtGen particle decay simulation
  package}'',} \textit{ Nucl. Instrum. Meth. A} \textbf{ 462} (2001) 152,
\href{http://dx.doi.org/10.1016/S0168-9002(01)00089-4}{\doi{10.1016/S0168-9002(01)00089-4}}.

\bibitem{Barberio:1993qi}
\hrefCMSnoop {} {E.~Barberio and Z.~Was, ``{PHOTOS --- A universal Monte Carlo
  for QED radiative corrections: version 2.0}'',} \textit{ Comput. Phys.
  Commun.} \textbf{ 79} (1994) 291,
  \href{http://dx.doi.org/10.1016/0010-4655(94)90074-4}{\doi{10.1016/0010-4655(94)90074-4}}.

\bibitem{Agostinelli:2002hh}
\hrefCMSnoop {} {{ GEANT} Collaboration, ``{GEANT4 --- A simulation
  toolkit}'',} \textit{ Nucl. Instrum. Meth. A} \textbf{ 506} (2003) 250,
\href{http://dx.doi.org/10.1016/S0168-9002(03)01368-8}{\doi{10.1016/S0168-9002(03)01368-8}}.

\bibitem{Lokhtin:2005px}
\hrefCMSnoop {} {I.~P. Lokhtin and A.~M. Snigirev, ``{A model of jet quenching
  in ultrarelativistic heavy ion collisions and high-\pt hadron spectra at
  RHIC}'',} \textit{ Eur. Phys. J. C} \textbf{ 45} (2006) 211,
  \href{http://dx.doi.org/10.1140/epjc/s2005-02426-3}{\doi{10.1140/epjc/s2005-02426-3}},
\href{http://www.arXiv.org/abs/hep-ph/0506189}{\texttt{ arXiv:hep-ph/0506189}}.

\bibitem{D'Enterria:2007xr}
\hrefCMSnoop {} {{ CMS} Collaboration, ``{CMS physics technical design report:
  Addendum on high density QCD with heavy ions}'',} \textit{ J. Phys. G}
  \textbf{ 34} (2007) 2307,
  \href{http://dx.doi.org/10.1088/0954-3899/34/11/008}{\doi{10.1088/0954-3899/34/11/008}}.

\bibitem{Roland:2006kz}
\hrefCMSnoop {} {{C.\,Roland (on behalf of the CMS Collaboration)}, ``{Track
  reconstruction in heavy ion collisions with the CMS silicon tracker}'',}
  \textit{ Nucl. Instrum. Meth. A} \textbf{ 566} (2006) 123,
  \href{http://dx.doi.org/10.1016/j.nima.2006.05.023}{\doi{10.1016/j.nima.2006.05.023}}.

\bibitem{Nakamura:2010zzi}
\hrefCMSnoop {} {{ Particle Data Group} Collaboration, ``{Review of particle
  physics}'',} \textit{ J. Phys. G} \textbf{ 37} (2010) 075021,
\href{http://dx.doi.org/10.1088/0954-3899/37/7A/075021}{\doi{10.1088/0954-3899/37/7A/075021}}.

\bibitem{Alessandro:2006yt}
\hrefCMSnoop {} {{ ALICE} Collaboration, ``{ALICE: Physics performance report,
  volume II}'',} \textit{ J. Phys. G} \textbf{ 32} (2006) 1295,
  \href{http://dx.doi.org/10.1088/0954-3899/32/10/001}{\doi{10.1088/0954-3899/32/10/001}}.

\bibitem{Buskulic:1993vi}
\hrefCMSnoop {} {{ ALEPH} Collaboration, ``{Measurement of the
  $\overline{\B}^0$ and $\B^-$ meson lifetimes}'',} \textit{ Phys. Lett. B}
  \textbf{ 307} (1993) 194,
  \href{http://dx.doi.org/10.1016/0370-2693(93)90211-Y}{\doi{10.1016/0370-2693(93)90211-Y}}.

\bibitem{Buskulic:1993viErratum}
\hrefCMSnoop {} {{ ALEPH} Collaboration, ``{Errata: Measurement of the
  $\overline{\B}^0$ and $\B^-$ meson lifetimes}'',} \textit{ Phys. Lett. B}
  \textbf{ 325} (1994) 537,
  \href{http://dx.doi.org/10.1016/0370-2693(94)90054-X}{\doi{10.1016/0370-2693(94)90054-X}}.

\bibitem{muon-pog}
\href {http://cdsweb.cern.ch/record/1279140/files/MUO-10-002-pas.pdf} {{ CMS}
  Collaboration, ``Performance of muon identification in \pp collisions at
  \sqrts = 7\TeV'',} {CMS Physics Analysis Summary} {MUO-2010/02}, (2010).

\bibitem{Faccioli:2012kp}
\hrefCMSnoop {} {P.~Faccioli and J.~Seixas, ``Observation of $\chi_c$ and
  $\chi_b$ nuclear suppression via dilepton polarization measurements'',}
\href{http://www.arXiv.org/abs/1203.2033}{\texttt{ arXiv:1203.2033}}.

\bibitem{Aamodt:2010jd}
\hrefCMSnoop {} {{ ALICE} Collaboration, ``{Suppression of charged particle
  production at large transverse momentum in central \PbPb Collisions at
  \sqrtsnn = 2.76 TeV}'',} \textit{ Phys. Lett. B} \textbf{ 696} (2011) 30,
  \href{http://dx.doi.org/10.1016/j.physletb.2010.12.020}{\doi{10.1016/j.physletb.2010.12.020}},
\href{http://www.arXiv.org/abs/1012.1004}{\texttt{ arXiv:1012.1004}}.

\bibitem{Affolder:1999wm}
\hrefCMSnoop {} {{ CDF} Collaboration, ``{Production of \PgUa\ mesons from
  $\chi_b$ decays in \ppbar collisions at \sqrts = 1.8\TeV}'',} \textit{ Phys.
  Rev. Lett.} \textbf{ 84} (2000) 2094,
  \href{http://dx.doi.org/10.1103/PhysRevLett.84.2094}{\doi{10.1103/PhysRevLett.84.2094}},
\href{http://www.arXiv.org/abs/hep-ex/9910025}{\texttt{ arXiv:hep-ex/9910025}}.

\end{thebibliography}\endgroup

\appendix
\section{Tables of Results}
\label{app:datatables}

\begin{table*}[htbp]
  \begin{center}
    \caption{Raw yield of inclusive \Jpsi as a function of \Jpsi
      rapidity and \pt in \PbPb and \pp collisions. For \PbPb, the raw
      yield is also included as a function of collision
      centrality. All quoted uncertainties are statistical.}
    \label{tab:inclyields}
    \begin{tabular}{cr@{--}l@{}rr@{$\,\pm\,$}lr@{$\,\pm\,$}l}
      \hline
      $|y|$ & \multicolumn{2}{c}{$\pt$} &
      centrality& 
      \multicolumn{4}{c}{Raw yield}\\
              &\multicolumn{2}{c}{[\GeVc{}]}&&\multicolumn{2}{c}{\PbPb}&\multicolumn{2}{c}{\pp}\\\hline
      \multirow{3}{*}{0.0--2.4} & 6.5&30  & \multirow{3}{*}{0--100\%} & 396&24 & 1026&35\\
                                & 6.5&10  &                           & 261&20 &  684&30\\
                                &  10&30  &                           & 138&14 &  342&19\\\hline
      0.0--1.2                  & 6.5&30  & 0--100\%                  & 174&16 &  462&36\\\hline
      \multirow{2}{*}{1.2--1.6} & 5.5&30  & \multirow{2}{*}{0--100\%} & 103&13 &  360&23\\
                                & 6.5&30  &                           & 90 &11 &  272&21\\\hline
      \multirow{2}{*}{1.6--2.4} & 3.0&30  & \multirow{2}{*}{0--100\%} & 446&56 & 1006&34\\
                                & 6.5&30  &                           & 150&15 &  329&19\\\hline
      \multirow{8}{*}{0.0--2.4}&\multicolumn{2}{c}{\multirow{8}{*}{\hspace*{-1.2em}6.5--30}}
                          &  0--10\% & 113&12 & \multicolumn{2}{c}{}\\
      \multicolumn{3}{c}{}& 10--20\% &  80&10 & \multicolumn{2}{c}{}\\
      \multicolumn{3}{c}{}& 20--30\% &  63&9  & \multicolumn{2}{c}{}\\
      \multicolumn{3}{c}{}& 30--40\% &  58&8  & \multicolumn{2}{c}{}\\
      \multicolumn{3}{c}{}& 40--50\% &  45&7  & \multicolumn{2}{c}{}\\
      \multicolumn{3}{c}{}&50--100\% &  37&6  & \multicolumn{2}{c}{}\\\cline{4-8}
      \multicolumn{3}{c}{}&  0--20\% & 193&16 & \multicolumn{2}{c}{}\\
      \multicolumn{3}{c}{}&20--100\% & 205&15 & \multicolumn{2}{c}{}\\\hline
    \end{tabular}
  \end{center}
\end{table*}

\begin{table*}[htbp]
  \begin{center}
    \caption{Raw yield of \PgUa\ as a function of \PgUa\ rapidity and
      \pt in \PbPb and \pp collisions. For \PbPb, the raw yield is
      also included as a function of collision centrality. All quoted
      uncertainties are statistical.}
    \label{tab:upsilonyields}
    \begin{tabular}{cr@{--}l@{}rr@{$\,\pm\,$}lr@{$\,\pm\,$}l}
      \hline
      $|y|$ & \multicolumn{2}{c}{$\pt$} &
      centrality& 
      \multicolumn{4}{c}{Raw yield}\\
              &\multicolumn{2}{c}{[\GeVc{}]}&&\multicolumn{2}{c}{\PbPb}&\multicolumn{2}{c}{\pp}\\\hline
      \multirow{4}{*}{0.0--2.4}  &  0&6.5 & \multirow{4}{*}{0--100\%} & 44&9  &  75&10\\
                                 &6.5&10  &                           & 18&5  &  15&5 \\
                                 & 10&20  &                           & 24&6  &  10&4 \\
                                 &  0&20  &                           & 86&12 & 101&12\\\hline
      0.0--1.2 & \multicolumn{2}{c}{\multirow{2}{*}{\hspace*{-0.5em}0--20}} & \multirow{2}{*}{0--100\%}
                                         & 48&9 & 66&9\\
      1.2--2.4 & \multicolumn{2}{c}{}  & & 40&8 & 34&7\\\hline
      \multirow{4}{*}{0.0--2.4}&\multicolumn{2}{c}{\multirow{4}{*}{\hspace*{-0.5em}0--20}}
                           & 0--10\%  & 24&7 & \multicolumn{2}{c}{}\\
      \multicolumn{3}{c}{} &10--20\%  & 30&7 & \multicolumn{2}{c}{}\\
      \multicolumn{3}{c}{} &20--100\% & 32&6 & \multicolumn{2}{c}{}\\
      \multicolumn{3}{c}{} & 0--20\%  & 54&9 & \multicolumn{2}{c}{}\\\hline
    \end{tabular}
  \end{center}
\end{table*}

\begin{table*}[htbp]
  \begin{center}
    \caption{Yield per unit of rapidity of inclusive \Jpsi divided by
      \taa and nuclear modification factor \raa as a function of \Jpsi
      rapidity, \pt, and collision centrality. The average \pt value
      for each bin is given. Listed uncertainties are statistical
      first, systematic second, and global scale third. The latter
      includes the uncertainties on the \pp integrated luminosity and,
      for centrality integrated bins, on \taa.}
    \label{tab:inclxsec}
    \begin{tabular}{cr@{--}l@{}rc@{}r@{$\,\pm\,$}l@{$\,\pm\,$}l@{$\,\pm\,$}lr@{$\,\pm\,$}l@{$\,\pm\,$}l@{$\,\pm\,$}l}
      \hline
      $|y|$ & \multicolumn{2}{c}{$\pt$} &
      centrality& $\langle\pt\rangle$ &
      \multicolumn{4}{c}{$\frac{1}{\taa}
        \cdot \frac{\mathrm{d}N}{\mathrm{d}y}$} & \multicolumn{4}{c}{\raa}\\
              &\multicolumn{2}{c}{[\GeVc{}]}&&[\GeVc{}]&\multicolumn{4}{c}{[nb]}&\multicolumn{4}{c}{}\\\hline
      \multirow{3}{*}{0.0--2.4} & 6.5&30  & \multirow{3}{*}{0--100\%} &  9.87 &  2.40&0.15&0.34&0.14 & 0.32&0.02&0.01&0.03\\
                                & 6.5&10  &                           &  8.11 &  2.05&0.15&0.30&0.12 & 0.32&0.03&0.02&0.03\\
                                & 10&30 &                           & 13.22 &  0.40&0.04&0.06&0.02 & 0.31&0.04&0.01&0.03\\\hline
      0.0--1.2                  & 6.5& 30  & 0--100\%                 & 10.92 &  2.76&0.26&0.43&0.16 & 0.29&0.04&0.02&0.02\\\hline
      \multirow{2}{*}{1.2--1.6} & 5.5&30  & \multirow{2}{*}{0--100\%} &  9.21 &  3.57&0.45&0.51&0.20 & 0.23&0.03&0.02&0.02\\
                                & 6.5&30  &                           &  9.65 &  2.29&0.28&0.33&0.13 & 0.28&0.04&0.02&0.02\\\hline
      \multirow{2}{*}{1.6--2.4} & 3.0&30  & \multirow{2}{*}{0--100\%} &  6.27 & 21.18&2.65&3.18&1.21 & 0.41&0.05&0.02&0.03\\
                                & 6.5&30  &                           &  8.92 &  2.22&0.21&0.32&0.13 & 0.40&0.05&0.01&0.03\\\hline
      \multirow{8}{*}{0.0--2.4}&\multicolumn{2}{c}{\multirow{8}{*}{\hspace*{-1.2em}6.5--30}}
                          &  0--10\% & 10.39 & 1.78&0.20&\multicolumn{2}{l}{\hspace{-0.548em}0.27} & 0.24&0.03&0.02&0.01\\
      \multicolumn{3}{c}{}& 10--20\% &  9.70 & 1.92&0.24&\multicolumn{2}{l}{\hspace{-0.548em}0.30} & 0.26&0.03&0.02&0.02\\
      \multicolumn{3}{c}{}& 20--30\% & 10.23 & 2.37&0.33&\multicolumn{2}{l}{\hspace{-0.548em}0.38} & 0.31&0.04&0.02&0.02\\
      \multicolumn{3}{c}{}& 30--40\% &  9.27 & 3.73&0.53&\multicolumn{2}{l}{\hspace{-0.548em}0.63} & 0.50&0.07&0.05&0.03\\
      \multicolumn{3}{c}{}& 40--50\% &  9.29 & 5.22&0.81&\multicolumn{2}{l}{\hspace{-0.548em}0.95} & 0.70&0.11&0.08&0.04\\
      \multicolumn{3}{c}{}&50--100\% &  9.64 & 4.67&0.80&\multicolumn{2}{l}{\hspace{-0.548em}0.97} & 0.62&0.11&0.10&0.04\\\cline{4-13}
      \multicolumn{3}{c}{}&  0--20\% &  9.27 & 1.84&0.15&\multicolumn{2}{l}{\hspace{-0.548em}0.28} & 0.25&0.02&0.02&0.02\\
      \multicolumn{3}{c}{}&20--100\% &  9.29 & 3.46&0.26&\multicolumn{2}{l}{\hspace{-0.548em}0.58} & 0.46&0.04&0.04&0.03\\\hline
    \end{tabular}
  \end{center}
\end{table*}

\begin{table*}[htbp]
  \begin{center}
    \caption{Yield per unit of rapidity of prompt \Jpsi divided by
      \taa and nuclear modification factor \raa as a function of \Jpsi
      rapidity, \pt, and collision centrality. The average \pt value
      for each bin is given. Listed uncertainties are statistical
      first, systematic second, and global scale third. The latter
      includes the uncertainties on the \pp integrated luminosity and,
      for centrality integrated bins, on \taa.}
    \label{tab:promptxsec}
    \begin{tabular}{cr@{--}l@{}rc@{}r@{$\,\pm\,$}l@{$\,\pm\,$}l@{$\,\pm\,$}lr@{$\,\pm\,$}l@{$\,\pm\,$}l@{$\,\pm\,$}l}
      \hline
      $|y|$&\multicolumn{2}{c}{$\pt$}&centrality&$\langle\pt\rangle$&\multicolumn{4}{c}{$\frac{1}{\taa}
        \cdot \frac{\mathrm{d}N}{\mathrm{d}y}$}&\multicolumn{4}{c}{\raa}\\
      &\multicolumn{2}{c}{[\GeVc{}]}&&[\GeVc{}]&\multicolumn{4}{c}{[nb]}&\multicolumn{4}{c}{}\\\hline
      \multirow{3}{*}{0.0--2.4} & 6.5&30  & \multirow{3}{*}{0--100\%} &  9.87 &  1.79&0.13&0.26&0.10 & 0.30&0.03&0.01&0.02\\
                                & 6.5&10  &                           &  8.11 &  1.56&0.14&0.23&0.09 & 0.30&0.03&0.02&0.02\\
                                & 10&30 &                           & 13.22 &  0.27&0.03&0.04&0.02 & 0.31&0.04&0.01&0.03\\\hline
      0.0--1.2                  & 6.5&30  & 0--100\%                  & 10.92 &  2.11&0.23&0.32&0.12 & 0.29&0.04&0.02&0.02\\\hline
      \multirow{2}{*}{1.2--1.6} & 5.5&30  & \multirow{2}{*}{0--100\%} &  9.21 &  2.95&0.44&0.45&0.17 & 0.24&0.04&0.02&0.02\\
                                & 6.5&30  &                           &  9.65 &  1.71&0.25&0.24&0.10 & 0.27&0.05&0.02&0.02\\\hline
      \multirow{2}{*}{1.6--2.4} & 3.0&30  & \multirow{2}{*}{0--100\%} &  6.27 & 17.78&2.35&2.60&1.01 & 0.40&0.05&0.02&0.03\\
                                & 6.5&30  &                           &  8.92 &  1.83&0.20&0.26&0.10 & 0.43&0.06&0.01&0.04\\\hline
      \multirow{8}{*}{0.0--2.4}&\multicolumn{2}{c}{\multirow{8}{*}{\hspace*{-1.2em}6.5--30}}
                          &  0--10\% & 10.39 &  1.18&0.17&\multicolumn{2}{l}{\hspace{-0.548em}0.18} & 0.20&0.03&0.01&0.01\\
      \multicolumn{3}{c}{}& 10--20\% &  9.70 &  1.29&0.21&\multicolumn{2}{l}{\hspace{-0.548em}0.20} & 0.22&0.04&0.02&0.01\\
      \multicolumn{3}{c}{}& 20--30\% & 10.23 &  2.18&0.33&\multicolumn{2}{l}{\hspace{-0.548em}0.35} & 0.37&0.06&0.03&0.02\\
      \multicolumn{3}{c}{}& 30--40\% &  9.27 &  2.97&0.48&\multicolumn{2}{l}{\hspace{-0.548em}0.50} & 0.51&0.09&0.05&0.03\\
      \multicolumn{3}{c}{}& 40--50\% &  9.29 &  3.88&0.75&\multicolumn{2}{l}{\hspace{-0.548em}0.70} & 0.66&0.13&0.08&0.04\\
      \multicolumn{3}{c}{}&50--100\% &  9.64 &  3.58&0.70&\multicolumn{2}{l}{\hspace{-0.548em}0.75} & 0.61&0.12&0.10&0.04\\\cline{4-13}
      \multicolumn{3}{c}{}&  0--20\% &  9.27 &  1.23&0.14&\multicolumn{2}{l}{\hspace{-0.548em}0.19} & 0.21&0.02&0.01&0.01\\
      \multicolumn{3}{c}{}&20--100\% &  9.29 &  2.84&0.25&\multicolumn{2}{l}{\hspace{-0.548em}0.47} & 0.48&0.05&0.05&0.01\\\hline
    \end{tabular}
  \end{center}
\end{table*}

\begin{table*}[htbp]
  \begin{center}
    \caption{Yield per unit of rapidity of non-prompt \Jpsi divided by
      \taa and nuclear modification factor \raa as a function of \Jpsi
      rapidity, \pt, and collision centrality. The average \pt value
      for each bin is given. Listed uncertainties are statistical
      first, systematic second, and global scale third. The latter
      includes the uncertainties on the \pp integrated luminosity and,
      for centrality integrated bins, on \taa.}
    \label{tab:nonpromptxsec}
    \begin{tabular}{cr@{--}l@{}rc@{}r@{$\,\pm\,$}l@{$\,\pm\,$}l@{$\,\pm\,$}lr@{$\,\pm\,$}l@{$\,\pm\,$}l@{$\,\pm\,$}l}
      \hline
      $|y|$ & \multicolumn{2}{c}{$\pt$} &
      centrality & $\langle\pt\rangle$ &
      \multicolumn{4}{c}{$\frac{1}{\taa}
        \cdot \frac{\mathrm{d}N}{\mathrm{d}y}$}& \multicolumn{4}{c}{\raa}\\
      &\multicolumn{2}{c}{[\GeVc{}]}&&[\GeVc{}]&\multicolumn{4}{c}{[nb]}&\multicolumn{4}{c}{}\\\hline

      0.0--2.4                  & 6.5&30 & 0--100\%                  & 9.87 & 0.60&0.09&0.09&0.03 & 0.38&0.07&0.02&0.03\\\hline
      \multirow{2}{*}{1.6--2.4} & 3.0&30 & \multirow{2}{*}{0--100\%} & 6.27 & 3.29&0.82&0.65&0.19 & 0.50&0.14&0.02&0.04\\
                                & 6.5&30 &                           & 8.92 & 0.39&0.12&0.06&0.02 & 0.31&0.11&0.01&0.03\\\hline
      \multirow{2}{*}{0.0--2.4}&\multicolumn{2}{c}{\multirow{2}{*}{\hspace*{-1.2em}6.5--30}}
                           & 0--20\%  & 9.27 & 0.59&0.12&\multicolumn{2}{l}{\hspace{-0.548em}0.10} & 0.37&0.08&0.02&0.02\\
      \multicolumn{3}{c}{} &20--100\% & 9.29 & 0.60&0.14&\multicolumn{2}{l}{\hspace{-0.548em}0.10} & 0.38&0.10&0.04&0.02\\\hline
    \end{tabular}
  \end{center}
\end{table*}

\begin{table*}[htbp]
  \begin{center}
    \caption{Yield per unit of rapidity of \PgUa\ divided by \taa and
      nuclear modification factor \raa as a function of \PgUa\
      rapidity, \pt, and collision centrality. The average \pt value
      for each bin is given. Listed uncertainties are statistical
      first, systematic second, and global scale third. The latter
      includes the uncertainties on the \pp integrated luminosity and,
      for centrality integrated bins, on \taa.}
    \label{tab:upsilonxsect}
    \begin{tabular}{cr@{--}l@{}rc@{}r@{$\,\pm\,$}l@{$\,\pm\,$}l@{$\,\pm\,$}lr@{$\,\pm\,$}l@{$\,\pm\,$}l@{$\,\pm\,$}l}
      \hline
      $|y|$ & \multicolumn{2}{c}{$\pt$} &
      \multicolumn{1}{@{}l@{}}{centrality}& $\langle\pt\rangle$ &
      \multicolumn{4}{c}{$\frac{1}{\taa}
        \cdot \frac{\mathrm{d}N}{\mathrm{d}y}$}& \multicolumn{4}{c}{\raa}\\
      &\multicolumn{2}{c}{[\GeVc{}]}&&[\GeVc{}]&\multicolumn{4}{c}{[nb]}&\multicolumn{4}{c}{}\\\hline

      \multirow{4}{*}{0.0--2.4}  & 0&6.5  & \multirow{4}{*}{0--100\%} & 3.03 & 0.293&0.057&0.051&0.02 & 0.44&0.10&0.06&0.04\\
                                 & 6.5&10 &                           & 8.04 & 0.093&0.028&0.017&0.01 & 0.91&0.38&0.13&0.08\\
                                 &10&20 &                           &13.17 & 0.066&0.016&0.011&0.004 & 1.77&0.76&0.24&0.15\\
                                 & 0&20 &                           & 6.79 & 0.485&0.066&0.084&0.03 & 0.63&0.11&0.09&0.05\\\hline
      0.0--1.2 & \multicolumn{2}{c}{\multirow{2}{*}{\hspace*{-0.5em}0--20}} & \multirow{2}{*}{0--100\%}
                                         & 6.44 & 0.495&0.091&0.086&0.03 & 0.54&0.12&0.08&0.04\\
      1.2--2.4 & \multicolumn{2}{c}{}  & & 6.60 & 0.498&0.097&0.088&0.03 & 0.85&0.25&0.12&0.07\\\hline
      \multirow{4}{*}{0.0--2.4}&\multicolumn{2}{c}{\multirow{4}{*}{\hspace*{-0.5em}0--20}}
                           & 0--10\%  & 6.65 & 0.347&0.096&\multicolumn{2}{l}{\hspace{-0.548em}0.069} & 0.45&0.14&0.08&0.03\\
      \multicolumn{3}{c}{} &10--20\%  & 6.88 & 0.643&0.144&\multicolumn{2}{l}{\hspace{-0.548em}0.118} & 0.84&0.21&0.13&0.05\\
      \multicolumn{3}{c}{} &20--100\% & 6.08 & 0.517&0.101&\multicolumn{2}{l}{\hspace{-0.548em}0.101} & 0.68&0.15&0.11&0.04\\
      \multicolumn{3}{c}{} & 0--20\%  & 6.85 & 0.467&0.081&\multicolumn{2}{l}{\hspace{-0.548em}0.093} & 0.61&0.13&0.11&0.04\\\hline
    \end{tabular}
  \end{center}
\end{table*} 

\begin{table*}[htbp]
  \begin{center}
    \caption{Cross section per unit of rapidity of inclusive \Jpsi as
      a function of rapidity and \pt in \pp collisions. The average
      \pt value for each bin is given. Listed uncertainties are
      statistical first, systematic second, and global scale
      third. The latter is the uncertainty on the \pp integrated
      luminosity.}
    \label{tab:inclxsecpp}
    \begin{tabular}{cr@{--}l@{}c@{}r@{$\,\pm\,$}l@{$\,\pm\,$}l@{$\,\pm\,$}l}
      \hline
      $|y|$ & \multicolumn{2}{c}{$\pt$} &
      $\langle\pt\rangle$ &
      \multicolumn{4}{c}{$\frac{\mathrm{d}\sigma}{\mathrm{d}y}$}\\
              &\multicolumn{2}{c}{[\GeVc{}]}&[\GeVc{}]&\multicolumn{4}{c}{[nb]}\\\hline
      \multirow{3}{*}{0.0--2.4} & 6.5&30  &  9.82 &  7.50&0.26&1.10&0.45\\
                                & 6.5&10  &  8.05 &  6.37&0.28&0.99&0.38\\
                                & 10&30   & 13.34 &  1.27&0.07&0.18&0.08\\\hline
      0.0--1.2                  & 6.5&30  & 10.81 &  9.45&0.74&1.38&0.57\\\hline
      \multirow{2}{*}{1.2--1.6} & 5.5&30  &  8.53 & 15.22&0.96&2.23&0.91\\
                                & 6.5&30  &  9.31 &  8.27&0.64&1.23&0.50\\\hline
      \multirow{2}{*}{1.6--2.4} & 3.0&30  &  6.15 & 51.52&1.74&7.33&3.09\\
                                & 6.5&30  &  8.98 &  5.54&0.33&0.79&0.33\\\hline
    \end{tabular}
  \end{center}
\end{table*}

\begin{table*}[htbp]
  \begin{center}
    \caption{Cross section per unit of rapidity of prompt \Jpsi as a
      function of rapidity, \pt in \pp collisions. The average \pt
      value for each bin is given. Listed uncertainties are
      statistical first, systematic second, and global scale
      third. The latter is the uncertainty on the \pp integrated
      luminosity.}
    \label{tab:promptxsecpp}
    \begin{tabular}{cr@{--}l@{}c@{}r@{$\,\pm\,$}l@{$\,\pm\,$}l@{$\,\pm\,$}l}
      \hline
      $|y|$&\multicolumn{2}{c}{$\pt$}&$\langle\pt\rangle$&
      \multicolumn{4}{c}{$\frac{\mathrm{d}\sigma}{\mathrm{d}y}$}\\
      &\multicolumn{2}{c}{[\GeVc{}]}&[\GeVc{}]&\multicolumn{4}{c}{[nb]}\\\hline
      \multirow{3}{*}{0.0--2.4} & 6.5&30 &  9.82 &  5.87&0.24&0.86&0.35\\
                                & 6.5&10 &  8.05 &  5.14&0.26&0.80&0.31\\
                                & 10&30  & 13.34 &  0.89&0.06&0.12&0.05\\\hline
      0.0--1.2                  & 6.5&30 & 10.81 &  7.40&0.64&1.08&0.44\\\hline
      \multirow{2}{*}{1.2--1.6} & 5.5&30 &  8.53 & 12.38&0.92&1.81&0.74\\
                                & 6.5&30 &  9.31 &  6.39&0.57&0.95&0.38\\\hline
      \multirow{2}{*}{1.6--2.4} & 3.0&30 &  6.15 & 44.67&1.78&6.35&2.68\\
                                & 6.5&30 &  8.98 &  4.26&0.31&0.61&0.26\\\hline
    \end{tabular}
  \end{center}
\end{table*}

\begin{table*}[htbp]
  \begin{center}
    \caption{Cross section per unit of rapidity of non-prompt \Jpsi as
      a function of rapidity and \pt in \pp collisions. The average
      \pt value for each bin is given. Listed uncertainties are
      statistical first, systematic second, and global scale
      third. The latter is the uncertainty on the \pp integrated
      luminosity.}
    \label{tab:nonpromptxsecpp}
    \begin{tabular}{cr@{--}l@{}c@{}r@{$\,\pm\,$}l@{$\,\pm\,$}l@{$\,\pm\,$}l}
      \hline
      $|y|$ & \multicolumn{2}{c}{$\pt$} &
      $\langle\pt\rangle$ &
      \multicolumn{4}{c}{$\frac{\mathrm{d}\sigma}{\mathrm{d}y}$}\\
      &\multicolumn{2}{c}{[\GeVc{}]}&[\GeVc{}]&\multicolumn{4}{c}{[nb]}\\\hline

      0.0--2.4                  & 6.5&30 & 9.82 & 1.60&0.16&0.35&0.10\\\hline
      \multirow{2}{*}{1.6--2.4} & 3.0&30 & 6.15 & 6.61&0.93&0.98&0.40\\
                                & 6.5&30 & 8.98 & 1.25&0.21&0.19&0.08\\\hline
    \end{tabular}
  \end{center}
\end{table*}

\begin{table*}[htbp]
  \begin{center}
    \caption{Cross section per unit of rapidity of \PgUa\ as a
      function of rapidity and \pt in \pp collisions. The average \pt
      value for each bin is given. Listed uncertainties are
      statistical first, systematic second, and global scale
      third. The latter is the uncertainty on the \pp integrated
      luminosity.}
    \label{tab:upsilonxsectpp}
    \begin{tabular}{cr@{--}l@{}c@{}r@{$\,\pm\,$}l@{$\,\pm\,$}l@{$\,\pm\,$}l}
      \hline
      $|y|$ & \multicolumn{2}{c}{$\pt$} &
      $\langle\pt\rangle$ &
      \multicolumn{4}{c}{$\frac{\mathrm{d}\sigma}{\mathrm{d}y}$}\\
      &\multicolumn{2}{c}{[\GeVc{}]}&[\GeVc{}]&\multicolumn{4}{c}{[nb]}\\\hline

      \multirow{4}{*}{0.0--2.4}  & 0&6.5  & 2.82 & 0.668&0.091&0.115&0.040 \\
                                 & 6.5&10 & 8.36 & 0.102&0.031&0.018&0.006 \\
                                 &10&20   &13.04 & 0.037&0.013&0.006&0.002 \\
                                 & 0&20   & 4.73 & 0.764&0.089&0.131&0.046 \\\hline
      0.0--1.2 & \multicolumn{2}{c}{\multirow{2}{*}{\hspace*{-0.5em}0--20}}
                                       & 5.18 & 0.921&0.128&0.157&0.055\\
      1.2--2.4 & \multicolumn{2}{c}{}  & 4.03 & 0.586&0.125&0.101&0.035\\\hline
    \end{tabular}
  \end{center}
\end{table*} 

\cleardoublepage \section{The CMS Collaboration \label{app:collab}}\begin{sloppypar}\hyphenpenalty=5000\widowpenalty=500\clubpenalty=5000\textbf{Yerevan Physics Institute,  Yerevan,  Armenia}\\*[0pt]
S.~Chatrchyan, V.~Khachatryan, A.M.~Sirunyan, A.~Tumasyan
\vskip\cmsinstskip
\textbf{Institut f\"{u}r Hochenergiephysik der OeAW,  Wien,  Austria}\\*[0pt]
W.~Adam, T.~Bergauer, M.~Dragicevic, J.~Er\"{o}, C.~Fabjan, M.~Friedl, R.~Fr\"{u}hwirth, V.M.~Ghete, J.~Hammer\cmsAuthorMark{1}, M.~Hoch, N.~H\"{o}rmann, J.~Hrubec, M.~Jeitler, W.~Kiesenhofer, M.~Krammer, D.~Liko, I.~Mikulec, M.~Pernicka$^{\textrm{\dag}}$, B.~Rahbaran, C.~Rohringer, H.~Rohringer, R.~Sch\"{o}fbeck, J.~Strauss, A.~Taurok, F.~Teischinger, P.~Wagner, W.~Waltenberger, G.~Walzel, E.~Widl, C.-E.~Wulz
% ...
\vskip\cmsinstskip
\dag:~Deceased\\
1:~~Also at CERN, European Organization for Nuclear Research, Geneva, Switzerland\\
% ...

\end{sloppypar}
\end{document}